\documentclass[sigconf]{acmart}

\usepackage[normalem]{ulem}

\usepackage[shortlabels]{enumitem}
\setlist[itemize]{leftmargin=*}
\setlist[enumerate]{leftmargin=*}

\usepackage{listings}
\usepackage{setspace}

\usepackage{amsmath,amsthm}
\usepackage{mathtools}
\usepackage{xcolor}
\usepackage{subcaption}
\usepackage{hyperref}
\newtheorem{theorem}{Theorem}[section]

\newtheorem{lemma}[theorem]{Lemma}
\newcommand{\yuliang}[1]{{\it\small\textcolor{red}{[[[ {#1}\ --yuliang]]]}}}

\renewcommand{\em}{\it}

\newcommand{\x}{$\times$}

\newcommand{\ignore}[1]{}

\def\cfigurec[#1,#2,#3]{
\begin{figure}[t]
\vspace*{-0mm}
\begin{center}
\includegraphics[width=3.5in]{#1} 
\caption[]{#2
} \label{#3}
\end{center}
\vspace*{-0.3in}
\end{figure}}

\def\cfigureb[#1,#2,#3]{
\begin{figure}
\vspace*{-0mm}
\begin{center}
\includegraphics[width=3in]{#1} 
\caption[]{#2
} \label{#3}
\end{center}
\vspace*{-8mm}
\end{figure}}

\def\cfigure[#1,#2,#3]{
\begin{figure}[t]
\vspace*{-0mm}
\begin{center}
\includegraphics[width=3.3in]{#1} 
\vspace*{-0.1in}
\caption[]{#2
} \label{#3}
\end{center}
\vspace*{-0.25in}
\end{figure}}

\def\twocfiguresh[#1,#2,#3,#4]{
\begin{figure}[!h]
\vspace*{-0mm}
\begin{center}
\includegraphics[width=3.5in]{#1} \\
(a)\\
\includegraphics[width=3.5in]{#2} \\
(b)\\
\caption[]{#3
} \label{#4}
\end{center}
\end{figure}}

\def\twocfigures[#1,#2,#3,#4]{
\begin{figure}[t]
\vspace*{-0mm}
\begin{center}
\includegraphics[width=3.5in]{#1} \\
(a)\\
\includegraphics[width=3.5in]{#2} \\
(b)\\
\caption[]{#3
} \label{#4}
\end{center}
\end{figure}}

\def\cfigurefour[#1,#2,#3]{
\begin{figure}
\vspace*{0mm}
\begin{center}

\includegraphics[width=4in]{#1} 
 
\vspace*{-3mm}\caption[]{#2
} \label{#3}
 
\vspace*{-5mm}
\end{center}
\end{figure}}

\def\cfiguretemp[#1,#2,#3]{
\begin{figure}
\vspace*{0mm}
\begin{center}

\includegraphics[width=3.5in]{#1} 
 
\vspace*{-3mm}\caption[]{#2
} \label{#3}
 
\vspace*{-5mm}
\end{center}
\vspace*{-2mm}
\end{figure}}

\def\wfigure[#1,#2,#3]{
\begin{figure*}[t]
\vspace*{0mm}
\begin{center}
 \includegraphics[width=\textwidth]{#1} 
\vspace*{-0.1in}
\caption[]{#2
} \label{#3}
 \end{center}
\vspace{-0.25in}
\end{figure*}}

\def\threefigure[#1,#2,#3,#4,#5]{
\begin{figure*}
\vspace*{0mm}
\begin{center}

\begin{tabular}{ccc}
\hspace{-0.15in}\includegraphics[width=2.48in]{#1} &
\hspace{-0.3in}\includegraphics[width=2.48in]{#2} &
\hspace{-0.3in}\includegraphics[width=2.48in]{#3} \\
(a) & (b) & (c) \\
\end{tabular}
\vspace*{-0.2in}
\caption[]{#4
} \label{#5}
\end{center}
\vspace*{-0.2in}
\end{figure*}}

\def\threefigureSC[#1,#2,#3,#4,#5]{
\begin{figure}
\vspace*{0mm}
\begin{center}

\begin{tabular}{ccc}
\hspace{-0.1in}\includegraphics[width=1.2in]{#1} &
\hspace{-0.1in}\includegraphics[width=1.2in]{#2} &
\hspace{-0.1in}\includegraphics[width=1.2in]{#3} \\
(a) & (b) & (c) \\
\end{tabular}
\caption[]{#4
} \label{#5}
\vspace*{-0.1in}
\end{center}
\vspace*{-0.2in}
\end{figure}}

\def\dcfigure[#1,#2,#3,#4,#5,#6]{
{
\begin{figure*}
\begin{center}
\begin{minipage}[c]{\columnwidth}{
\includegraphics[width=\columnwidth]{#1} 
\vspace*{0mm}\caption[]{#2} \label{#3} \
}\end{minipage}\hspace*{\columnsep}\
\begin{minipage}[c]{\columnwidth}{
\includegraphics[width=\columnwidth]{#4} 
\vspace*{0mm}\caption[]{#5}\label{#6} \
}\end{minipage}
\end{center}
\end{figure*}
}
}

\def\tableByTable[#1,#2,#3,#4,#5,#6]{
{
\begin{table*}
\begin{center}
\begin{minipage}[c]{3in}{
\centering
{#1}
\vspace*{0mm}\tabcaption[]{#2}\label{#3} \
}\end{minipage}\hspace*{\columnsep}\
\begin{minipage}[c]{3in}{
\centering
{#4}
\vspace*{0mm}\tabcaption[]{#5}\label{#6} \
}\end{minipage}
\end{center}
\end{table*}
}
}

\def\figureByTable[#1,#2,#3,#4,#5,#6]{
{
\begin{figure*}
\begin{center}
\begin{minipage}[c]{3in}{
\centering
\includegraphics[width=\textwidth]{#1}
\vspace*{0mm}\figcaption[]{#2} \label{#3} \
}\end{minipage}\hspace*{\columnsep}\
\begin{minipage}[c]{3.3in}{
\centering
{#4}
\vspace*{0mm}\tabcaption[]{#5}\label{#6} \
}\end{minipage}
\end{center}
\end{figure*}
}
}

\def\tableByFigure[#1,#2,#3,#4]{
{
\begin{figure*}[htp]
\begin{center}
\begin{minipage}[c]{2.8in}{
\begin{center}
\centering
{#1}
\end{center}
}\end{minipage}\hspace*{\columnsep}\
\begin{minipage}[c]{3.8in}{
\begin{center}
\centering
\includegraphics[width=3.8in]{#2}
\end{center}
}\end{minipage}
\end{center}
\vspace*{-0.0in}\caption[]{#3}\label{#4}
\end{figure*}
}
}

\def\doublecfigure[#1,#2,#3,#4]{
{
\begin{figure}
\begin{center}
\begin{minipage}[c]{1.5in}{
\begin{center}
\includegraphics[width=1.5in]{#1}
\end{center}
}\end{minipage}\hspace*{1em}\
\begin{minipage}[c]{1.5in}{
\begin{center}
\includegraphics[width=1.5in]{#2}
\end{center}
}\end{minipage}
\vspace*{0mm}\caption[]{#3} \label{#4} \
\end{center}
\end{figure}
}
}
\def\doublecfigurec[#1,#2,#3,#4]{
{
\begin{figure}
\begin{tabular}{cc}
\hspace{-0.1in}\includegraphics[width=1.7in]{#1} & \hspace{-0.1in}\includegraphics[width=1.7in]{#2} \\
(a) & (b) \\
\end{tabular}
\vspace*{0mm}\caption[]{#3} \label{#4} \
\end{figure}
}
}

\def\doubleunevencfigure[#1,#2,#3,#4]{
{
\begin{figure*}[t]
\begin{center}
\begin{tabular}{cc}
\includegraphics[width=2in]{#1} &
\includegraphics[width=4.5in]{#2} \\
(a) & (b) \\
\end{tabular}
\caption[]{#3} \label{#4} \
\end{center}
\end{figure*}
}
}

\def\doubleunevencfigureB[#1,#2,#3,#4]{
{
\begin{figure*}[t]
\begin{center}
\begin{tabular}{cc}
\includegraphics[width=3in]{#1} &
\includegraphics[width=3.5in]{#2} \\
(a) & (b) \\
\end{tabular}
\vspace*{0mm}\caption[]{#3} \label{#4} \
\end{center}
\end{figure*}
}
}

\def\tripleunevencfigure[#1,#2,#3,#4,#5]{
{
\begin{figure*}[t]
\begin{center}
\begin{tabular}{ccc}
\includegraphics[width=1.5in]{#1} &
\includegraphics[width=3.5in]{#2} &
\includegraphics[width=2in]{#3} \\
(a) & (b) & (c) \\
\end{tabular}
\vspace*{0mm}\caption[]{#4} \label{#5} \
\end{center}
\end{figure*}
}
}

\def\triplecfigure[#1,#2,#3,#4,#5]{
{
\begin{figure}
\begin{center}
\begin{minipage}[c]{3.5in}{
\begin{center}
\includegraphics[width=3.5in]{#1}\\(a)
\end{center}
}\end{minipage}\\
\begin{minipage}[c]{3.5in}{
\begin{center}
\includegraphics[width=3.5in]{#2}\\(b)
\end{center}
}\end{minipage}\\
\begin{minipage}[c]{3.5in}{
\begin{center}
\includegraphics[width=3.5in]{#3}\\(c)
\end{center}
}\end{minipage}
\vspace*{0mm}\caption[]{#4} \label{#5} \
\end{center}
\end{figure}
}
}

\def\qcfigure[#1,#2,#3,#4,#5,#6]{
{
\begin{figure*}
\vspace*{0.2in}\
\begin{center}
\begin{minipage}[c]{3in}{
\includegraphics[width=3in]{#1} 
\vspace*{-3mm}
}
\end{minipage}\hspace*{0.5in}\
\begin{minipage}[c]{3in}{
\includegraphics[width=3in]{#2} 
\vspace*{-3mm}
}\end{minipage}

\begin{minipage}[c]{3in}{
\includegraphics[width=3in]{#3} 
\vspace*{-3mm}
}
\end{minipage}\hspace*{0.5in}\
\begin{minipage}[c]{3in}{
\includegraphics[width=3in]{#4} 
\vspace*{-3mm}
}\end{minipage}
\end{center}
\caption[]{#5}\label{#6}
\end{figure*}
}
}

\def\twfigure[#1,#2,#3,#4,#5]{
{
\begin{figure*}
\vspace*{0.2in}\
\begin{center}
\begin{minipage}[c]{6.5in}{
\includegraphics[width=6.5in]{#1} 
\vspace*{-3mm}
}
\end{minipage}

\begin{minipage}[c]{6.5in}{
\includegraphics[width=6.5in]{#2} 
\vspace*{-3mm}
}\end{minipage}

\begin{minipage}[c]{6.5in}{
\includegraphics[width=6.5in]{#3} 
\vspace*{-3mm}
}
\end{minipage}
\end{center}
\caption[]{#4}\label{#5}
\end{figure*}
}
}

\def\dwfigure[#1,#2,#3,#4]{
{
\begin{figure*}
\vspace*{0.2in}\
\begin{center}
\begin{minipage}[c]{6.5in}{
\includegraphics[width=6.5in]{#1} 
\vspace*{-3mm}
}
\end{minipage}

\begin{minipage}[c]{6.5in}{
\includegraphics[width=6.5in]{#2} 
\vspace*{-3mm}
}\end{minipage}

\end{center}
\caption[]{#3}\label{#4}
\end{figure*}
}
}

\def\dssfigure[#1,#2,#3,#4,#5,#6]{
{
\begin{figure*}
\vspace*{0.2in}\
\begin{center}
\begin{minipage}[c]{4in}{
\includegraphics[width=4in]{#1}
\vspace*{-3mm}\caption[]{#2} \label{#3} \
}\end{minipage}\hspace*{0.5in}\
\begin{minipage}[c]{2in}{
\includegraphics[width=2in]{#4}
\vspace*{-3mm}\caption[]{#5}\label{#6} \
}\end{minipage}
\end{center}
\vspace*{-0.4in}\
\end{figure*}
}
}

\def\dsfigure[#1,#2,#3,#4,#5,#6]{
{
\begin{figure*}
\vspace*{0.2in}\
\begin{center}
\begin{minipage}[c]{3in}{
\includegraphics[width=3in]{#1}
\vspace*{-3mm}\caption[]{#2} \label{#3} \
}\end{minipage}\hspace*{0.5in}\
\begin{minipage}[c]{3in}{
\hspace*{0.5in}\
\includegraphics[height=3in]{#4}
\vspace*{-3mm}\caption[]{#5}\label{#6} \
}\end{minipage}
\end{center}
\vspace*{-0.4in}\
\end{figure*}
}
}

\def\dsyfigure[#1,#2,#3,#4,#5,#6]{
{
\begin{figure*}
\vspace*{0.2in}\
\begin{center}
\begin{minipage}[c]{2.5in}{
\includegraphics[height=2.5in]{#1}
\vspace*{-3mm}\caption[]{#2} \label{#3} \
}\end{minipage}\hspace*{0.5in}\
\begin{minipage}[c]{2.5in}{
\includegraphics[height=2.5in]{#4}
\vspace*{-3mm}\caption[]{#5}\label{#6} \
}\end{minipage}
\end{center}
\vspace*{-0.4in}\
\end{figure*}
}
}

\def\dyfigure[#1,#2,#3,#4,#5,#6]{
{
\begin{figure*}
\vspace*{0.2in}\
\begin{center}
\begin{minipage}[c]{3in}{
\includegraphics[height=3in]{#1} 
\vspace*{-3mm}\caption[]{#2} \label{#3} \
}\end{minipage}\hspace*{0.5in}\
\begin{minipage}[c]{3in}{
\includegraphics[height=3in]{#4} 
\vspace*{-3mm}\caption[]{#5}\label{#6} \
}\end{minipage}
\end{center}
\vspace*{-0.4in}\
\end{figure*}
}
}

\def\dyoldfigure[#1,#2,#3,#4,#5,#6]{
{
\begin{figure*}
\vspace*{0.2in}\
\begin{center}
\begin{minipage}[c]{3in}{
\epsfysize=2.0in\
\hspace{0.5in}\
\epsfbox{#1}
\vspace*{-3mm}\caption[]{#2} \label{#3} \
}\end{minipage}\hspace*{0.25in}\
\begin{minipage}[c]{3in}{
\epsfysize=2.0in\
\hspace{0.5in}\
\epsfbox{#4}
\vspace*{-3mm}\caption[]{#5}\label{#6} \
}\end{minipage}
\end{center}
\vspace*{-0.4in}\
\end{figure*}
}
}

\def\cfiguredouble[#1,#2,#3,#4]{
\begin{figure}
\vspace*{0.2in}\
\begin{center}
\begin{minipage}[c]{1.5in}{
\begin{center}
\epsfxsize=1.5in\
\epsfbox{#1}\\(a)
\end{center}
}
\end{minipage}
\hspace*{0.1in}\
\begin{minipage}[c]{1.5in}{
\begin{center}
\epsfxsize=1.5in\
\vspace{0.1in}\epsfbox{#2}\\(b)
\end{center}
}
\end{minipage}
\vspace*{-0.10in} \caption[]{#3}\label{#4}
\end{center}
\vspace*{-0.4in}\
\end{figure}
}

\def\wpfigure[#1,#2,#3,#4]{
\begin{figure*}
\vspace*{4mm}
\begin{center}

\includegraphics[width=#4]{#1} 

\vspace*{-3mm}\caption[]{#2
} \label{#3}

\vspace*{-5mm}
\end{center}
\end{figure*}}

\def\wprfigure[#1,#2,#3,#4,#5]{
\begin{figure*}
\vspace*{4mm}
\begin{center}

\includegraphics[width=#4, angle=#5]{#1} 

\vspace*{-3mm}\caption[]{#2
} \label{#3}

\vspace*{-5mm}
\end{center}
\end{figure*}}

\def\DoubleFigureWSlide[#1,#2,#3,#4,#5,#6,#7,#8,#9]{
\begin{figure*}
\vspace*{#9}
\begin{center}
\begin{minipage}{#4}
\includegraphics[width=#4]{#1}
\vspace*{-3mm}\caption{#2
}\label{#3}
\end{minipage}
\hspace{2em}
\begin{minipage}{#8}
\includegraphics[width=#8]{#5}
\vspace*{-3mm}\caption{#6
}\label{#7}
\end{minipage}
\vspace*{-5mm}
\end{center}
\end{figure*}
}

\def\DoubleFigureW[#1,#2,#3,#4,#5,#6,#7,#8]{
\begin{figure*}
\vspace*{0in}
\begin{center}
\begin{minipage}{#4}
\includegraphics[width=#4]{#1}
\vspace*{-3mm}\caption{#2
}\label{#3}
\end{minipage}
\hspace{2em}
\begin{minipage}{#8}
\includegraphics[width=#8]{#5}
\vspace*{-3mm}\caption{#6
}\label{#7}
\end{minipage}
\vspace*{-5mm}
\end{center}
\end{figure*}
}

\def\DoubleFigureWHack[#1,#2,#3,#4,#5,#6,#7,#8]{
\begin{figure*}
\vspace*{0in}
\begin{center}
\begin{minipage}{3in}
\includegraphics[width=#4]{#1}
\vspace*{-3mm}\caption{#2
}\label{#3}
\end{minipage}
\hspace{2em}
\begin{minipage}{3in}
\includegraphics[width=#8]{#5}
\vspace*{-3mm}\caption{#6
}\label{#7}
\end{minipage}
\vspace*{-5mm}
\end{center}
\end{figure*}
}

\def\ddcfigure[#1,#2,#3,#4]{
\begin{figure*}
\vspace*{0.2in}\
\begin{center}
\begin{minipage}[c]{\columnwidth}{\begin{center}
\includegraphics[width=\columnwidth]{#1}\\(a) \end{center}
}\end{minipage}\hspace{.2in}\
\begin{minipage}[c]{\columnwidth}{\begin{center}
\includegraphics[width=\columnwidth]{#2}\\(b)\end{center}
}\end{minipage} \caption[]{#3}\label{#4}
\end{center}
\end{figure*}
}

\def\ddcfigureSlide[#1,#2,#3,#4,#5]{
\begin{figure*}
\vspace*{#5}\
\begin{center}
\begin{minipage}[c]{3in}{
\includegraphics[height=3in]{#1} 
}\end{minipage}\hspace{0.5in}\
\begin{minipage}[c]{3in}{
\includegraphics[height=3in]{#2} 
}\end{minipage}\vspace*{-0.10in} \caption[]{#3}\label{#4}
\end{center}
\vspace*{-0.4in}\
\end{figure*}
}

\def\cxfigure[#1,#2,#3]{
\begin{figure}
\vspace*{4mm}
\begin{center}
 
\epsfxsize=2.5in\
\epsfbox{#1}\
 
\vspace*{-0.10in}\caption[]{#2
} \label{#3}
 
\vspace*{-5mm}
\end{center}
\vspace*{-2mm}
\end{figure}}

\setcopyright{acmcopyright}
\copyrightyear{2022}
\acmYear{2022}

\acmConference[To Appear: SIGMOD '22]{To Appear: SIGMOD '22: The 41st ACM SIGMOD International Conference on Management of Data}{June 12--17, 2022}{Philadelphia, PA}
\acmBooktitle{To Appear: SIGMOD '22: The 41st ACM SIGMOD International Conference on Management of Data,
	June 12--17, 2022, Philadelphia, PA}
\setcopyright{none}

\newcommand{\otto}[1]{{\color{blue}{#1}}}
\newcommand{\hungwei}[1]{{{#1}}}

\newcommand{\HDB}{HippogriffDB}
\newcommand{\YDB}{YDB}
\newcommand{\TCUDB}{TCUDB}

\newcommand{\QOB}{query-over-block}

\newcommand{\vvspace}[1]{\vspace{#1}}

\newcommand{\rev}[1]{{ #1}}

\setcopyright{none} 
\begin{document}

\title{\TCUDB{}: Accelerating Database with Tensor Processors}

\author{Yu-Ching Hu}
\affiliation{%
  \institution{University of California, Riverside}
}
  \email{yhu130@ucr.edu}
\author{Yuliang Li}
\affiliation{%
  \institution{Megagon Labs}
}
  \email{yuliang@megagon.ai}
\author{Hung-Wei Tseng}
\affiliation{%
  \institution{University of California, Riverside}
}
  \email{htseng@ucr.edu}



\begin{abstract}
The emergence of novel hardware accelerators 
has powered the tremendous growth of machine learning in recent years. 
These accelerators deliver incomparable performance gains in processing high-volume matrix operators, 
particularly matrix multiplication, a core component of neural network training and inference.
In this work, we explored opportunities of accelerating database systems using NVIDIA's Tensor Core Units (TCUs).
We present \TCUDB{}, a TCU-accelerated query engine processing a set of query operators including
natural joins and group-by aggregates as matrix operators within TCUs.
Matrix multiplication was considered inefficient in the past; however,
this strategy has remained largely unexplored in conventional GPU-based databases, which primarily rely on
vector or scalar processing.
We demonstrate the significant performance gain of \TCUDB{} in a range of real-world applications
including entity matching, graph query processing, and matrix-based data
analytics.
\TCUDB{} achieves up to 288\x{} speedup compared to a baseline GPU-based
query engine. 
%
%
\end{abstract}

\maketitle

\section{Introduction}
\label{sec:introduction}
The enormous demand for artificial intelligence (AI) and machine learning (ML) workloads
has driven the development and integration of accelerators containing
instructions operating on two-dimensional tensors (i.e., matrices). Examples include
NVIDIA's Tensor Core Units (TCUs)~\cite{markidis2018nvidia},  Google's Tensor Processing Units
(TPUs)~\cite{sato2017depth}, and Apple's Neural Processing Units (NPUs)~\cite{npu}. Improving matrix algebra through matrix units (MXUs), 
which popular AI/ML models heavily rely on, 
drastically increases the 
orders of magnitude speedup and energy
efficiency.
This is particularly true when compared with conventional scalar processors (e.g., CPUs) and 
vector processors (e.g., graphical processing units [GPUs]). 

In this work, we explore opportunities of integrating Tensor Core Units (TCUs) 
into a database engine's architecture.
Despite being originally designed for AI/ML workloads, tensor processors
also hold potential performance improvements for database engines.
This is due to both the increasing demand for native support of 
linear algebra queries (e.g., matrix multiplication itself)  in SQL 
DB engines~\cite{hellerstein2012madlib,aberger2018levelheaded,AIDA,scalaleLA,dolmatova2020relational}
and the observation that a large number of regular query operators
can be cast into matrix multiplication. For example, one can show that 
the most commonly used natural joins \cite{amossen2009faster,deep2020fast} and 
group-by aggregates can be encoded as matrix multiplication, which enables
TCUs to deliver exceptional performance.

However, the presence of these AI/ML accelerators, or more generally matrix
processors, does not provide a drop-in upgrade to the query engine's
performance. Three major challenges must be addressed.

\vspace*{1mm}
\noindent
\textbf{Challenges. } First, the conventional GPU databases primarily
implement the physical operators (e.g., the partitioned hash join algorithm \cite{kaldewey2012gpu})
in a non-matrix-friendly manner. These algorithms and operators typically do not operate on tensors directly.
As a result, it is hard to modify them with the intent of taking advantage of TCUs' computation power.

Second, although DB operators such as joins can theoretically be encoded 
as matrix multiplications, executing all of them as dense multiplication
might not always be beneficial. For example, 
the underlying data distributions can cause the two operands to be 
sparse matrices, which require a different data organization
and APIs to achieve the best performance.
\ignore{\yuliang{added the challenge for sparse MM}}

Next, a DB engine with TCUs must prevent itself from generating erroneous query results because of the low-precision nature of the tensor processors.
The current tensor processors are limited in precision as AI/ML applications are error-tolerant because NVIDIA's
TCUs only support 16-bit floating-point numbers while Google's TPUs only work
on at most 8-bit integers.
Moreover, these tensor processors share the same data movement overhead with other hardware accelerators
while additionally suffering from the data transformation overhead (i.e., table $\rightarrow$ tensor). 
A higher precision requirement means introducing more data movement and transformation overhead.
As a result, the proposed system must maintain a balance between two factors.

\vspace*{1mm}
\noindent
\textbf{\TCUDB{}. }
\wfigure[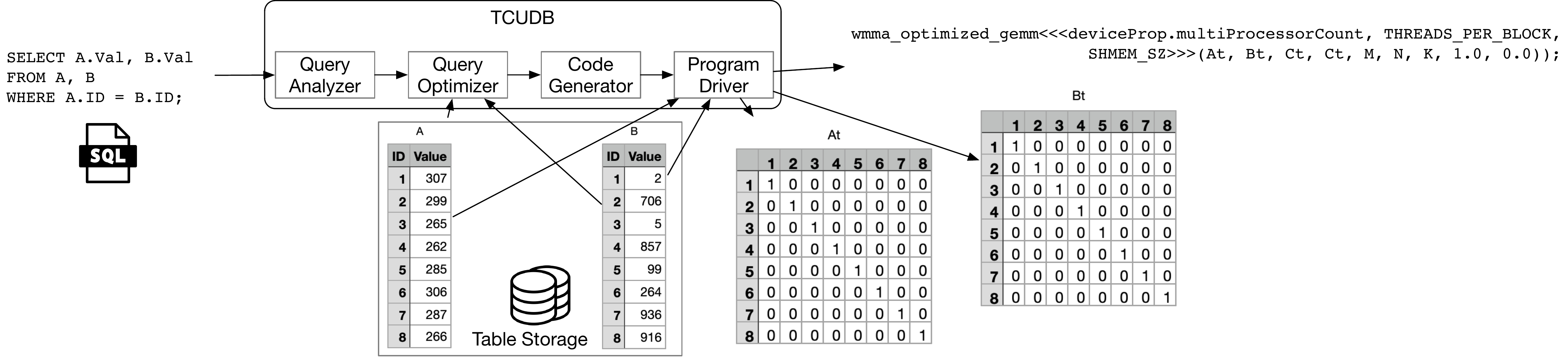,
{An overview of \TCUDB{}'s workflow.},fig:TCUDB]
This paper presents \TCUDB{}, an analytic database query engine that 
explores the potential of tensor processors to
accelerate analytic query workloads using TCUs by tackling the
aforementioned challenges. Figure~\ref{fig:TCUDB} provides an overview of
the system architecture of  \TCUDB{}. 
\TCUDB{} extends the common architecture of GPU-accelerated databases
\cite{govindaraju2004fast,walkowiak2010exploring,wu2012kernel,yuan2013yin,bress2013time,Wang:2014:CAQ:2732967.2732976,wu2014multipredicate,li2016hippogriffdb,paul2016gpl,shanbhag2020study}
as a way to further accommodate executing query operators with TCU acceleration in
the query analyzer, the query optimizer, the code generator, and the program driver.

To address the challenge of executing queries using matrix operations, we
re-engineered a set of query operators 
that are theoretically feasible to be mapped to tensor/matrix algebra operations
for \TCUDB{}. 
The query operators cover a large set of commonly used ones including natural joins and group-by aggregates.
As shown in Figure \ref{fig:TCUDB},
\TCUDB{} features a code generator for generating executable code 
mapping input tables to tensor format and processes the query as matrix multiplication via WMMA or cuBLAS API calls. Depending on the data sparsity,
\TCUDB{} provides the option of sparse tensor encoding with
sparse matrix multiplication. We developed the TCU-SpMM 
operator \ignore{via cuSPARSE API calls} to support sparse matrix multiplication with TCU acceleration.
\ignore{\yuliang{added the last two sentences}}
Then, the \TCUDB{} query analyzer is capable of generating query plans, which use these TCU-accelerated
physical operators. 

To resolve the challenge of limited precision and overhead in
modern tensor processors, \TCUDB{}'s query optimizer carefully gauges the
parameters in precision, data movement overhead, data transformation overhead, and
computation throughput --- as using lower data precision yields
lower data movement overhead and higher computation throughput, but also takes higher risks 
of leading into unacceptable answers as well as higher data transformation
overhead. \TCUDB{} presents an adaptive mixed-precision query optimization
that dynamically selects the most appropriate precision in delivering the
desired level of accuracy using the shortest end-to-end latency to handle
queries. 

\noindent
\textbf{Contributions. }
By presenting, implementing and evaluating \TCUDB{}, this paper makes the following contributions:
\begin{itemize}
\item We explored the space of opportunities of optimizing a GPU-accelerated analytic query engine by leveraging TCUs.
In our initial investigation, we found that TCU delivers >5$\times$ performance gains for 
matrix multiplication compared to the conventional CUDA cores in GPUs. 
This finding contradicts the conventional wisdom that considers
matrix multiplication a slow operator because of its high computational complexity.
As such, TCUs provide new opportunities to optimize processing analytic queries 
as matrix multiplication.
\item Next, we identified a collection of query patterns that can potentially  be accelerated by TCUs.
The query patterns include the most commonly used SQL operators in analytic queries 
such as joins and group-by aggregates (e.g., \texttt{SUM} and \texttt{COUNT}). 
We demonstrate simple algorithms for transforming relational tables into matrix format
and translating SQL operator into one or more matrix multiplication operators.
Our algorithmic design is generic as it can be generalized to multi-way joins and aggregation
over joins.
\item We designed and implemented \TCUDB{}, a TCU-accelerated analytic database engine. 
On top of a traditional GPU database,
\TCUDB{} features a query optimizer that identifies
(1) the most efficient TCU query plan and (2) the best GPU/CPU-based plan and decides which plan to execute
via cost estimation. If a TCU-accelerated plan is selected, \TCUDB{} leverages a code generator to
rewrite (parts of) the query into C programs that invoke NVIDIA's CUDA API.
To the best of our knowledge, \TCUDB{} is the first analytic database engine with TCU-accelerated built-in.
\item We evaluated \TCUDB{} on 4 real-world use cases: 
(1) linear algebra (LA) queries, 
(2) entity matching (EM),
(3) graph analytics, and
        (4) analytic queries such as the star-schema benchmark. \ignore{\yuliang{added the 4th}}
\TCUDB{} demonstrates an outstanding performance advantage over a GPU-based engine
(\YDB{}),
by achieving up to 288\x{} speedup. Our results also highlight 
the necessity of the query optimizer and \TCUDB's scalability advantage in future GPU architecture. 
\end{itemize}

\section{Background and Motivation}
\label{sec:background}

This section describes the background of the conventional query processing on a GPU and the motivation inspired by the characteristic of Tensor Core Units (TCUs). By comparing to the traditional vector processing model, we demonstrate the tensor processing model in a database system that can deliver better performance on linear algebra queries in terms of computing capability and scalability. 
\cfigure[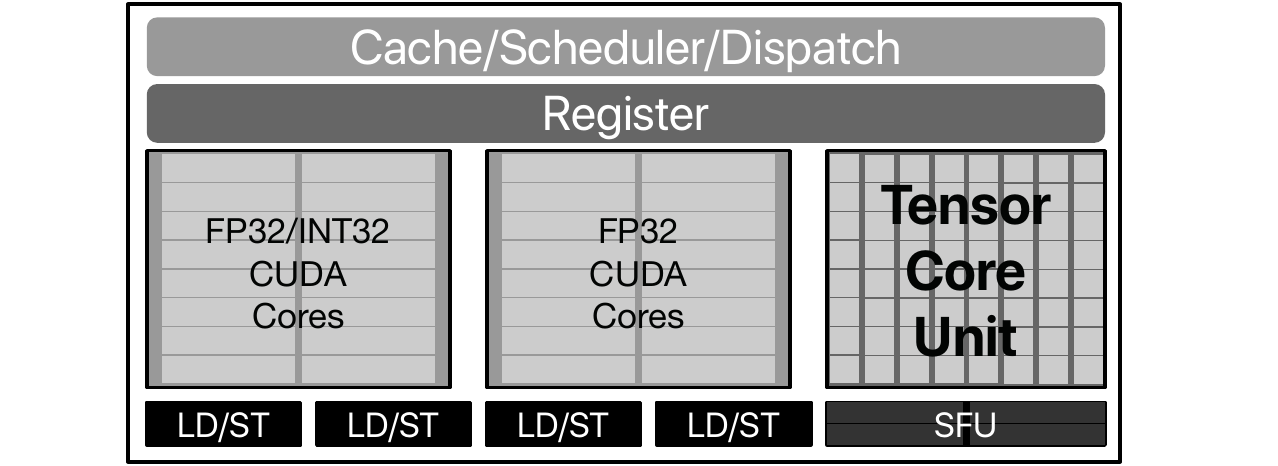,
{The GA102 Streaming Multiprocessor (SM) architecture in GeForce RTX 30-series
GPUs.},fig:GA102SM]
\ignore{
\begin{figure}
\includegraphics[width=50mm,scale=0.75]{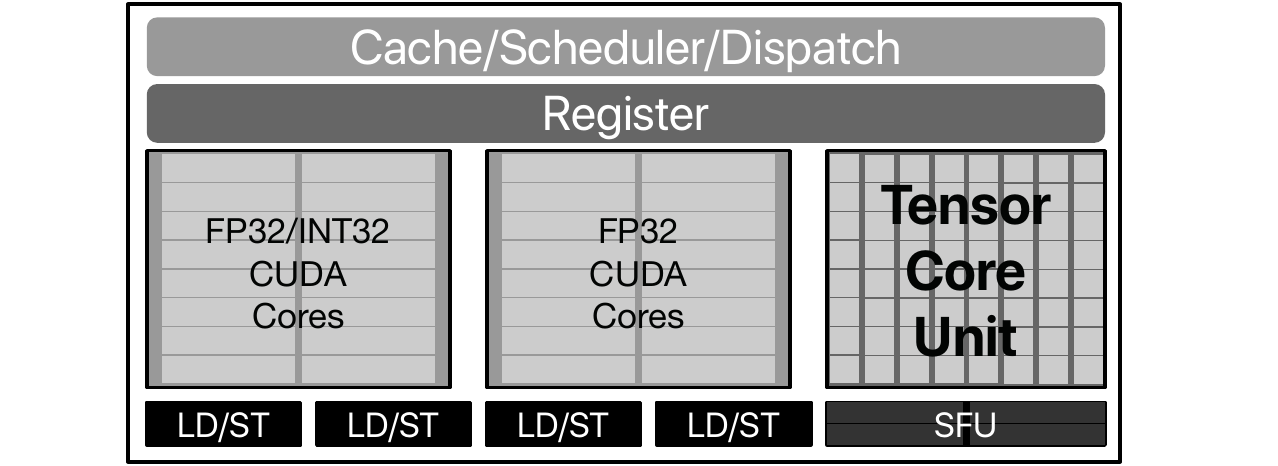}
\caption{\small{The GA102 Streaming Multiprocessor (SM) architecture in GeForce RTX 30-series GPUs.}}
\label{fig:GA102SM}
\end{figure}
}
\subsection{Tensor Core Units (TCUs)}
\label{sec:tcu_architecture}
As deep neural networks heavily rely on operations using matrix
multiplications (e.g., convolution), recent hardware accelerators feature
matrix units (MXUs) in their microarchitectures to significantly boost the
performance in machine learning (ML) workloads. Famous examples include NVIDIA's
Tensor Core Units (TCUs), Google's Tensor Processing Units (TPUs), and Apple's 
Neural Engine. 

This paper selects TCUs as the underlying accelerators for
the following reasons: (1) Programmability: TCUs expose their low-level C++
API to programmers such as highly optimized cuBLAS APIs or customizable WMMA (Warp Matrix Multiply-Accumulate) APIs, giving programmers complete freedom in implementing
algorithms and integrating with existing systems. By contrast, their counterparts
are only programmable through domain-specific languages tailored for ML. (2)
Accessibility: TCUs are now standardized components in NVIDIA's GPU
architectures, ranging from high-end server solutions, gaming solutions, to
embedded solutions. Conversely, high-performance TPUs are only accessible
through Google's cloud services and Apple's NPUs are only available on their
machines. (3) Flexibility: Tensor cores together with other ALUs on the GPU
supports multiple data precision with various operations. Other ML
accelerators only support limited precision. 
\cfigure[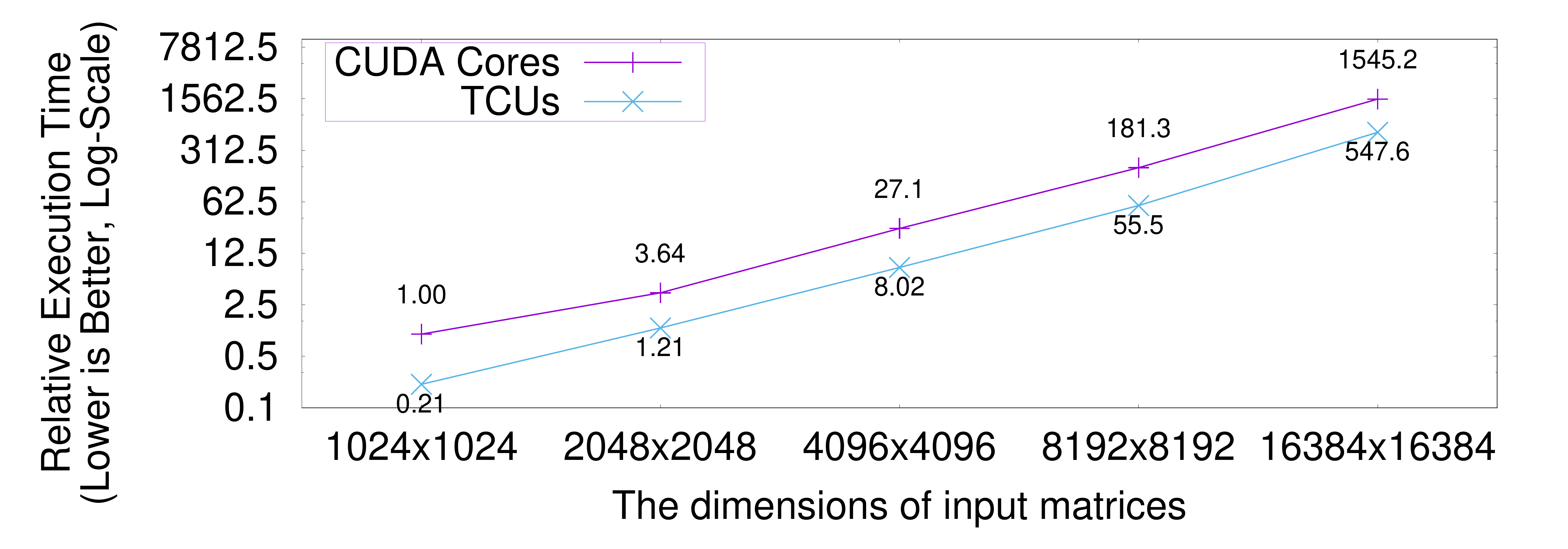,{The performance of performing matrix
multiplications using conventional CUDA cores and TCUs.},fig:matmul_comparison]

TCUs are currently available as separated functional units from 
conventional vector floating-point and integer ALUs within the current generation 
of streaming multiprocessors (SM) as Figure~\ref{fig:GA102SM} depicts. 
Figure~\ref{fig:matmul_comparison} compares the latency of multiplying
matrices with different sizes, ranging from 1024\x{}1024 inputs matrices to
16384\x{}16384 ones, using conventional vector processing
units (CUDA cores) and TCUs, on NVIDIA's RTX 3090 GPU. The results show that TCUs consistently outperform
CUDA cores by up to a 5\x{} speedup. \hungwei{By translating the latency to TFLOPs, we measured a peak 
of 63 TFLOPs on TCUs and 19 TFLOPs using mixed precision on CUDA cores only.}

Despite the significant speedup in matrix operations, TCUs still have 
limited precision drawbacks seen in other AI/ML accelerators in a way that
TCUs only support at most 16-bit numbers as inputs and incur additional
overhead in casting data into the desired 16-bit formats. 
Being separated functional
units within an SM and the nature that an SM can only perform a single type
of operations simultaneously, a compute kernel can activate either
conventional vector units or TCUs, but not both of them due to the power
constraints and the hardware architecture. Therefore, if programmers do not
specifically enable TCUs and rewrite algorithms to perform matrix
multiplications, a GPU program cannot automatically take advantage of 
TCUs. Instead, it wastes the rich speedup that the TCUs can provide.


\subsection{GPU-accelerated Database System Architecture (GPUDB)}
Prior to the introduction of TCUs in GPU architectures, database systems have
exploited the potential of using the massive amount of vector processing units
within GPUs to accelerate query processing~\cite{bress2013time, bakkum2010accelerating, yuan2013yin, Volk2010GPUBasedSQ}. 
The rich thread-level parallelism from these vector processing units
delivers better performance on easily parallelizable operations (e.g., arithmetic
computation). 
\cfigure[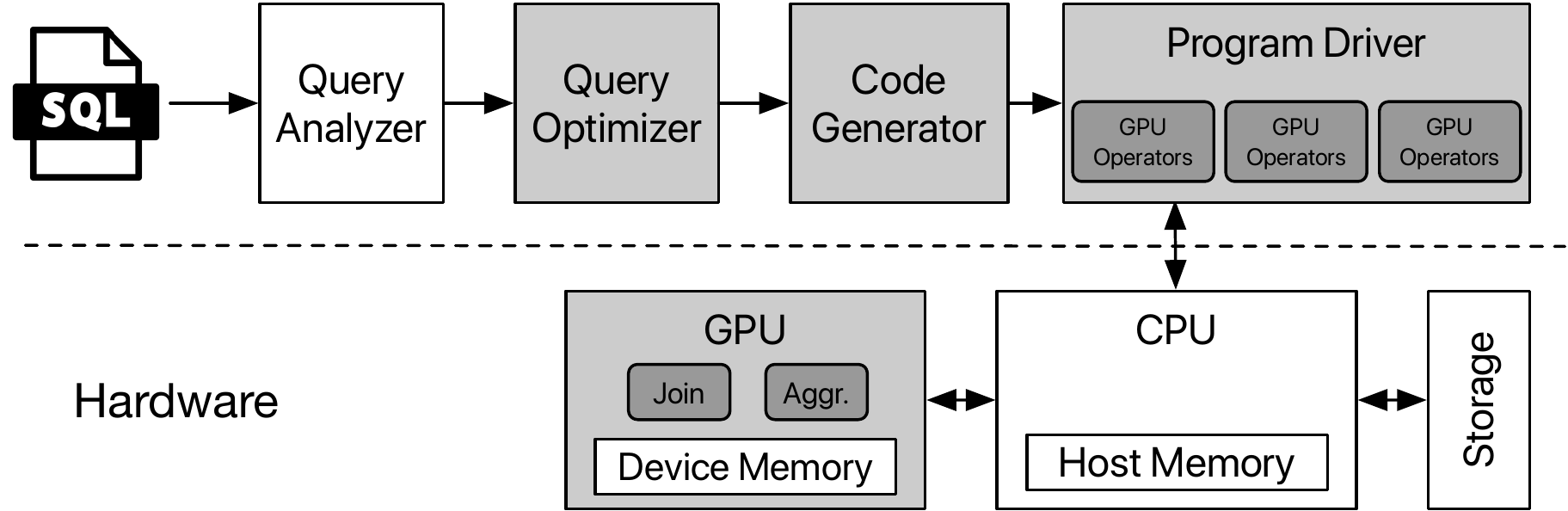,
{Typical GPU-accelerated database architecture.},fig:GPUDB_arch]
\ignore{
\begin{figure}
\includegraphics[width=50mm,scale=0.90]{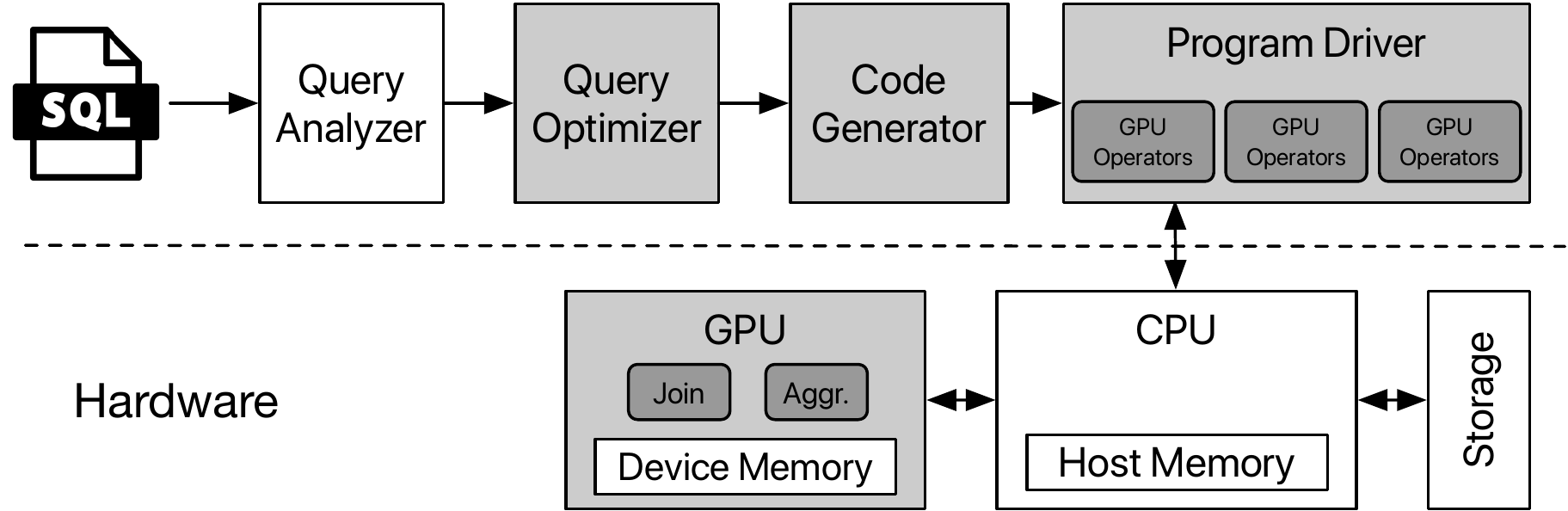}
\caption{\small{Typical GPU-accelerated database architecture.}}
\label{fig:GPUDB_arch}
\end{figure}
}
Figure~\ref{fig:GPUDB_arch} shows the architecture of a typical GPUDB system that
Yinyang DB (\YDB{})~\cite{YDB, yuan2013yin} and GPUQP~\cite{he2009coprocessing}
adopt. 
Upon receiving a query, the GPU-accelerated DB will go through the following stages:
(1) Query plan generation: the query parser translates SQL query into query plan tree and 
the query optimizer analyzes the costs and
benefits of query plans to determine the most efficient
implementation (i.e., the cheapest plan) as the physical query plan.
(2) Code generation: the query engine is in charge of the query execution flow by generating the back-end
system-level code (e.g., program driver) that maps the selected query plan to utilize CPU and GPU
cores. According to the type of target queries, different GPU kernels are implemented to execute relational database operators.
(3) Data movements: data movements involve loading table data to the host main memory from back-end storage, 
moving essential data from the host main memory 
to GPU device memory and copying results back to the host main memory.

In the aforementioned database system architecture, data movement between 
GPU and CPU usually dominates the execution time~\cite{he2009coprocessing}
and cancels out the performance gain in the computation part. Therefore, GPU database architecture 
should make full use of an in-memory technique such as keeping all tables in GPU RAM~\cite{ghodsnia2012GPURAM} 
to mitigate the I/O bottleneck. There is no common-use GPU algorithm suitable for all database systems; the 
challenge is to identify which operators can leverage the GPU and combine it with traditional database query processing.
Additionally, the data storage format also affects the performance of data
movement. Due to the GPU memory access pattern, 
column-store~\cite{abadi2009column, abadi2008column, ghodsnia2012GPURAM} 
helps to exploit coalesced memory as well as reduce data volume going through 
the PCIe bus by only sending the needed data. 
\subsection{The Missing Opportunities of GPU Databases in TCUs}
\label{sec:opportunities}
Before the emergence of TCUs, conventional wisdom assumed that matrix multiplication is an 
inefficient operation. Therefore, state-of-the-art GPUDB systems
are designed in
favor of vector processing, yet completely avoid the usage of matrix
multiplications. Without redesigning application algorithms and data layout, existing
GPUDB systems cannot reap the benefits of TCUs. 
\ignore{
As conventional computing resources are optimized for pairs of scalar and vector data, programmers tend to design/develop the system and layout data based on scalar or vector processing models to fully utilize hardware units. 

For example, the state-of-the-art GPU-aware database systems, \YDB{} (GPUDB)~\cite{\YDB{}, yuan2013yin} and GPUQP~\cite{he2009coprocessing}.
}


\begin{figure}
\vspace{4mm}
\begin{lstlisting}[
           label=lst:mm,
           language=SQL,
           showspaces=false,
           basicstyle=\ttfamily\small,
           columns=fullflexible,
  		     frame=single,
  		     breaklines=true,
           numbers=left,
           numberstyle=\tiny,
           commentstyle=\color{gray}
        ]
-- Matrix multiplication query:
SELECT A.col_num, B.row_num, SUM(A.val * B.val) as res
FROM A, B
WHERE A.row_num = B.col_num
GROUP BY A.col_num, B.row_num;
\end{lstlisting}
\caption{Example matrix multiplication query.}
\vspace{-0.2in}
\label{fig:ex_mmquery}
\end{figure}

The query in Figure~\ref{fig:ex_mmquery} provides an example of how
an existing GPUDB misses the potential of using TCUs. 
The result of this query is essentially a list of triples of ($row\_num$, $col\_num$,
$val$) with unique combinations of $row\_num$, $col\_num$ and the $val$ in
each triple is the sum of the pairwise multiplications on $val$ fields from
a record in table $A$ with its $row\_num$ matching another record's 
$col\_num$ from table $B$. 
This is essentially an SQL query that performs matrix multiplication on elements from 
two tables $A$ and $B$. 
This query can be implemented through one matrix multiplication if we can
layout the matching elements in matrices appropriately. 

However, conventional GPUDB query processing algorithms are designed at the operator level
with each operator as a kernel function running on GPUs. To execute the above 
query, conventional GPUDB uses operators to build hash tables for $A$ and $B$, scanning 
both tables, performing $\mathtt{HashJoin}$, and aggregating the final result. Among these 
GPU operators, $\mathtt{HashJoin}$ where performs join operation in a
pairwise, vectorized fashion to find matching 
tuples between two hash tables usually takes the most time during the query execution. 
The aggregation operator is second to $\mathtt{HashJoin}$, which is also
time-consuming in accumulating the computation result using vector
operations. As the above computation
only requires vector inner-products, the generated GPU kernel code will
never enable TCUs. 

\section{TCU-accelerated query patterns}
\label{sec:theory}


As mentioned above, TCUs can potentially improve the performance of an analytic query by executing (parts of)
the query as matrix multiplication. Next, to achieve this goal, 
we start by identifying a number of query patterns that \TCUDB{} can execute as matrix
multiplications. 





\subsection{Two-way natural join}
\label{sec:2wayjoin}
The first supported query pattern is the simple 2-way join. For example,
given two tables $\mathtt{A}$ and $\mathtt{B}$ with two attributes $(\mathtt{ID}, \mathtt{Val})$,
consider the following query:
\begin{lstlisting}[
        	language=SQL,
        	showspaces=false,
        	basicstyle=\ttfamily\small,
        	columns=fullflexible,
  			frame=single,
  			breaklines=true,
        	numbers=left,
            numberstyle=\tiny,
        	commentstyle=\color{gray}
        ]
-- Q1:
SELECT A.Val, B.Val 
FROM A, B
WHERE A.ID = B.ID;
\end{lstlisting}
To process this query as a matrix operation, we first need to convert the two tables into a matrix format.
Suppose table $\mathtt{A}$ contains $n$ tuples $\{a_1, \dots, a_n\}$ and table $\mathtt{B}$ contains $m$
tuples $\{b_1, \dots, b_m\}$ where each $a_i$ and $b_i$ are unique row IDs. Let $\mathtt{dom}(\mathtt{A.ID})$
and $\mathtt{dom}(\mathtt{B.ID})$ be the domains of the \texttt{ID} column of $\mathtt{A}$ and $\mathtt{B}$
respectively. Let $\mathtt{dom(ID)}$ to be the union of the two domains $\mathtt{dom(A.ID)} \cup \mathtt{dom(B.ID)}$
having $k$ distinct values $\{v_1, \dots, v_k\}$. 
To compute the join, we construct a $n \times k$ matrix $\mathtt{mat(A)}$ and a $m \times k$ matrix $\mathtt{mat(B)}$
where 
\begin{align*}
\mathtt{mat(A)}_{ij} = 1 \text{ if } a_i.\mathtt{ID}=v_j \text{, otherwise 0 } & ; \\
\mathtt{mat(B)}_{ij} = 1 \text{ if } b_i.\mathtt{ID}=v_j \text{, otherwise 0 } & .
\end{align*}
The result of the join $\mathtt{A} \bowtie \mathtt{B}$ is then the $n$ by $m$ matrix 
$$\mathtt{C} = \mathtt{mat(A)} \times \mathtt{mat(B)^T} . $$
It is easy to show that a tuple $(a_i, b_j)$ is in the join result if and only if $\mathtt{C}_{ij} > 0$.

Alternatively, when the domains $\mathtt{dom(A.Val)}$ and \hungwei{$\mathtt{dom(B.Val)}$} are small, one can also construct 
$\mathtt{mat(A)}$ and $\mathtt{mat(B)}$ as the adjacency matrices
where $\mathtt{mat(A)}_{ij} = 1$ if $(u_i, v_j) \in \mathtt{A}$ (and respectively for $\mathtt{mat(B)}$) otherwise 0.
The number of rows of $\mathtt{mat(A)}$ and $\mathtt{mat(B)}$ will be $|\mathtt{dom(A.Val)}|$ and 
 $|\mathtt{dom(B.Val)}|$ respectively. 

Note that in this query pattern, the single attributes $\mathtt{A.ID}$, $\mathtt{A.Val}$, $\mathtt{B.ID}$
and $\mathtt{B.Val}$ can be generalized to sets of multiple attributes. 
The attribute sets $\mathtt{*.ID}$ and $\mathtt{*.Val}$
can potentially overlap thus it is general enough to cover all cases of 2-way natural join.



\subsection{Multi-way joins}

Next, we extend the querying capability with matrix multiplication
to multi-way joins. Consider the following snippet of a 3-way join query
where the 3 input tables are $\mathtt{A(ID_1, Val)}$, $\mathtt{B(ID_1, ID_2, Val)}$, and 
$\mathtt{C(ID_2, Val)}$ respectively.
\begin{lstlisting}[
        	language=SQL,
        	showspaces=false,
        	basicstyle=\ttfamily\small,
        	columns=fullflexible,
  			frame=single,
  			breaklines=true,
        	numbers=left,
            numberstyle=\tiny,
        	commentstyle=\color{gray}
        ]
-- Q2:
SELECT A.Val, B.Val, C.Val
FROM A, B, C
WHERE A.ID_1 = B.ID_1 AND B.ID_2 = C.ID_2;
\end{lstlisting}

As in conventional join processing, we assume a join order of $\mathtt{A} \rightarrow \mathtt{B} \rightarrow \mathtt{C}$.
To evaluate this join, one needs to (1) first compute $\mathtt{A} \bowtie \mathtt{B}$ as
$\mathtt{mat(A)} \times \mathtt{mat(B)^T}$, (2) convert the resulting $n$ by $m$ matrix back to table format and 
(3) compute the join with table $\mathtt{C}$ as a second matrix operator. 
By repeating step (2) and (3) to convert intermediate results to tables,
we can generalize this algorithm from 3-way joins to multi-way joins.

To avoid unnecessary data transfer from GPU memory to the host, in step (2),
one can perform the matrix-table conversion
with a CUDA-enabled $\mathtt{nonzero}(\cdot)$ operator~\cite{pytorch}. Formally, given a matrix $M$, 
$\mathtt{nonzero}(M)$ computes $\{(i, j) | M_{ij} > 0\}$. Next, to perform the second join,
let 
\begin{itemize}
\item $n'$ be the size of $\mathtt{nz = nonzero(mat(A) \times mat(B)^T)}$, 
\item $m'$ be the size of table $\mathtt{C} = \{c_1, \dots, c_{m'}\}$ and
\item $k'$ be the size of $\mathtt{dom(B.ID_2) \cup dom(C.ID_2)} = \{u_1, \dots, u_{k'}\}$.
\end{itemize}
We denote by $\mathtt{nz}_i$ the $i$-th pair of the $\mathtt{nz}$ array.
Next, we construct a $n'$ by $k'$ matrix
$\mathtt{mat(AB)}$ and a $m'$ by $k'$ matrix $\mathtt{mat(C)}$ where 
\begin{align*}
\mathtt{mat(AB)}_{ij} &= 1 \text{ if } b_{i'}.\mathtt{ID_2} = u_j \text{ for } \mathtt{nz}_i = (\_, i') \text{, otherwise 0} ; \\
\mathtt{mat(C)}_{ij} &= 1 \text{ if } c_i.\mathtt{ID_2} = u_j \text{, otherwise 0. }
\end{align*}
The result of the 3-way join is then $\mathtt{mat(AB) \times \mathtt{mat(C)^T}}$.

There is an exception case where the intermediate matrix-table conversion can be omitted. 
When $\mathtt{B.Val} = \emptyset$ (i.e., relation $\mathtt{B}$ is projected out entirely),
the result of the join can be simplified as 
$$ \mathtt{mat(A) \times mat(B)^T \times mat(C)^T }$$
where $\mathtt{mat(B)}$ is a $k$ by $k'$ matrix constructed as 
$\mathtt{B}_{ij} = 1$ if $(v_i, u_j) \in \mathtt{B}$ otherwise 0.

Similar to the 2-way join case, the method can be generalized to multi-way joins 
consisting of multiple join and/or return attributes.





\subsection{Group-by aggregates over joins}
\label{sec:gb_agg}
A simple yet useful extension of the above
two query patterns with joins is to add group-by aggregates. For example, 
over the same schema $\mathtt{(ID, Val)}$ of the previous 2-way join case:
\begin{lstlisting}[
        	language=SQL,
        	showspaces=false,
        	basicstyle=\ttfamily\small,
        	columns=fullflexible,
  			frame=single,
  			breaklines=true,
        	numbers=left,
            numberstyle=\tiny,
        	commentstyle=\color{gray}
        ]
-- Q3:
SELECT SUM(A.Val), B.Val
FROM A, B
WHERE A.ID = B.ID
GROUP BY B.Val;
\end{lstlisting}

A naive method to evaluate this query is to first evaluate the natural join in the TCU-optimized manner,
convert the matrix result to the table format, and then compute the group-by and SUM aggregate with CPU or GPU-based methods.
We propose the following method that avoids any unnecessary intermediate computation via 2 matrix operations.
First, we construct the two input matrices.
For the matrix dimensions, we let 
\begin{itemize}
\item $n$ be the size of $\mathtt{A}$, 
\item $m$ be the size of $\mathtt{dom(B.Val)} = \{u_1, \dots, u_m\}$, and 
\item $k$ be the size of $\mathtt{dom(A.ID)} \cup \mathtt{dom(B.ID)} = \{v_1, \dots, v_k\}$. 
\end{itemize}
We construct a $n$ by $k$ matrix $\mathtt{mat(A)}$ and a $m$ by $k$ matrix where
\begin{align*}
\mathtt{mat(A)}_{ij} &= a_i.\mathtt{Val} \text{ if } a_i.\mathtt{ID} = v_j  \text{, otherwise 0} ; \\
\mathtt{mat(B)}_{ij} &= 1 \text{ if } (u_i, v_j) \in \mathtt{B} \text{, otherwise 0. }
\end{align*}

Next, the query result can be computed as
$$ \mathbf{1}^{1 \times n} \times \mathtt{mat(A)} \times \mathtt{mat(B)^T}  $$
where $\mathbf{1}^{1 \times n} $ is an $1 \times n$ matrix consisting of only ones.
We can show the following:
\begin{lemma}{(Q3, informal)}
For every tuple $(a_i^{\mathtt{sum}}, b_i)$ and for
    $ M = \mathbf{1}^{1 \times n} \times \mathtt{mat(A)} \times \mathtt{mat(B)^T}$,
$(a_i^{\mathtt{sum}}, b_i)$ is in the query result of Q3 if and only if
$M_{i,1} = a_i^{\mathtt{sum}}$.
\end{lemma}

Intuitively, we leverage the first multiplication with $\mathtt{mat(B)^T}$ to compute the join.
By filling the input matrices $\mathtt{mat(A)}$ with actual values instead of 0's or 1's,
we keep track of those values in the intermediate matrix product $\mathtt{mat(A)} \times \mathtt{mat(B)^T}$. 
The multiplication with $\mathbf{1}^{1 \times n}$
then serves as a reduction operator that sums up all columns of $\mathtt{mat(A)} \times \mathtt{mat(B)^T}$.

In addition to $\mathtt{SUM}$, we are able to apply the same method to support the $\mathtt{COUNT}$ and $\mathtt{AVG}$
aggregate functions. For $\mathtt{COUNT}$, when we construct $\mathtt{mat(A)}$, we simply need to set 
$\mathtt{mat(A)}_{ij}$ to 1 for $a_i.\mathtt{ID} = v_j$ (instead of $a_i.\mathtt{Val}$). We can obtain $\mathtt{AVG}$
by dividing $\mathtt{SUM}$ by $\mathtt{COUNT}$.

For aggregate queries without \texttt{GROUP BY}, such as
\begin{lstlisting}[
        	language=SQL,
        	showspaces=false,
        	basicstyle=\ttfamily\small,
        	columns=fullflexible,
  			frame=single,
  			breaklines=true,
        	numbers=left,
            numberstyle=\tiny,
        	commentstyle=\color{gray}
        ]
-- Q4:
SELECT SUM(A.Val * B.Val)
FROM A, B
WHERE A.ID = B.ID;
\end{lstlisting}
we set $\mathtt{mat(A)}_{ij} = a_i.\mathtt{Val}$ for $a_i.\mathtt{ID} = v_j$ and 
$\mathtt{mat(B)}_{ij} = b_i.\mathtt{Val}$ for $b_i.\mathtt{ID} = v_j$ and compute the sum as
$ \mathtt{mat(A)} \times \mathtt{mat(B)^T} \times \mathbf{1}^{m \times 1} $
with an additional reduction by multiplying $\mathbf{1}^{1 \times n}$.

\subsection{Other supported operators}

The above query patterns can also be extended with the \texttt{ORDER BY} clause to sort the results in 
ASC/DESC order by a certain column. Instead of sorting after the multiplication operators,
we preserved the specified order in the input matrices (e.g., $\mathtt{mat(A)}$ and $\mathtt{mat(B)}$)
so that the result matrix is naturally sorted.

Another class of supported query pattern is the non-equi join such as:
\begin{lstlisting}[
        	language=SQL,
        	showspaces=false,
        	basicstyle=\ttfamily\small,
        	columns=fullflexible,
  			frame=single,
  			breaklines=true,
        	numbers=left,
            numberstyle=\tiny,
        	commentstyle=\color{gray}
        ]
-- Q5:
SELECT A.Val, B.Val 
FROM A, B
WHERE A.ID < B.ID;
\end{lstlisting}
We can compute this query by slightly adjusting the translation for Q1 by setting
$\mathtt{mat(A)}_{ij} = 1$ for $a_i.\mathtt{Val} < v_j$. The same method applies to the other comparison operators
$\{<, >, \leq, \geq, \neq\}$.

Last but not least, for the query pattern that is of the semantics of matrix multiplication 
as Figure \ref{fig:ex_mmquery} shows, we can directly map the query to the corresponding matrix operation. 

\smallskip
\noindent
\hungwei{
\textbf{Beyond the supported patterns. } For queries that do not match exactly with any of the supported query patterns,
as part of the query optimization workflow (Figure \ref{fig:tcudb_query_optimizer}),
\TCUDB{} relies on pattern matching to identify subqueries 
that can be TCU-accelerated from the input query's AST. 
We note that there are common subqueries that are beyond the expressiveness of the TCU platform,
such as aggregation with MIN/MAX or arithmetic operators such as addition and division.
The limited expressiveness is mainly due to  NVIDIA's current TCU programming interface which only supports matrix multiply-accumulate.
However, since the underlying hardware is powerful enough to perform the aforementioned operators,
we anticipate a more flexible programming interface in the future so that \TCUDB{} can support a wide range of query patterns.
}

\section{\TCUDB{}: A TCU-Accelerated DB Engine}\label{sec:tcudb}

To leverage TCUs for queries in relational database systems, this paper presents \TCUDB{}, a DB engine 
that identifies, optimizes, evaluates and implements aforementioned query patterns in
Section~\ref{sec:theory}. This section provides an overview of the design of
\TCUDB{}'s extensions and discusses the optimizations on a TCU-accelerated
query plan.

\subsection{Overview}
\label{sec:overview}
\TCUDB{} implements the system architecture in Figure~\ref{fig:TCUDB} 
to execute queries on TCUs using the following major components.

\smallskip
\noindent\textbf{Query Optimizer}
In a system with TCUs presented, the query plan \ignore{\yuliang{remember to drop ``optimal''}} in
exercising a query is from either (1) the most efficient TCU-accelerated query 
plan or (2) the most efficient conventional CPU/GPU-based plan, depending on 
which one can deliver the lowest cost (i.e., the shortest end-to-end latency). 
\TCUDB{} leverages existing infrastructure in GPUDB to evaluate the second option 
but extends the query optimizer in creating, optimizing and evaluating the latency 
of TCU-accelerated query plans. 

\smallskip
\noindent\textbf{Program Driver}
\TCUDB{} extends the program driver to additionally contain a set of library functions 
that implement operators mentioned in Section~\ref{sec:theory} using TCUs.
These functions invoke NVIDIA's CUDA C++ Warp Matrix Multiply and Accumulate (WMMA)
or cuBLAS API functions to achieve the series of computation that each operator
requires. These operators also present interfaces in various data types to
support the demand for the most efficient query plan. 

\smallskip
\noindent\textbf{Code Generator}
If \TCUDB{} selects a TCU-accelerated query plan to exercise an incoming
query, the code generator will rewrite the query as C code and dynamically
compile the code to execute the selected query plan. 
The \TCUDB{} code extension is responsible for creating the input matrices,
calling operator functions in corresponding data types and remapping the output 
from the operator outcome. 

Among these three intensively extended modules, the query optimizer is the most
critical component as it serves as the core controlling the use of TCUs as well as
 code generation for queries. In the rest of this section, we will focus on
\ignore{the discussion of our design and optimizations} the query optimizer. 

\subsection{\TCUDB{} query optimizer}
\label{sec:query_optimizer}

\begin{figure}[!ht]
\includegraphics[width=0.48\textwidth]{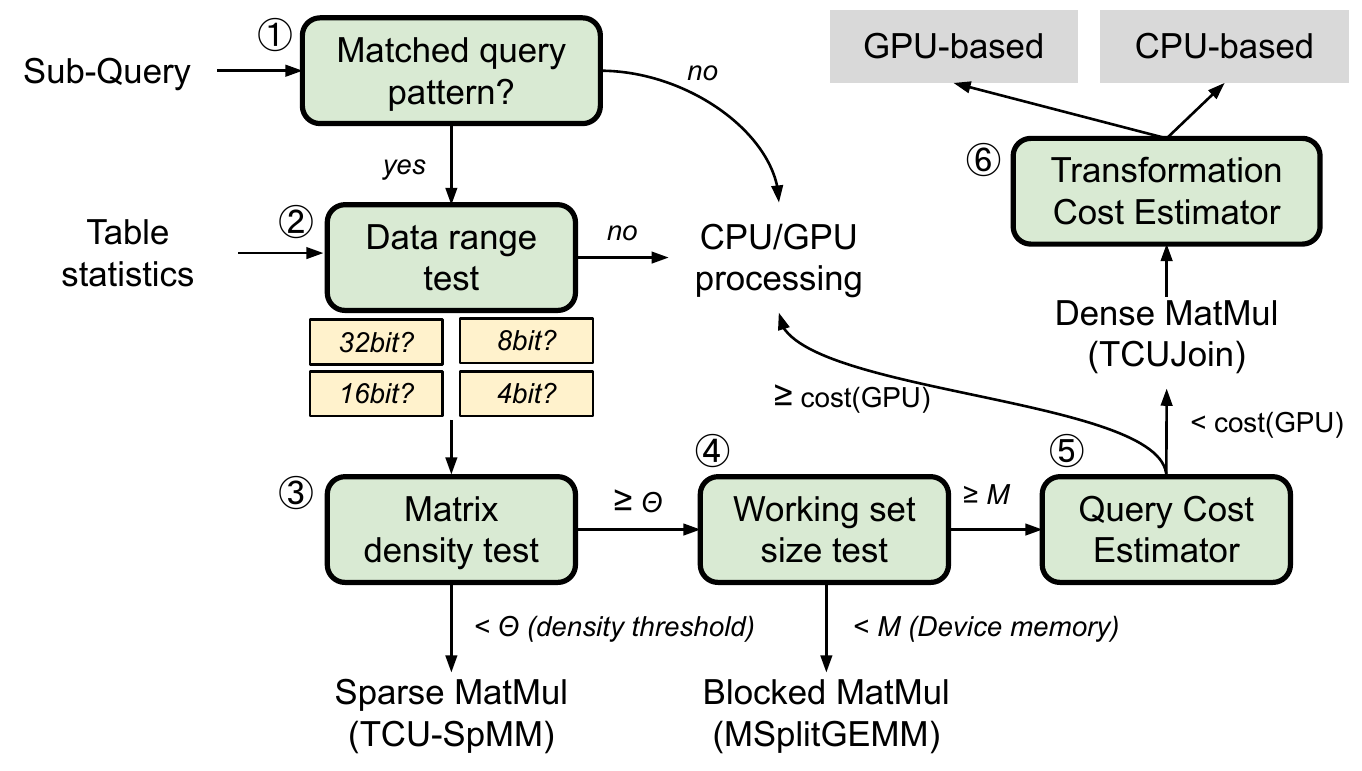}
\caption{The workflow of the \TCUDB{} query optimizer.}
\label{fig:tcudb_query_optimizer}
\end{figure}

\hungwei{
Figure \ref{fig:tcudb_query_optimizer} shows the workflow 
of the \TCUDB{} query optimizer.
The optimizer takes a subquery from the query AST as input
and performs  a series of tests to 
determine whether the subquery should be executed with TCU and how.
The optimizer first checks if the subquery falls in one of the supported query patterns.
Next, it performs the data range feasibility test (Section \ref{sec:test})
to decide if particular data types can provide sufficient precision to the query.
After that, the input tables may also result in matrices too large to fit in the GPU's device memory
or sparse matrices for which dense multiplication algorithm is sub-optimal.
For these cases, the optimizer estimates the working set sizes and matrix density
from statistics pre-computed from input tables.
\TCUDB{} applies blocked matrix multiplication (MSplitGEMM, Section \ref{sec:msplitGemm}) and sparse matrix multiplication (TCU-SpMM, Section \ref{sec:sparseMM}) 
respectively.
Finally, the optimizer estimates the query execution cost with TCU
and tests whether the cost is lower than the estimated cost with CPU/GPU 
(Section \ref{sec:cost}).
If any of the tests fail, \TCUDB{} falls back to the standard CPU or GPU-based
query execution.

Note that the query cost estimator needs to take into account the 
data transformation cost which consists of both computation and data movement overhead. 
If the original table size plus the working set size fits in the device memory,
\TCUDB{} can transform tables into matrix format within GPU to save the overhead
of transforming data within CPU and moving large matrices into the GPU device.
}

\subsubsection{Feasibility Test}
\label{sec:test}
Even though a query contains patterns matching identified patterns in
Section~\ref{sec:theory}, a query may still be unfeasible for TCUs due to
the limitations of TCUs in input precision and data types. If applying TCUs would result in
loss of precision or lead to unwanted outcomes, \TCUDB{} should not use TCUs to evaluate 
the incoming query. 
\ignore{
two critical limitations of the current TCU architectures. First, as TCUs are
limited in input precision and data types, if applying TCUs would result in
loss of precision or lead to inaccurate outcomes, \TCUDB{} should be use TCUs to evaluate the incoming query. 
Second, as TCUs can only access a GPU's device memory, the TCU-accelerated physical operators 
will not perform efficiently if the total size of
input/output matrices surpasses the available device memory. 
}

Therefore, \TCUDB{} must perform a feasibility test for each query that contains
qualified patterns by evaluating the input data ranges, identifying the most
compact inputs/outputs data types and estimating the working set
sizes for operators within a query. 
To facilitate this process, \TCUDB{} adds
metadata to each database table to contain three values for each column,
including (1) the minimum value, (2) the maximum value, and
(3) the number of distinct values. 

If the operator works with the numerical computation on the input data values
directly, \TCUDB{} first uses the minimum and maximum values along with the raw data types of the
operator's input data. If the input data can be represented by 
TCU-compatible data types, including 16-bit half floating-point (\texttt{half}), 
8-bit integers (\texttt{int8}), and 4-bit integers (\texttt{int4}), this
stage will also determine the most compact data type. \hungwei{However, if
the dataset cannot leverage any TCU-compatible data type, the feasibility test
will suggest that the system not use TCUs in the incoming query. The database
system can use other available options (e.g., a CPU-based or a pure GPU-based query engine)
instead.}

\hungwei{
The number of records, the number of distinct values and the maximum/minimum
values of each column also help the feasibility test to identify the case where the result
value can surpass the range of 16-bit numbers and potentially lead to
errors. Let $m_{1}$ represents the maximum of the maximum value and the
absolute value of the minimum value within a column of $n$ elements in one
of the input matrix and that of a row with $n$ elements is $m_{2}$ for another input matrix, the feasibility test
can conservatively estimate the maximum value in the resulting matrix as $m_{1} \times m_{2}
\times n$. If the maximum result value falls beyond the range of TCUs 16-bit
number ranges, \TCUDB{} will use query executors based on other hardware components
instead. }

\subsubsection{Cost estimation of query plans}
\label{sec:cost}

The cost of a TCU-accelerated operator contains:\\
\ignore{
\vspace*{-\lineskip}
\begin{itemize}
\item[(1)] The data transformation cost ($\mathtt{DT\_{op}}$) which equals  
the latency for creating input matrices to perform the TCU-accelerated operators from the input tables,
\item[(2)] the data movement overhead ($\mathtt{DM\_{op}}$) for copying data between the host main memory or 
data storage to the GPU's device memory, and 
\item[(3)] the computation time ($\mathtt{CT\_{op}}$), the actual running time that the TCUs spend on executing the generated TCU code. 
\end{itemize}
\vspace*{-\lineskip}
}
(1) the data transformation cost $\mathtt{DT\_{op}}$ which equals  
the latency for creating input matrices to perform the TCU-accelerated operators from the input
tables,\\
(2) the data movement overhead $\mathtt{DM\_{op}}$ for copying data between the host main memory or 
data storage to the GPU's device memory, and \\
(3) the computation time $\mathtt{CT\_{op}}$, the actual running time that the TCUs spend on executing the generated TCU code. 

\hungwei{Depending on the estimated working set size of the query, the data transformation
process of \TCUDB{} can take place using the CPU or the GPU. The costs of 
$\mathtt{DT\_{op}}$ and $\mathtt{DM\_{op}}$ vary according to the approach.

\paragraph{CPU-based data transformation}
The most general data transformation approach in \TCUDB{} uses the host main
memory and CPU to prepare inputs for the designated TCU-accelerated
operator. This approach fills input matrices for a TCU-accelerated operator
using methods described in Section~\ref{sec:theory} and works regardless of the
estimated working set size of the query. }

Consider the example of the 2-way natural join.
To create the input matrices for an operator, \TCUDB{} typically needs to
scan through qualified/valid records for the operator and convert the values
into the desired matrix representations. 
The data transformation cost is linear to the number of qualified/valid records. 
Let $A$ and $B$ be two input tables (which can also be intermediate results from subqueries)
of size $m$ and $n$ respectively.
Assume the throughput of the host system in scanning the raw data is a constant $\alpha$.
If their matrix representations $\mathtt{mat(A)}$ and $\mathtt{mat(B)}$ are not yet created, the scan operator
will take $\mathtt{DT\_{op}} \approx \alpha \cdot (m + n)$ in transforming input data to the desired
matrices. The cost can also be $\alpha \cdot m$ or $\alpha \cdot n$ if either matrix is already created.


In this approach, the data 
movement overhead is controlled by (1) the volume of transformed matrices or input data 
and (2) the available bandwidth between the GPU and the host processor
denoted by $\mathtt{Bandwidth_{GPU/host}}$. 
If $A$ is of dimension $M\times K$ with $\mathtt{type\_A}$ and
$B$ is of dimension $K \times N$ with $\mathtt{type\_B}$, the data movement cost can be estimated by
\begin{equation}
\label{eq:move}
\mathtt{DM\_{op}} \approx  \dfrac{MK \cdot \mathtt{sizeof(type\_A)} + NK \cdot
\mathtt{sizeof(type\_B)}}{\mathtt{Bandwidth_{GPU/host}}} .
\end{equation}

\paragraph{GPU-assisted data transformation}
To optimize the data transformation overhead $\mathtt{DT\_{op}}$, 
the query plan may perform the data transformation on the GPU to leverage its massive parallelism 
to convert thousands of pairs of values simultaneously \hungwei{into matrix format. In other words, 
we can take advantage of the GPU's parallelism to speed up the data transformation operation as well 
as avoid the additional data movement that copies the transformed matrix from the host memory to the 
GPU device memory.} \hungwei{In contrast to the CPU-based approach,
the data movement occurs before the data transformation in the GPU-assisted approach
as the raw data must be present in the GPU's device memory in advance for the
transformation to begin. Therefore, \TCUDB{} can only use GPU-assisted data
transformation when both the estimated working set size and the volume of necessary
raw data (e.g., columns from the selected table) for transformation can fit
in GPU's device memory. Leveraging the same 2-way natural join example, 
\TCUDB{} can estimate the corresponding $\mathtt{DM\_{op}}$ using 
Equation~\ref{eq:move_gpu} as:}
\begin{equation}
\label{eq:move_gpu}
\mathtt{DM\_{op}} \approx \dfrac{M \cdot \mathtt{sizeof(type\_A)} + N \cdot
\mathtt{sizeof(type\_B)}}{\mathtt{Bandwidth_{GPU/host}}} .
\end{equation}
\hungwei{
where $M$ and $N$ are the numbers of elements in the raw data columns of the
joined columns and $\mathtt{(type\_A)}$ and $\mathtt{(type\_B)}$
are the raw data types of both columns before transformation. 

In terms of $\mathtt{DT\_{op}}$, the GPU-based scan operator still takes 
$\approx \alpha \cdot (m + n)$ operations in transforming input data to the desired
matrices -- but a GPU can perform $\mathtt{p}$ of these in parallel if the
GPU has $\mathtt{p}$ vector processors available. In modern GPU
architectures, $\mathtt{p}$ is typically more than 2,000. The $\mathtt{DT\_{op}}$
in GPU-assisted approach is estimated as $\mathtt{DT\_{op}} \approx \frac{\alpha \cdot (m +
n)}{p}$. Notice that the GPU-based approach needs to move raw data in Equation~\ref{eq:move_gpu},
\TCUDB{} still needs to evaluate the summation of $\mathtt{DM\_{op}}$ and $\mathtt{DT\_{op}}$
to determine the most appropriate data transformation method. 
} 
\ignore{
\otto{On the other hand}, moving the data transformation process to the GPU also
results in two negative effects on data movement overhead $\mathtt{DM\_{op}}$. First,
it requires the system to move raw data into the GPU. As the raw data are
usually represented with more precise data types than \texttt{half} and
contain more records than the filtered ones, the data volume can potentially
increase and make $\mathtt{DM\_{op}}$ longer. Second, 
 this approach increases the pressure
of GPU device memory and leads to additional data swap ins/outs if the total
volume of the raw data and transformed data is larger than the available GPU memory. 
This is because the GPU needs to host both the raw data and the transformed data. \otto{To mitigate the GPU memory footprint and data movement overhead, \TCUDB{} moves columnstore raw data and constructs the input matrices in an adaptive way that \TCUDB{} dynamically selects a more compact data type based on the metadata such as maximum/minimum values discussed in Section 4.2.1.}
The values of $\mathtt{DT\_{op}}$ and $\mathtt{DM\_{op}}$ vary according to the data
types, input tables and the resulting matrices, \TCUDB{} evaluates all 
possible combinations that perform the data transformation process on different locations
before making a decision. 
}

\paragraph{Computation cost}
Finally, the dimensions of the transformed input matrices
also determine the TCU computation time. 
\hungwei{
Using the number of records, the number of distinct values and the most compact 
data type derived from the feasibility test, \TCUDB{} can estimate the required 
device memory and the density of input matrices for the operator. Based on
the estimation, \TCUDB{} can potentially take three different approaches in
performing an operator. \\
(1) If all inputs and outputs fit within the device memory, \TCUDB{}
simply needs to copy all inputs into the device memory and invokes the
matrix multiplication function once. \\
(2) In case the working set size exceeds
the available device memory, \TCUDB{}'s query plan will need to apply
the blocked and pipeline matrix multiplication algorithm~\cite{lam1991cache,
MSplitGEMM} to move parts of input and output data as well as perform matrix 
multiplications block-by-block. (Section~\ref{sec:msplitGemm})\\
(3) If the densities of input matrices are
lower than a certain threshold (an architecture-dependent value), \TCUDB{}
will use sparse matrix multiplications instead. (Section~\ref{sec:sparseMM})}


Since each pair of values in
input matrices requires 2 operations for multiplication and accumulation, 
the computation time in
the simplest case where all input matrices fit in the device memory
can be estimated by 
\begin{equation}
\label{eq:cost}
 \mathtt{CT\_{op}} \approx M N K \times \dfrac{2}{\mathtt{peak\_TCU\_TFLOPS}} 
\end{equation}
where $\mathtt{peak\_TCU\_TFLOPS}$ is the TCUs' peak number of floating-point operations per second
(FLOPS).
\hungwei{
If the query results in inputs larger than device memory, \TCUDB{} still
leverages Equation~\ref{eq:cost} to estimate the cost but replaces $\mathtt{peak\_TCU\_TFLOPS}$
with the measured FLOPS from the blocked/pipelined matrix multiplications.
For the cases where input matrices are sparse, \TCUDB{} estimates the
computation costs not only using the FLOPS from our sparse matrix
multiplication implementation but also multiplying the cost by the density
of inputs. }

The final cost estimation is then the summation of the above three terms 
$ \mathtt{DT\_{op}} + \mathtt{DM\_{op}} + \mathtt{CT\_{op}}$.
\TCUDB{} then compares this estimated cost with the estimated cost of the other
CPU/GPU-based operators to decide whether to use TCUs. 
\TCUDB{}  
obtain the most up-to-date 
estimations for $\mathtt{Bandwidth_{GPU/host}}$
and $\mathtt{peak\_TCU\_TFLOPS}$ by checking the execution time of previous
queries.


Note that there can be more than one TCU-accelerated plan because the system
can choose a higher or lower-precision data type, which can change the decision of
whether to perform 
transformation operator within the GPU or not.
%

\subsubsection{Handling large datasets. }
\label{sec:msplitGemm}
\hungwei{Due to the limited device memory capacity (e.g., 24 GBs in our
case), the input matrices of \TCUDB{}'s operators cannot fit in
the GPU's device memory if the datasets are extremely large and dense.
Once \TCUDB{} catches such a case during the feasibility test, \TCUDB{} will 
consider applying a blocked matrix multiplication algorithm for the
corresponding query operators. The blocked matrix multiplication algorithm
works by fetching a submatrix from the system main memory as a
multiplicand, gradually fetching other same-sized submatrices as the
multiplier, and aggregating the result to the corresponding submatrix in the
result matrices. 

\TCUDB{}'s implementation of blocked matrix multiplication extends
MSplitGEMM~\cite{MSplitGEMM} to support blocked matrix multiplications using
TCUs. Both \TCUDB{}'s implementation and MSplitGEMM exploit pipeline
parallelism by creating multiple streams in fetching input submatrices,
performing matrix multiplication and accumulation, and writing back results
simultaneously. \TCUDB{}'s implementation uses TCUs for matrix multiplication and accumulation
instead of conventional GPU cores. 
During the periodical microbenchmark tests, \TCUDB{} also performs a series
of tests to figure out the optimal size of submatrices that balances the
latency of each stage in the pipeline to maximize the computation
throughput. The measured throughput using these optimal parameters will also
be used as the metrics for evaluating the costs of large and dense inputs in
Section~\ref{sec:cost}. }

\subsubsection{Handling sparse matrices. }
\label{sec:sparseMM}
\ignore{\yuliang{rewrite this subsection. Pls check.}}

Due to the current capability of TCU hardware in handling sparse matrices,
conventional TCU operators that assume dense matrices as their inputs may not always 
outperform a GPU plan when the input matrices to a TCU-accelerated operator are
very sparse. Therefore,
\TCUDB{} implements a TCU-accelerated sparse matrix multiplication
(TCU-SpMM) operator that 
\begin{itemize}
\hungwei{\item transforms an input into a compressed sparse row matrix format (CSR)}
\item partitions an input matrix into 16\x{}16 submatrices,
\item skips submatrices containing all 0s,
\item multiplies the rest using TCUs and accumulates 
results~\cite{ZACHARIADIS2020106848}.
\end{itemize}
By doing so, the TCU-SpMM operator can still leverage TCU's computation power
but on a much smaller number of submatrices pairs when the input matrices are large and sparse.

To determine whether a TCU-SpMM-based plan should replace the dense multiplication plan,
\TCUDB{} needs to estimate the cost similar to the regular cases with dense matrices.
We estimate the total cost by multiplying the estimated dense operator cost
by the inputs' densities.
In addition, the TCU-SpMM-based operator
requires scanning inputs to construct/partition a matrix and filter those all-0-submatrices.
\TCUDB{} estimates this part of the cost with a simple linear function with respect
to the input size.


Finally, the query optimizer of \TCUDB{} still needs to evaluate plans
using the GPU-based HashJoin cost model~\cite{yuan2013yin}, 
in particular sparse matrix multiplication on conventional CUDA cores \hungwei{to determine whether a TCU-SpMM-based plan is more efficient}.

\section{Experimental Results}
\label{sec:result}
Leveraging TCUs' capabilities in optimizing matrix algebra, \TCUDB{} delivers up to
14\x{} speedup over a conventional GPU-based DB engine for 
the sample queries that Section~\ref{sec:theory} describes. Inspired
by the result, we experimented with \TCUDB{} in 
real-world application query workloads with inputs as large as 24~GBs. 
In summary, \TCUDB{} achieves
up to 7.52\x{} speedup in matrix multiplications, 
up to 3.96\x{} speedup for analytic queries in the star schema benchmark,
up to 288\x{} speedup in entity matching queries, and
up to 4.22\x{} speedup for the core of the PageRank algorithm. 
The comparison of \TCUDB{} performance on different GPU 
architectures also reveals the strong potential of TCU-accelerated DB engines in the future.
\ignore{\yuliang{modified a bit.}}

\threefigure[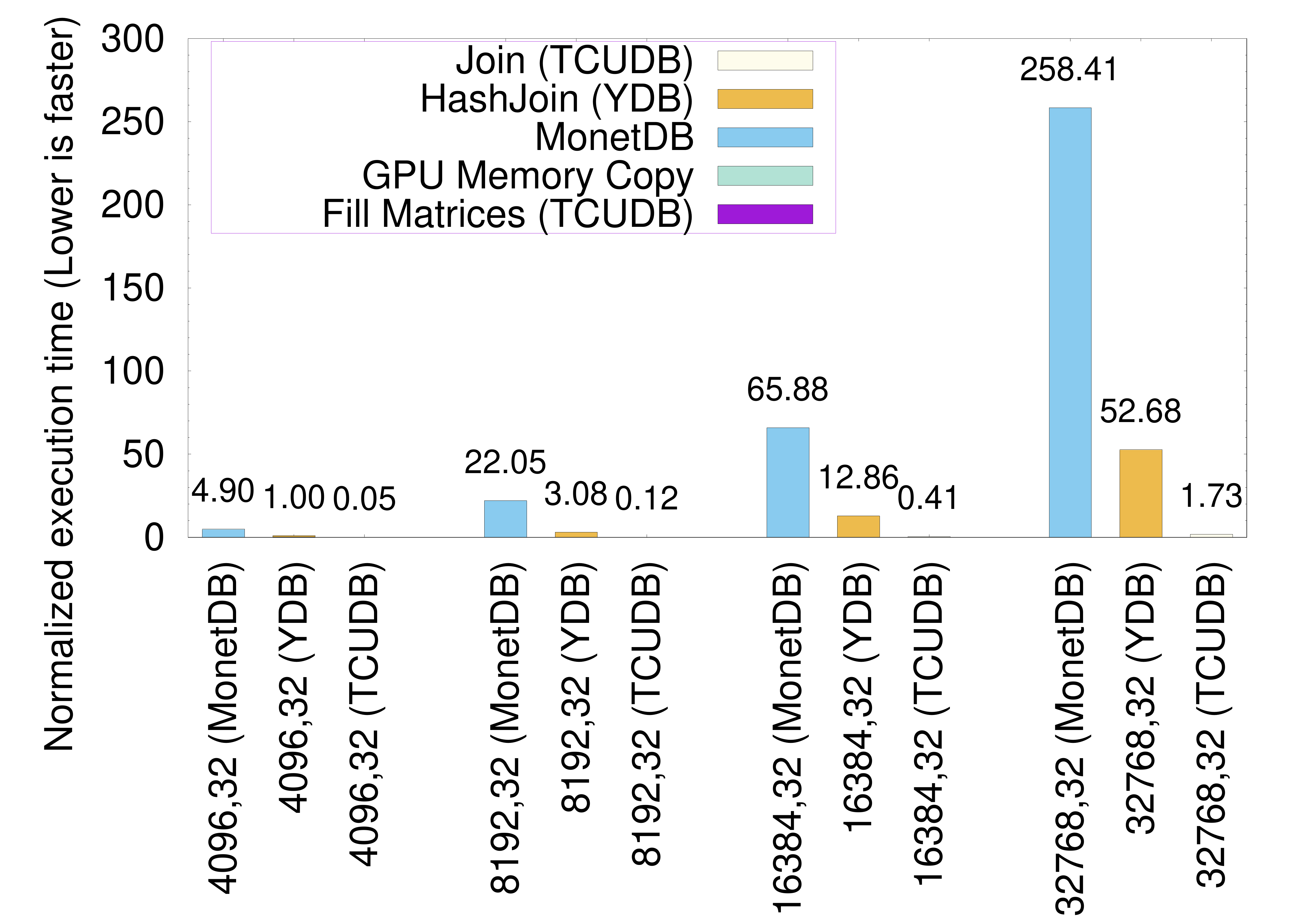,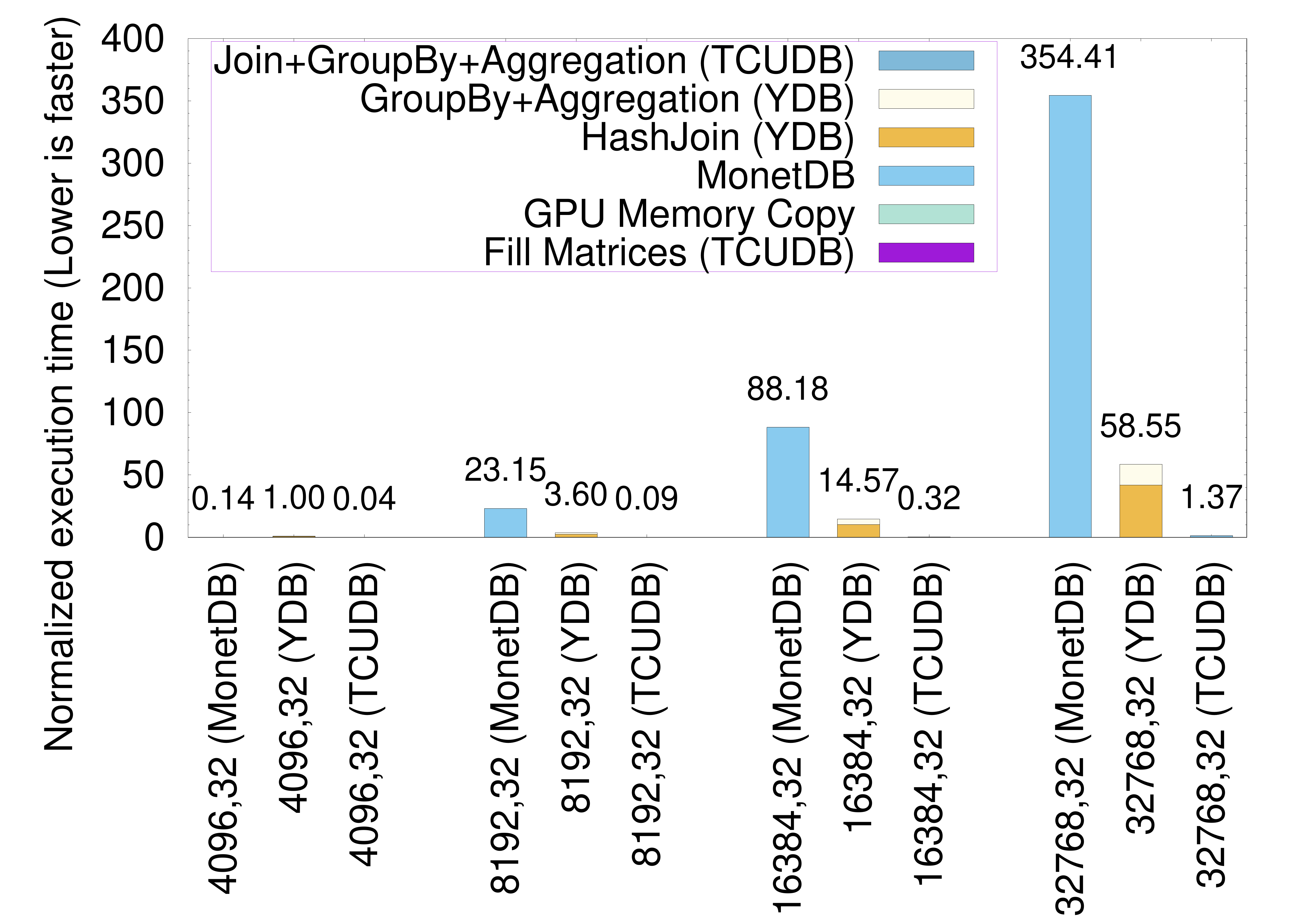,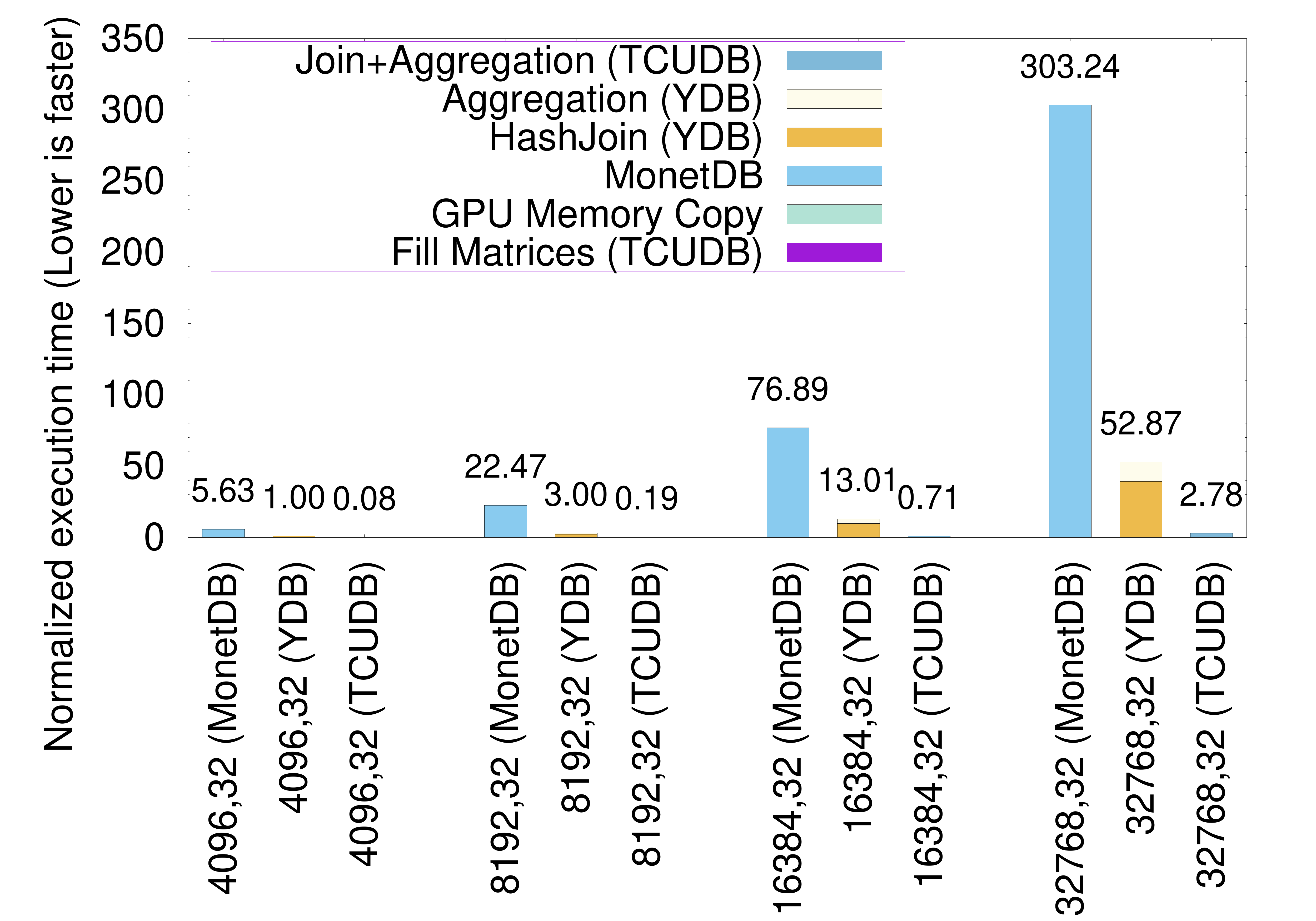,{The relative execution time of running
(a) Q1,
(b) Q3, and (c) Q4 with various number of records and 32 distinct values in the target
attribute on \TCUDB{}, \YDB{}, and MonetDB.},fig:micro_all]
\threefigure[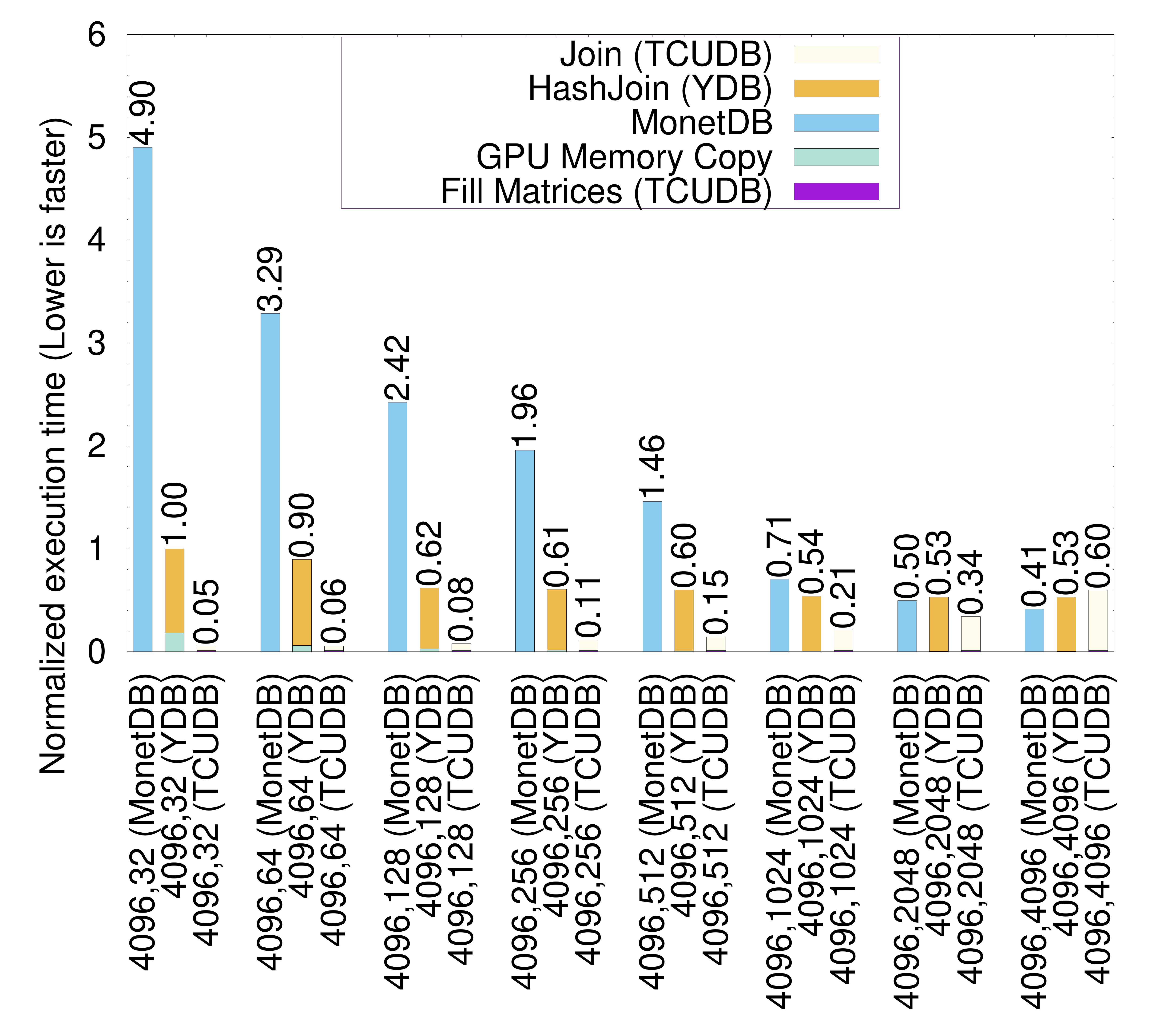,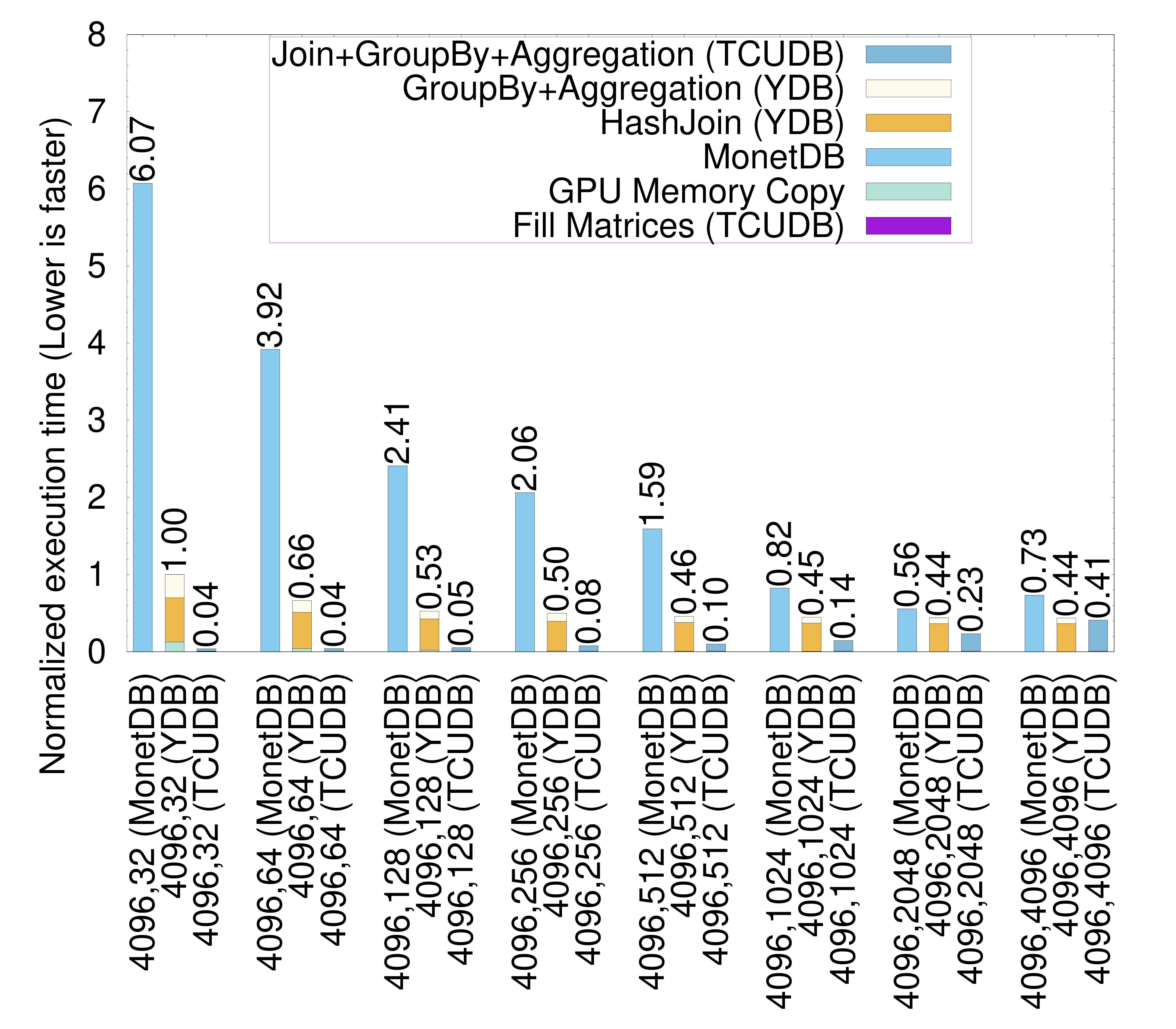,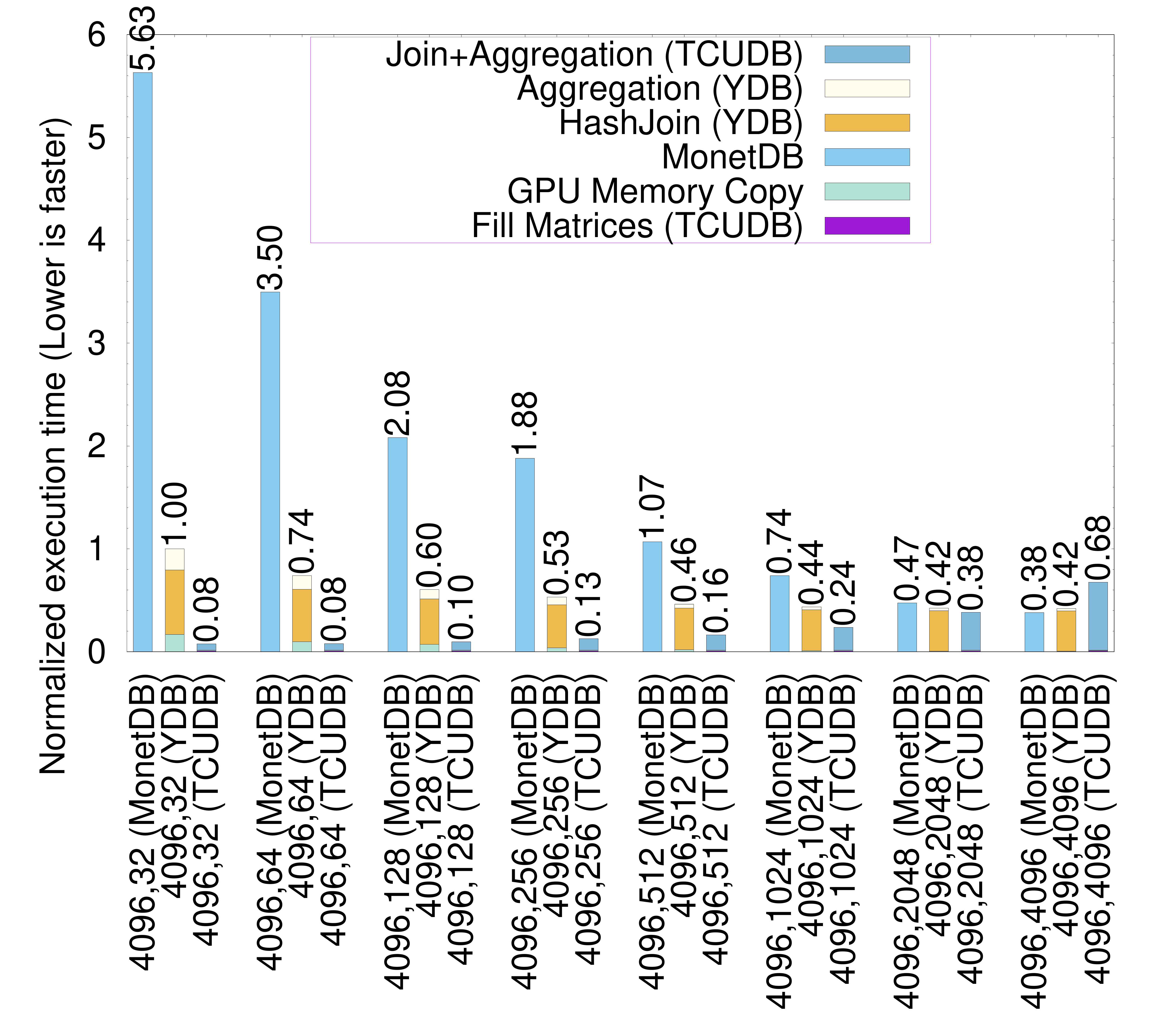,{The relative execution time of running
(a) Q1,
(b) Q3, and (c) Q4 with 4096 records and various distinct values in the target
attribute on \TCUDB{}, \YDB{}, and MonetDB.},fig:micro_k]

\subsection{Experimental Methodology}
\label{sec:method}
\ignore{
We built \TCUDB{} and a testbed that contains an Intel i7 processor and an NVIDIA RTX 3090 GPU. 
We evaluate \TCUDB{} using \otto{[FXIME]} workloads and compare the performance with one 
state-of-the-art data analytics. This section describes the experimental platform, case 
studies, metrics and the system we compare \TCUDB{} with.}

We conducted experiments on a machine with an Intel Core i7-7700K processor, 32 GB DDR4 
DRAM. The processor contains 4 cores and each processor core runs at 4.2 GHz by default. 
The GPU in our experiments is an NVIDIA GeForce RTX 3090 GPU based on Ampere
architecture. This GPU contains 24 GB GDDR6X 
device memory and 328 Tensor Cores and attaches to a PCIe 3.0 x16 slot. The TCU-accelerated 
operator library in \TCUDB{} is implemented using a NVIDIA CUDA Toolkit 11.2.
The system runs a Linux 4.15.0 kernel with the NVIDIA driver version in
460.32.03. 
We compared \TCUDB{} with a state-of-the-art GPU execution engine for warehouse-style
queries, \YDB{}~\cite{yuan2013yin} and a pure CPU-based execution engine,
MonetDB~\cite{boncz2005monetdb}, as reference designs.

\subsection{Microbenchmark}
\label{sec:microbenchmark}
To allow query optimizers to select the right query
plans, the
database engine must obtain samples of executing workloads using
TCU-accelerated operations. Upon installing \TCUDB{} in the system
or when the system detected any change in hardware configurations, \TCUDB{} 
will perform a one-time sampling process that runs a set of microbenchmark
workloads to collect critical timing information for query optimizations. 

During the sampling process, \TCUDB{} will execute three main queries, Q1,
Q3 and Q4 from Section~\ref{sec:theory}, with various-sized, random-generated input
datasets. \TCUDB{} does not evaluate Q2 and Q5 as they are essentially combinations of
other queries. The sampling process also helps us to classify the cases
where \TCUDB{} is superior to the conventional GPU-accelerated engine and identify the source of
performance gain/loss in \TCUDB{}. With large system main memory and
aggressive file system caching by operating systems as well as
the underlying high-performance NVMe SSD, we have not observed significant disk load
time in each DB engine's initialization phase.

\hungwei{As MonetDB is a full-fledged system, we excluded the additional steps/overheads by measuring only the time to execute the physical plan for a fair comparison. (We use the \textsf{``--timer=performance''} option and disable the resulting output to report the runtime part only.)} 

Figure~\ref{fig:micro_all} and Figure~\ref{fig:micro_k} present
a subset of
microbenchmark results from the sampling process on the default testbed
described in Section~\ref{sec:method}. We label the x-axis of each sample in
this figure with two parts in the configuration. The first part is the
parameters for the query, $M$, $K$ and $N$, that represent the
sizes of the input matrices for each evaluated
operator where one matrix has the dimension of $M\times K$ and the other is
$K\times N$. To save space, we only present the cases when $M=N$ and label
each configuration with their values of $M$ and $K$ as $M, K$ in these
figures. 
The second part is the DB engine (i.e, \TCUDB{}, \YDB{}, or MonetDB).
The vertical axis in each figure shows the aggregated
execution time in each step of running these queries, normalized to the
total time when running the same query using \YDB{}, the conventional GPU-accelerated
engine, with $M=N=4096$ and $K=32$.

Figure~\ref{fig:micro_all}(a) shows the performance of Q1 for \TCUDB{}, \YDB{}
and MonetDB from 
input sizes 4096 to 32768. Both \TCUDB{} and \YDB{} significantly
outperform MonetDB for this query. 
\TCUDB{} outperforms \YDB{} in most configurations.
The advantage of \TCUDB{} is especially significant when datasets grow.
\TCUDB{} outperforms \YDB{} by 14\x{} for the case of (32768, 32) and 9.3\x{}
for (16384, 32), but only 1.18\x{} for (4096,32). Observing the breakdown of
execution time in Figure~\ref{fig:micro_all}(a), we found the major speedup comes
from the significant reduction of computation time from the TCU-accelerated
join operator, despite the additional overhead in filling and transforming datasets into the desired matrices
for \TCUDB{}. 

\ignore{
\TCUDB{} starts losing its advantages after the number of distinct values
is more than 128. 
However, the slowdown of \TCUDB{} simply comes from the fixed-cost
activation overhead of Tensor Cores in CUDA runtime that future architecture
or updates of the CUDA system can improve, but not the limitation of Tensor
Cores themselves. In fact, the initialization overhead were improved by
100\x{} after we updated the CUDA runtime from 10.2 to 11.1. Later
experiments in Section~\ref{sec:PageRank} also showed the overhead can be
avoided if appropriate system-level API is presented. Excluding this
initialization overhead, \TCUDB{}'s join operator is actually 9.67\x{}
faster than \YDB{}'s GPU HashJoin. 
}
Figure~\ref{fig:micro_k}(a) varies the number of distinct values that
affect the sparsity of input matrices in Q1 for \TCUDB{}'s join operator. 
As the number of distinct values becomes larger, the performance advantage of \TCUDB{}'s join operator
over \YDB{} and MonetDB begins to shrink.  
Because the sizes of one dimension of both input
matrices for the \TCUDB{} join operator in Q1 depends on the number of distinct values
from the chosen attribute to perform matching, matching on an
attribute with more distinct values will lead to computation on larger but
sparse matrices. 
In contrast, \YDB{}'s and MonetDB's $\mathtt{HashJoin}$ algorithm produces
smaller vectors as the chance (i.e., total number) of records sharing 
a single value reduces if the number of distinct values increases. Therefore,
even though \YDB{}'s and MonetDB's $\mathtt{HashJoin}$ operator needs to work on more pairs of vectors,
each pair of vectors have smaller dimensions.  
However, \TCUDB{}'s join operator still outperforms \YDB{} and MonetDB in all cases
until the number of distinct values reaches 4096. This profiling
result suggests that \TCUDB{} select a GPU-hash-join-based or sparse-matrix-based
implementation if the density of input matrices is below 0.04\% on our
testbed. 
Figure~\ref{fig:micro_all}(b) presents the performance of running Q3 using
\TCUDB{}, \YDB{} and MonetDB. Q3 evaluates the group-by and aggregations over join query. 
Unlike the conventional GPU-accelerated DB engine where group-by and aggregations are
separate operations after the hash join, \TCUDB{} can implement the whole Q3
using just one matrix multiplication. As a result, the execution time of
using \TCUDB{} of executing Q3 remains similar to executing Q1 when the
input parameters are the same. However, \YDB{} or MonetDB always have to perform the
additional group-by operations and leads to a longer execution time than
performing Q1 for the same inputs. Therefore, the performance advantage of
\TCUDB{} becomes more significant for Q3. For (32768, 32), \TCUDB{} can
outperform \YDB{} by 45\x{}. 

When we increase the number of distinct values as in
Figure~\ref{fig:micro_k}(b), \TCUDB{} becomes less advantageous, similar to
the phenomenon in Q1. However,
as \TCUDB{} still uses single-matrix-multiplication-based Join/Aggregation/GroupBy operation
to perform operations where \YDB{} or MonetDB needs multiple-step HashJoin and GroupBy/Aggregation
operators, \TCUDB{} still outperforms \YDB{} and MonetDB in all cases.



Figure~\ref{fig:micro_all}(c) presents the relative execution time of Q4 on
\TCUDB{}, \YDB{} and MonetDB. \YDB{} and MonetDB will perform Q4 using $HashJoin$ and then an aggregate
query but without a group-by operator. Therefore, the overall execution time
in each configuration of \YDB{} and MonetDB is less than Q3 because of the elimination of
group-by operator. However, again, \TCUDB{} still implements this operator using
single matrix multiplication on the transformed input matrices. 
Therefore, \TCUDB{} achieves 19\x{} speedup for (32768, 32). 

As in Q1 and Q3, \TCUDB{} becomes less advantageous when we increase the number of distinct values as in
Figure~\ref{fig:micro_k}(c). Because the amount of operations in \YDB{} and
MonetDB for Q4
is fewer than Q3, we still see \TCUDB{} falls short when the number of distinct
reaches 4096 and suggest an alternative plan for cases where input
matrix densities are below 0.04\%.

\subsection{Analytic queries: Star Schema Benchmark}
\label{sec:tpch}
\begin{figure*}[t]
\vspace*{-0mm}
\begin{center}
\begin{tabular}{cccc}
\hspace{-0.15in}\includegraphics[width=1.9in]{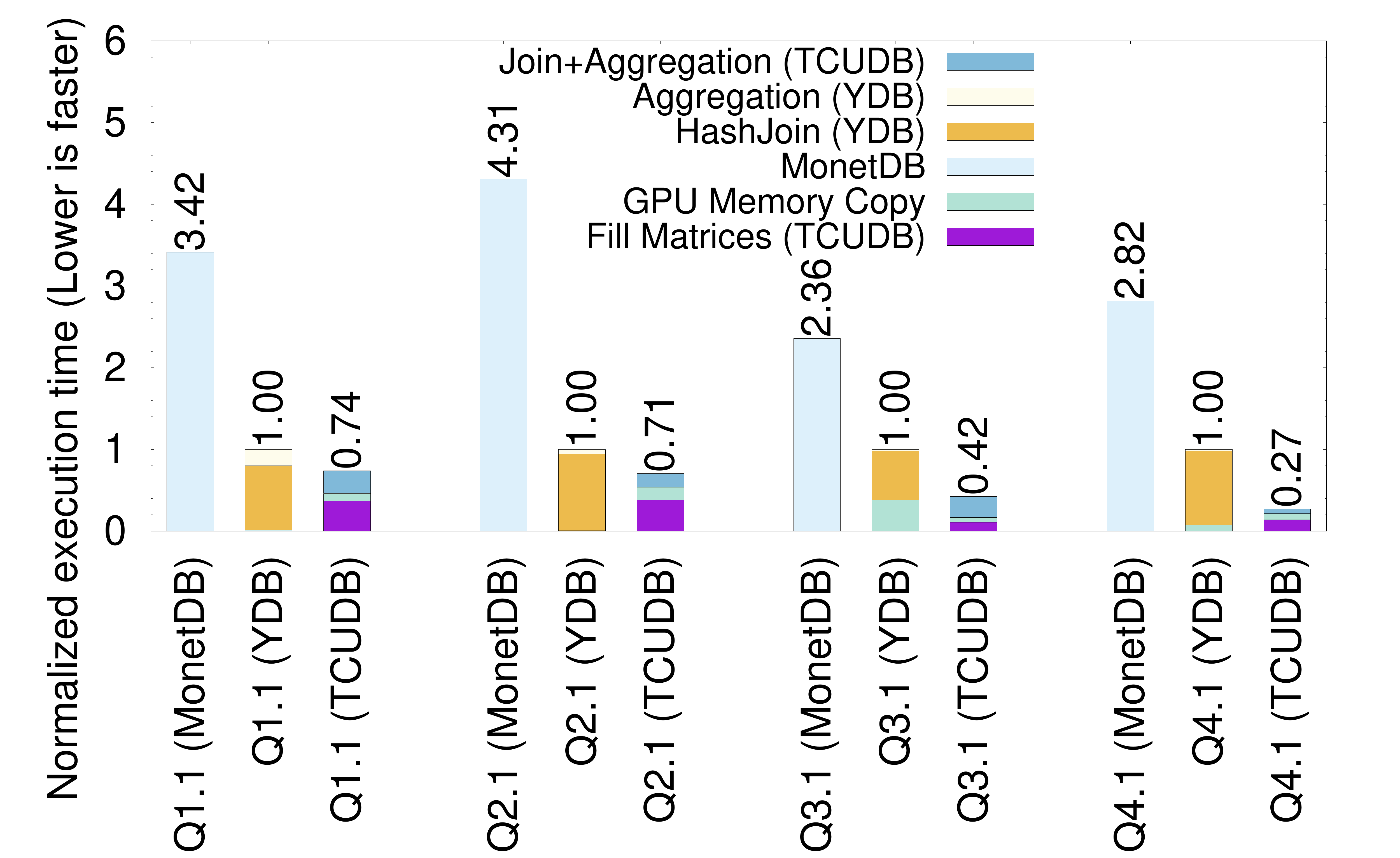} &
\hspace{-0.25in}\includegraphics[width=1.9in]{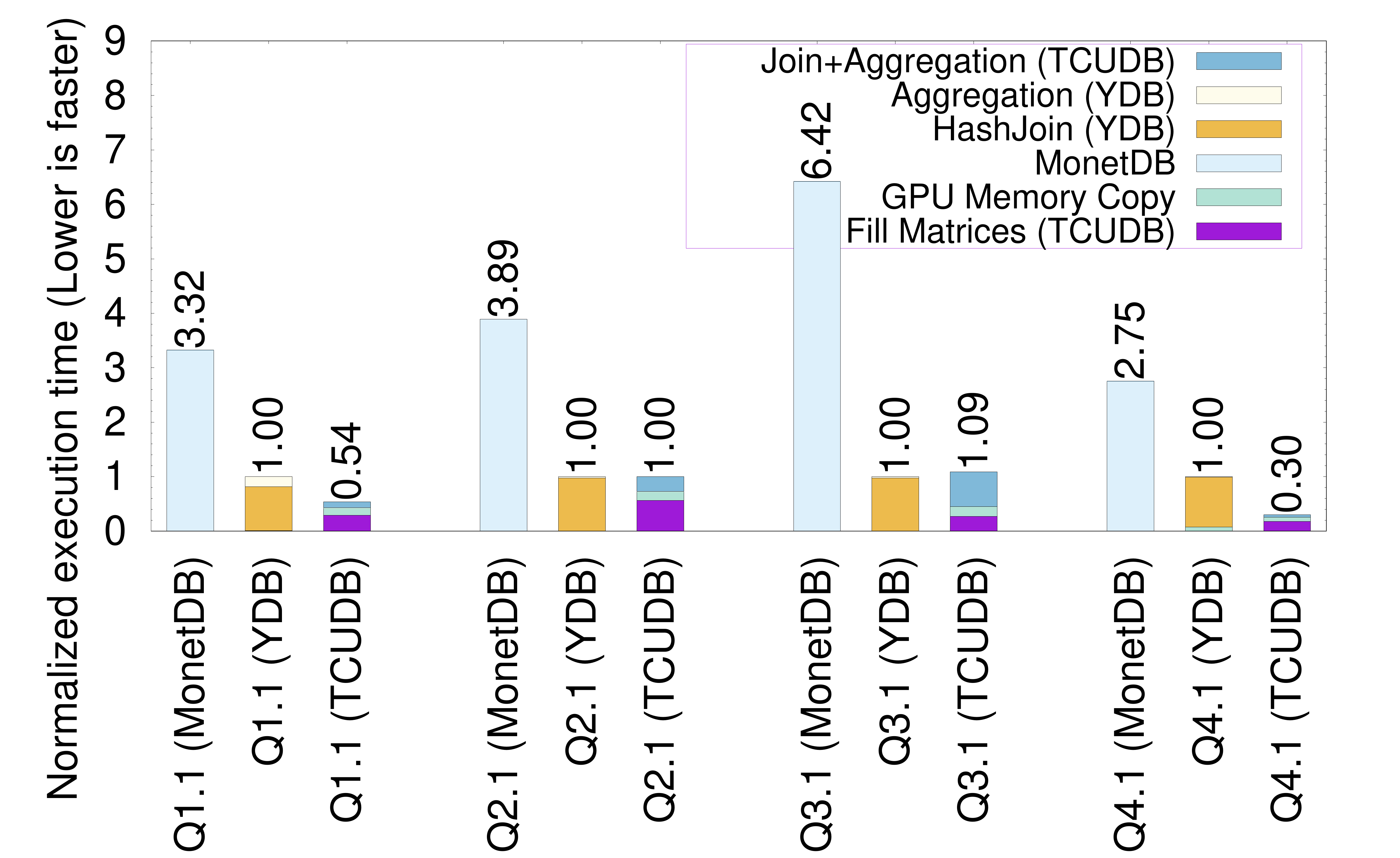} &
\hspace{-0.25in}\includegraphics[width=1.9in]{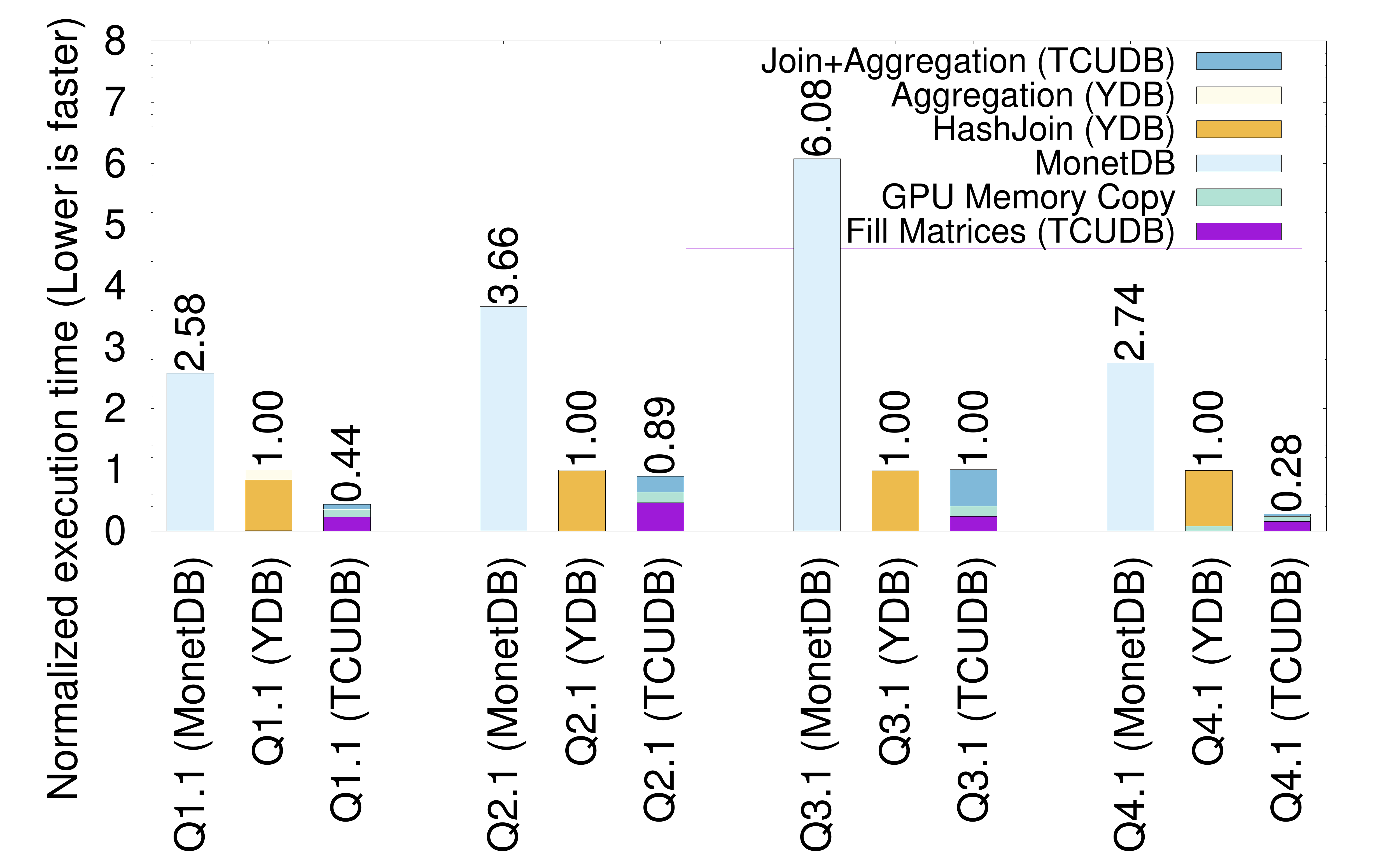} &
\hspace{-0.25in}\includegraphics[width=1.9in]{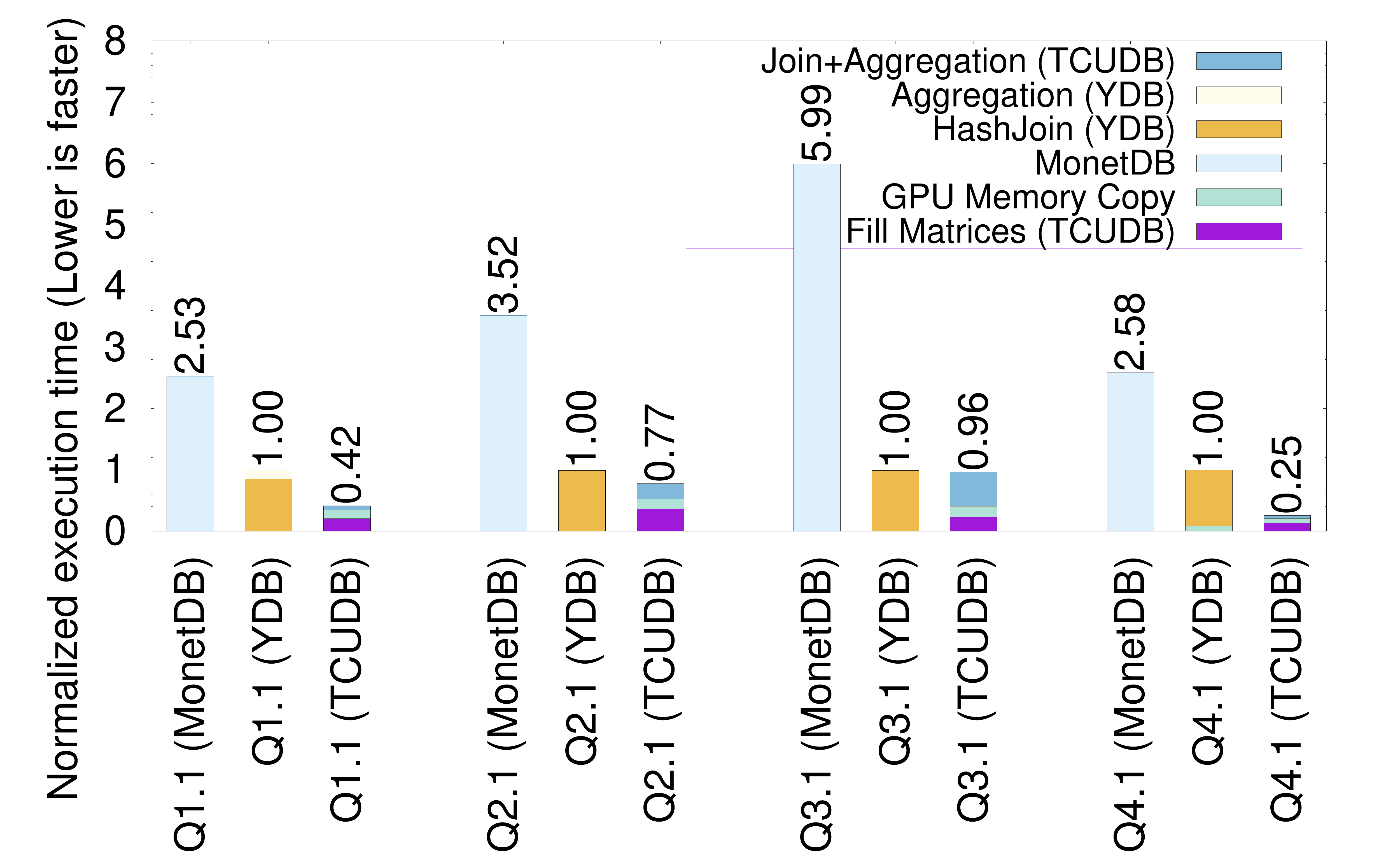} \\
(a) & (b) & (c) & (d) \\
\end{tabular}
\vspace*{-0.1in}
\caption{The relative runtime of star schema benchmark on \TCUDB{} compared to 
MonetDB and YDB running the same query as the baseline with scaling factor
(a) 1, (b) 2, (c) 4 and (d) 8.}
\label{fig:ssb}
\end{center}
\vspace*{-0.2in}
\end{figure*}

We evaluate the performance of \TCUDB{} 
on the popular Star Schema Benchmark
(SSB)~\cite{o2009star}, a benchmark suite modeling the data warehouse
workloads. 
SSB is widely used in benchmarking analytic engines due to
its realistic modeling of data warehousing workloads. 
The database form a star schema consisting of
one fact table (\texttt{lineorder}) and four
dimension tables (\texttt{supplier, customer, date} and \texttt{part})
connected to the fact table by foreign keys.

The benchmark provides 13 queries in 4 flights. \TCUDB{} supports all the 13 
SSB queries. Figure~\ref{fig:ssb} compares the performance of \TCUDB{}, \YDB{} and 
MonetDB in running SSB queries with scaling factors varying from 1 to 8
resulting in data sizes from 0.7GB to 5.6GB. 

Figure \ref{fig:ssb} summarizes the results.
\TCUDB{} outperforms both \YDB{} and MonetDB in all evaluated SSB workloads
with up to 3.96\x{} speedup when running Q4.1 with scaling factor as 8. Even
with the worst performing SSB Q3.1, \TCUDB{} still maintains the same level
of performance as \YDB{}. These promising results show that \TCUDB{} has the potentials
of being integrated into real-world analytic engines.

\subsection{Case studies: matrix multiplication, entity matching, and PageRank}
\label{sec:case_studies}

In
addition to individual operators, we also evaluated three representative use
cases, matrix multiplication, entity matching and PageRank to demonstrate
\TCUDB{}'s capabilities in handling intensive operations and large datasets. 


\cfigure[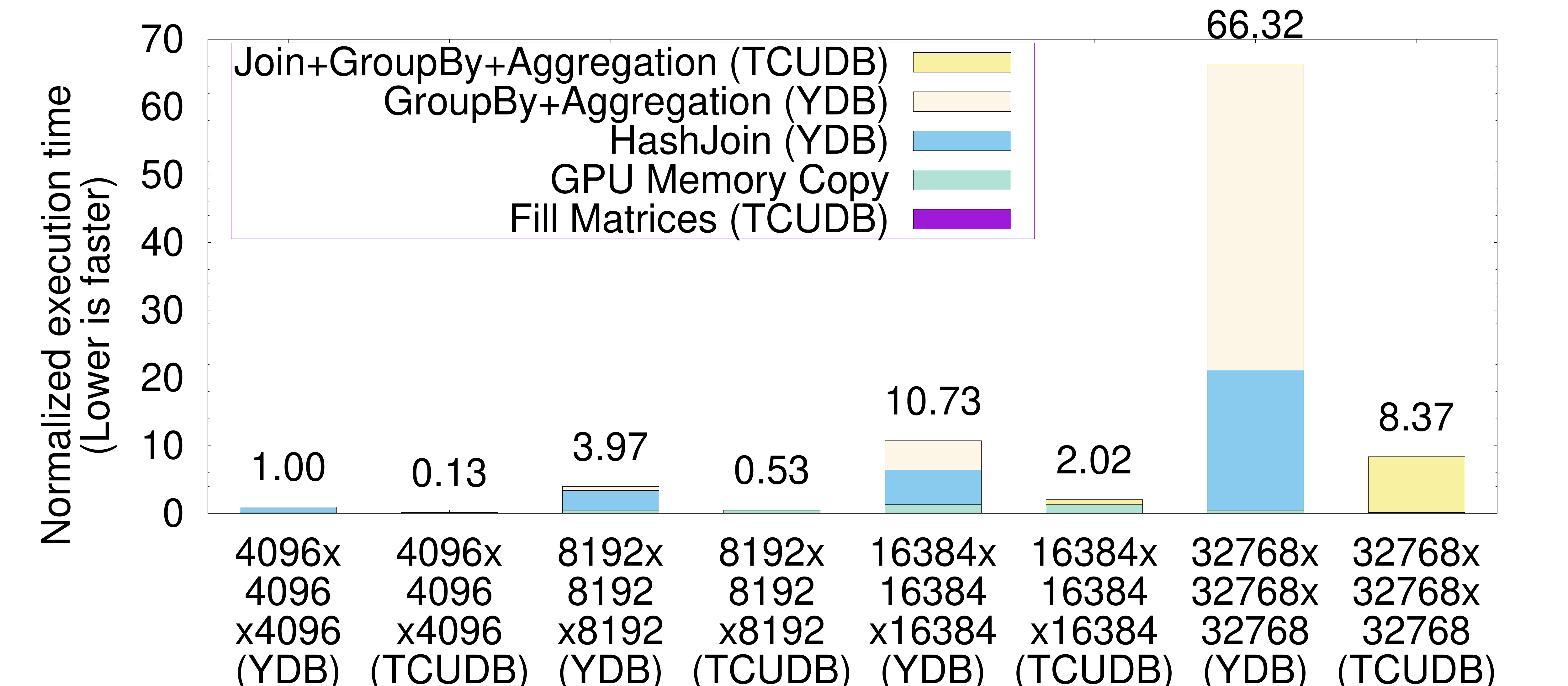,{The relative execution time and 
breakdown of matrix multiplication query on \TCUDB{} and YDB.},fig:case_MM_breakdown]

\ignore{

\begin{figure}[t]
\includegraphics[width=70mm,scale=0.9]{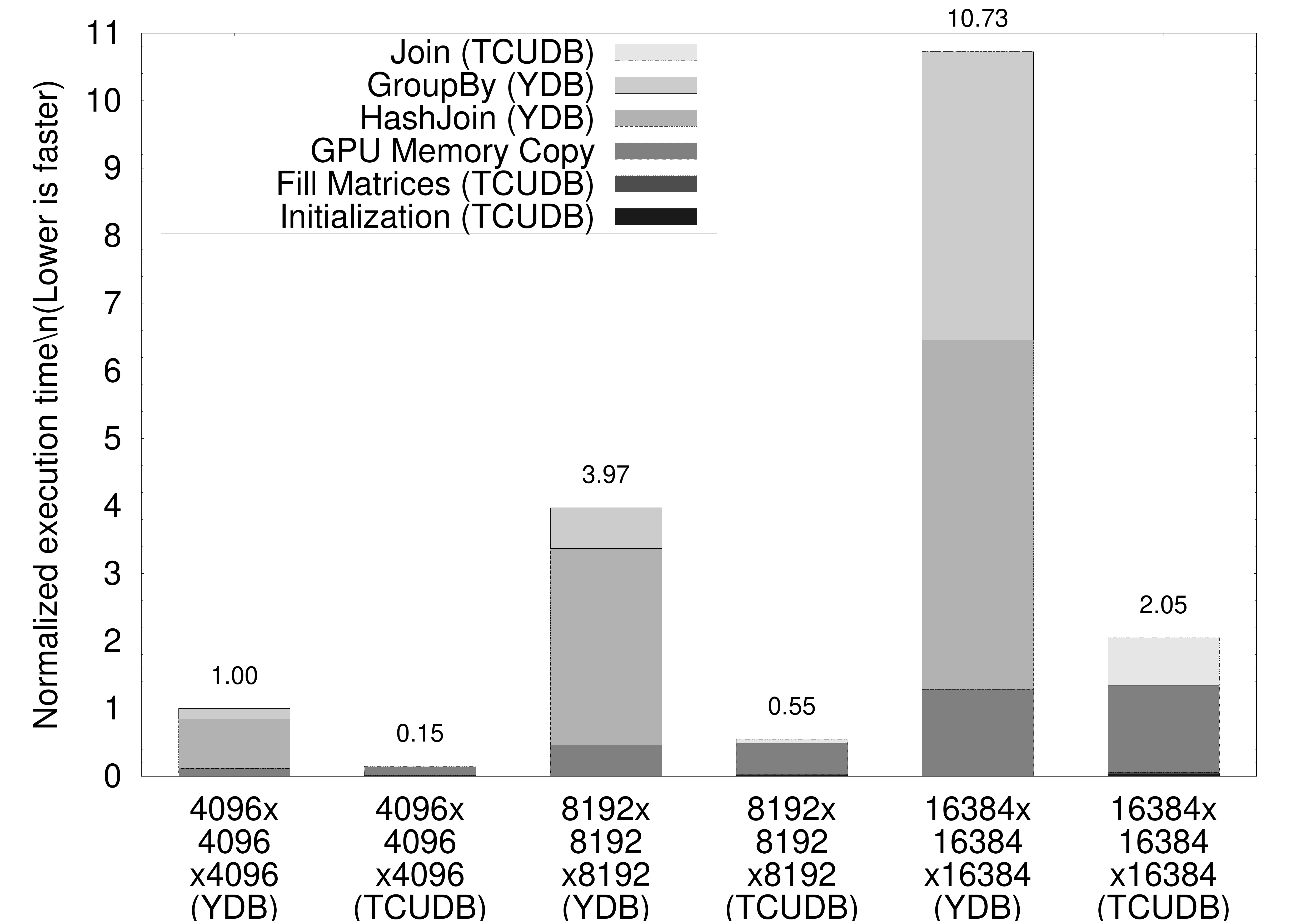}
\caption{\small{The performance of matrix multiplication query on \TCUDB{} and YDB.}}
\label{fig:case_MM}
\end{figure}

\begin{figure}[ht]
  \begin{subfigure}{0.2\textwidth}
    \centering
    \includegraphics[width=1\linewidth]{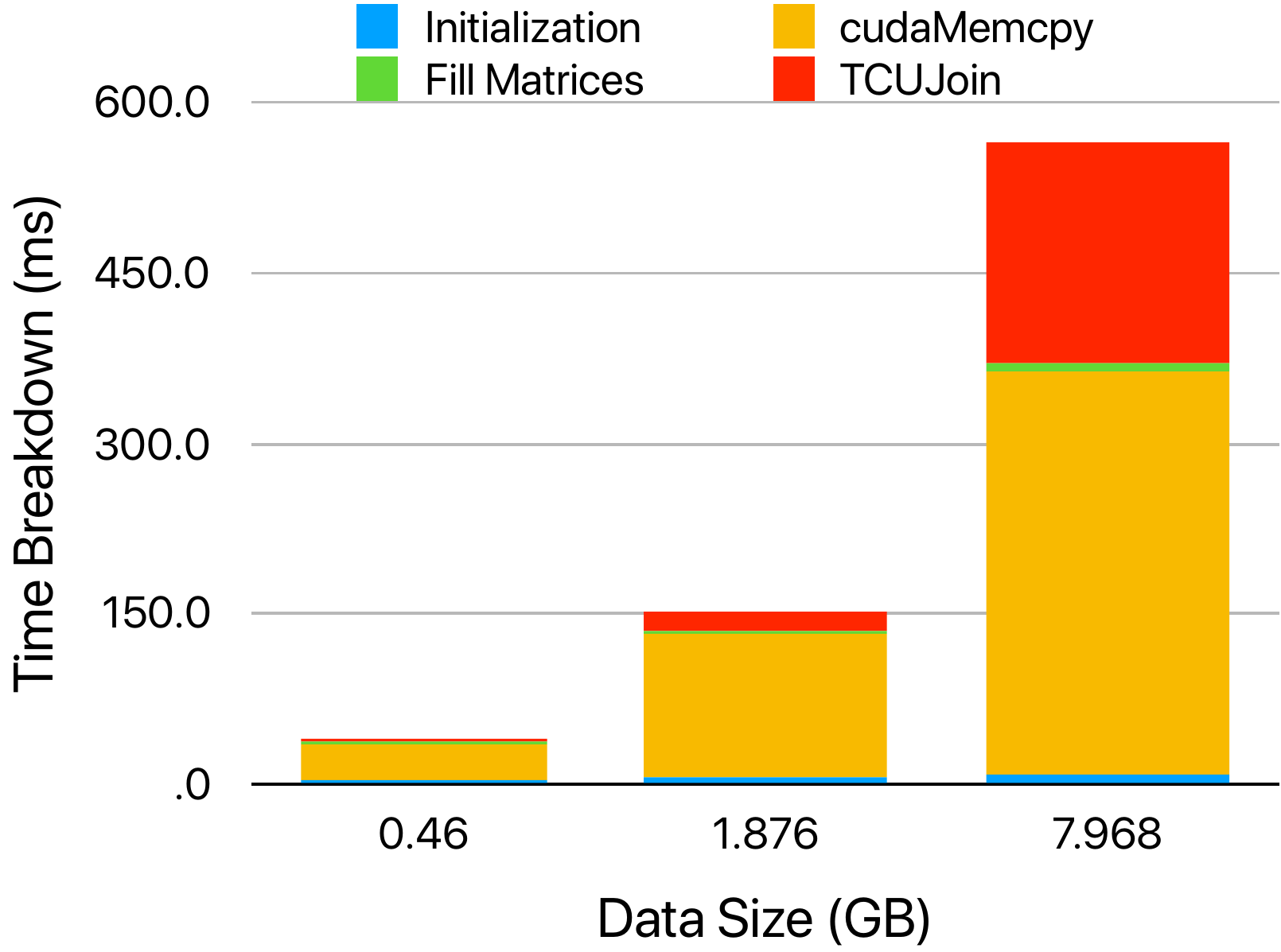}
    \vspace{-0.05in}
    \caption{\TCUDB{}.} \label{fig:case_MM_tcu}
  \end{subfigure}
  \hspace*{0.05in}
  \begin{subfigure}{0.2\textwidth}
    \centering
    \includegraphics[width=1\linewidth]{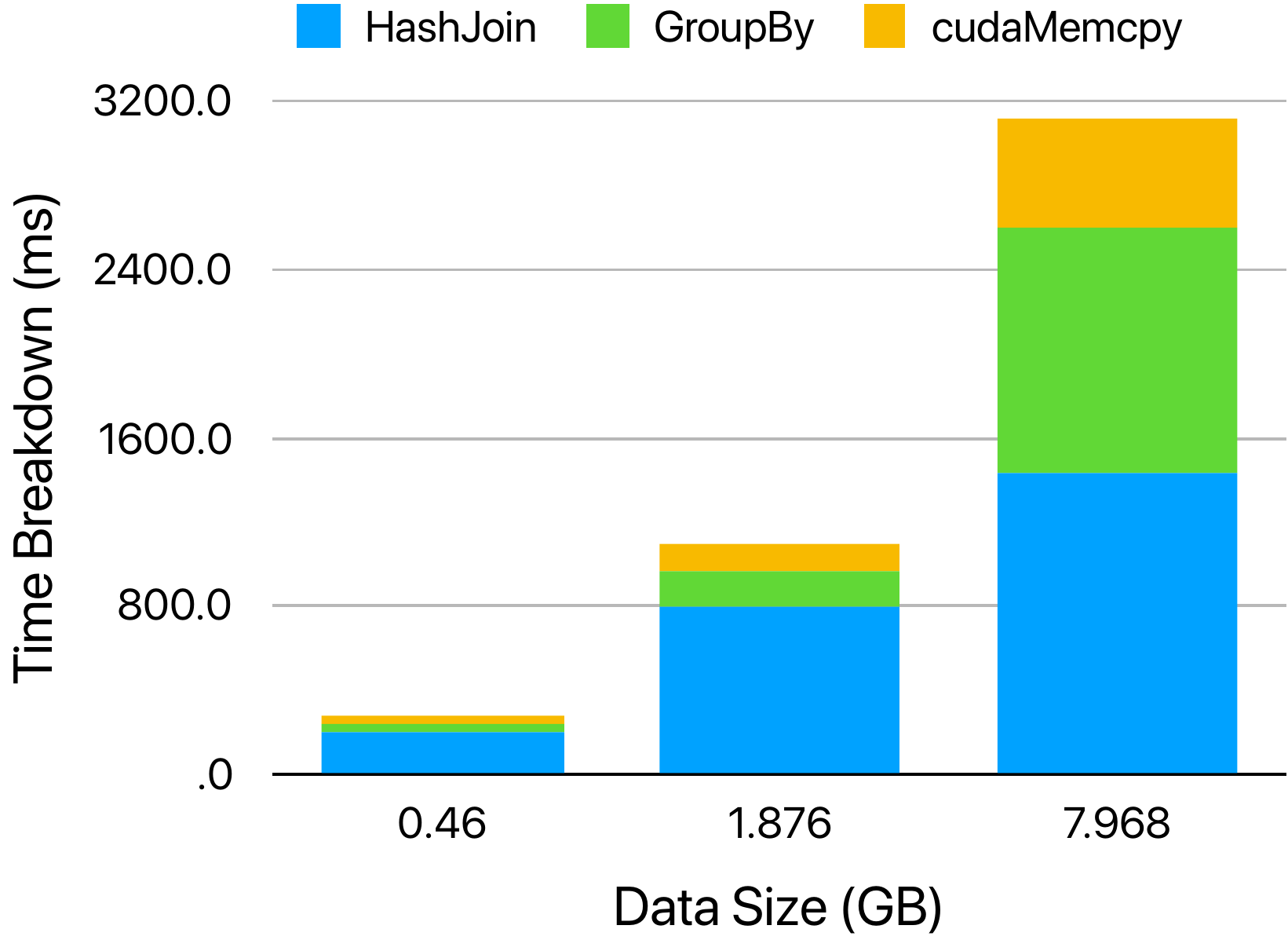}
    \vspace{-0.05in}
    \caption{YDB.} \label{fig:case_MM_gpu}
  \end{subfigure}
\caption{The execution time breakdown.}
\vspace{-0.05in}
\label{fig:time_breakdown}
\end{figure}
}

\subsubsection{Matrix Multiplication}
\label{sec:mm}

Matrix multiplication was once considered inefficient for relational databases.
With the help of hardware-accelerated matrix multiplications, \TCUDB{} can
make queries containing complex linear algebra operations more efficient. 
We use a query in Figure~\ref{fig:ex_mmquery} to demonstrate this use case. 
We create two tables $A$ and $B$ where each record in both tables has three attributes $(row\_num,\ col\_num,\ val)$
as the input. We generate the synthetic dataset according to this schema with
input matrices of dimensions up to 32768\x{}32768 and data
volume up to 24~GB, approximately 2.14 billion records.
\ignore{
Upon receiving the query, \TCUDB{} parses the query, identifies the pattern of
Section~\ref{sec:theory} in the query. \TCUDB{} then prepares input matrices for the
corresponding operator by filling the value of each tuple into matrices based 
on $(row\_num,\ col\_num)$.
As TCUs perform multiplication and aggregation in one-pass, \TCUDB{} does not need to handle SUM() separately. 
To achieve the $\mathtt{GroupBy}$ operator, we can simply utilize the coordinates of the resulting matrix.
\yuliang{not sure what the last sentence means.}
}

Figure~\ref{fig:case_MM_breakdown} presents the relative execution time and breakdown of performing matrix multiplication on
\TCUDB{} and YDB, using YDB with each table containing 4096\x{}4096 records
as the baseline. We did not include MonetDB's result in these Figures as
MonetDB cannot finish these queries within a reasonable amount of time and
present MonetDB's results in Figure~\ref{fig:case_MM_breakdown} would render
the results of \TCUDB{} and YDB invisible. When the dataset contains fewer than 16384\x{}16384
records, the input matrices that \TCUDB{} creates for the TCU's $\mathtt{Join+Aggregation+GroupBy}$ operator
completely fit in the GPU's device memory.   
\TCUDB{} consistently outperforms YDB and delivers up to 7.51\x{} speedup. 
When the dataset contains 32768\x{}32768 records for each table, \TCUDB{} must
partition the input matrices into submatrices, use the block algorithm, and pipeline the
swapping in/out of submatrices to perform the Join/Aggregation/GroupBy operator.
\TCUDB{} still performs multiplication and aggregation of submatrices using
TCUs. Even with the overhead of data exchanges in the blocked  
Join/Aggregation/GroupBy operator, \TCUDB{} is still able to
outperform \YDB{} by 7.92\x{} for the case of 32768\x{}32768 records for each
table.
As datasets fit in the system's main memory as well as the operating system's aggressive 
caching and the help of high-speed NVMe SSD, the data load time from storage is
relatively insignificant in these experiments.  
The data movement (\texttt{cudaMemcpy}) time is the most timing critical stage for
\TCUDB{}. However, the amount of time is comparable to \TCUDB{} and YDB because both
engines only transfer the required data to the device memory. 
The most time-consuming 
parts for YDB are $\mathtt{HashJoin}$ and $\mathtt{GroupBy}$ operations because code using 
conventional CUDA cores needs to iterate tables row by row. YDB spends up to 14\x{}
(in the case of 16384\x{}16384 records in each table) more 
execution time in $\mathtt{HashJoin}$ and $\mathtt{GroupBy}$ than \TCUDB{}'s single
Join/Aggregation/GroupBy operator.

\begin{table}[t]
\centering
\scriptsize
\hspace*{-0.2in}
\begin{tabular}{|r||r|r|r|r|r|}
\hline
	 &	2048&	4096&	8192&	16384	&32768\\
	 &	\x{}2048&	\x{}4096&\x{}8192&\x{}16384&\x{}32768\\
	 &	\x{}2048&	\x{}4096&\x{}8192&\x{}16384&\x{}32768\\
\hline
$x$ = 0, 1 	&	0	&	0	 &	0&	0& 0\\
$-2^{7} \leq x < 2^{7}$ 	&	0	&	0	 &	0.00076\%& 0.00076\%
& 0.00076\%\\
$-2^{15} \leq x < 2^{15}$  &	0.00114\%&	0.00450\%&	0.00908\%& 0.00908\%
& 0.00908\%\\
$-2^{31} \leq x < 2^{31}$ &	0.00122\%& 0.00451\%&0.00909\%&0.00909\% & 0.00909\%\\
\hline
\end{tabular}
\caption{The mean absolute percentage error rates (MAPE) of matrix multiplication
queries with various value ranges.}
\vspace*{-0.1in}
\label{table:error}
\end{table}

\ignore{
\wfigure[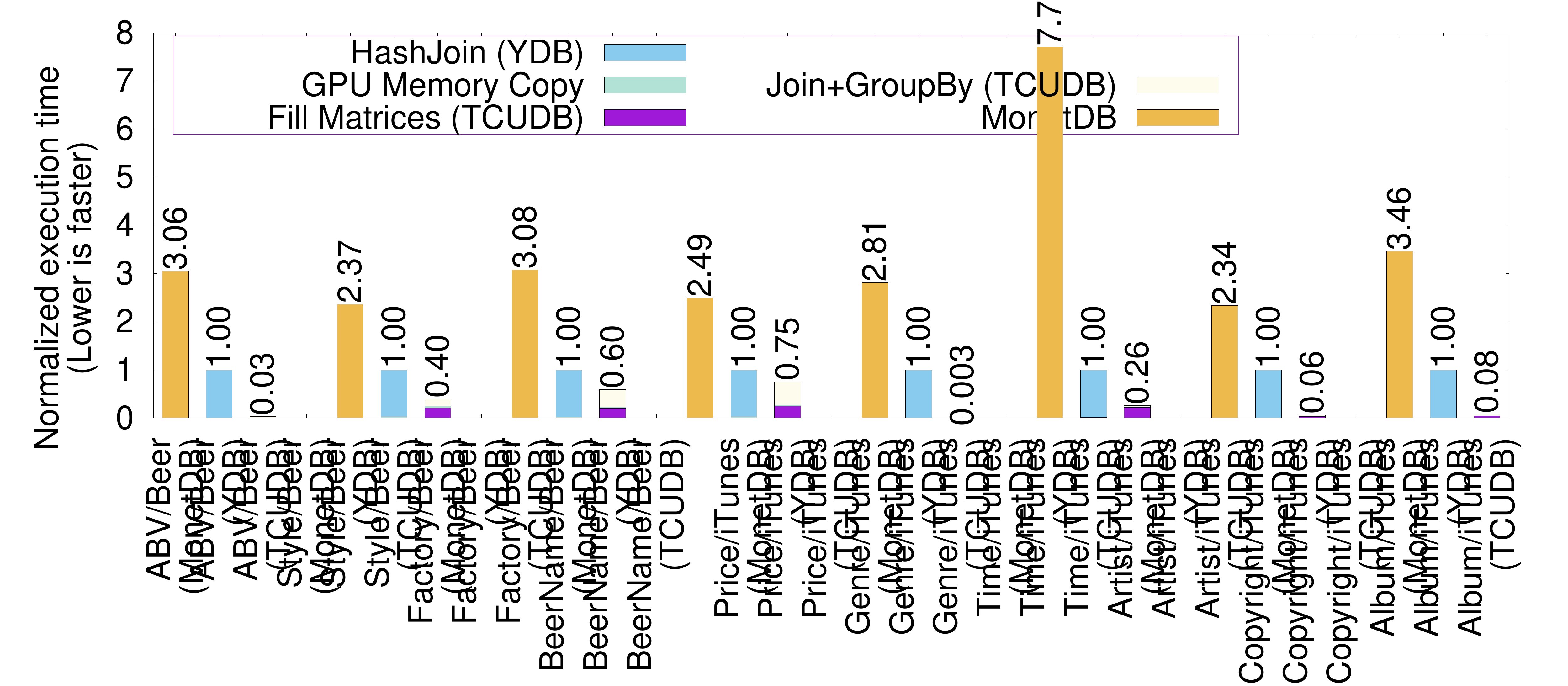,{The relative runtime of 
the EM-blocking queries on \TCUDB{} using the default Deepmatcher
datasets, compared to YDB running the same query as the
baseline.},fig:case_EM]
\cfigure[Figures/case_EM_breakdown.pdf,{The relative runtime of 
the EM-blocking queries on \TCUDB{} using the default Deepmatcher
datasets, compared to YDB running the same query as the
baseline.},fig:case_EM]
\cfigure[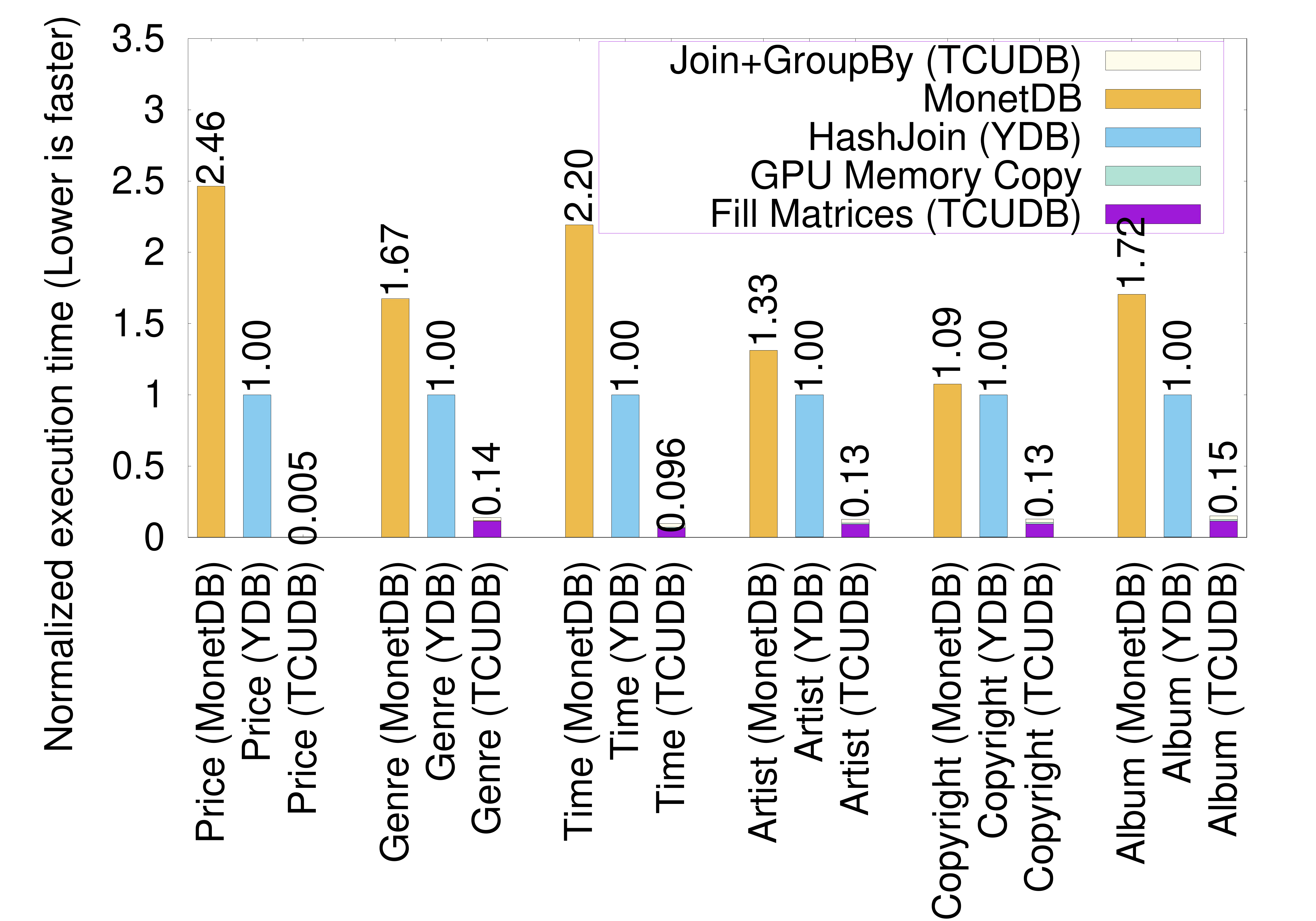,{The relative runtime of 
the EM-blocking queries on \TCUDB{} using the scaled Deepmatcher
datasets, compared to YDB running the same query as the
baseline.},fig:case_EM_scaleup]}
\begin{figure*}[t]
\vspace*{-0mm}
\begin{center}
\begin{tabular}{ccc}
\hspace{-0.15in}\includegraphics[width=2.5in]{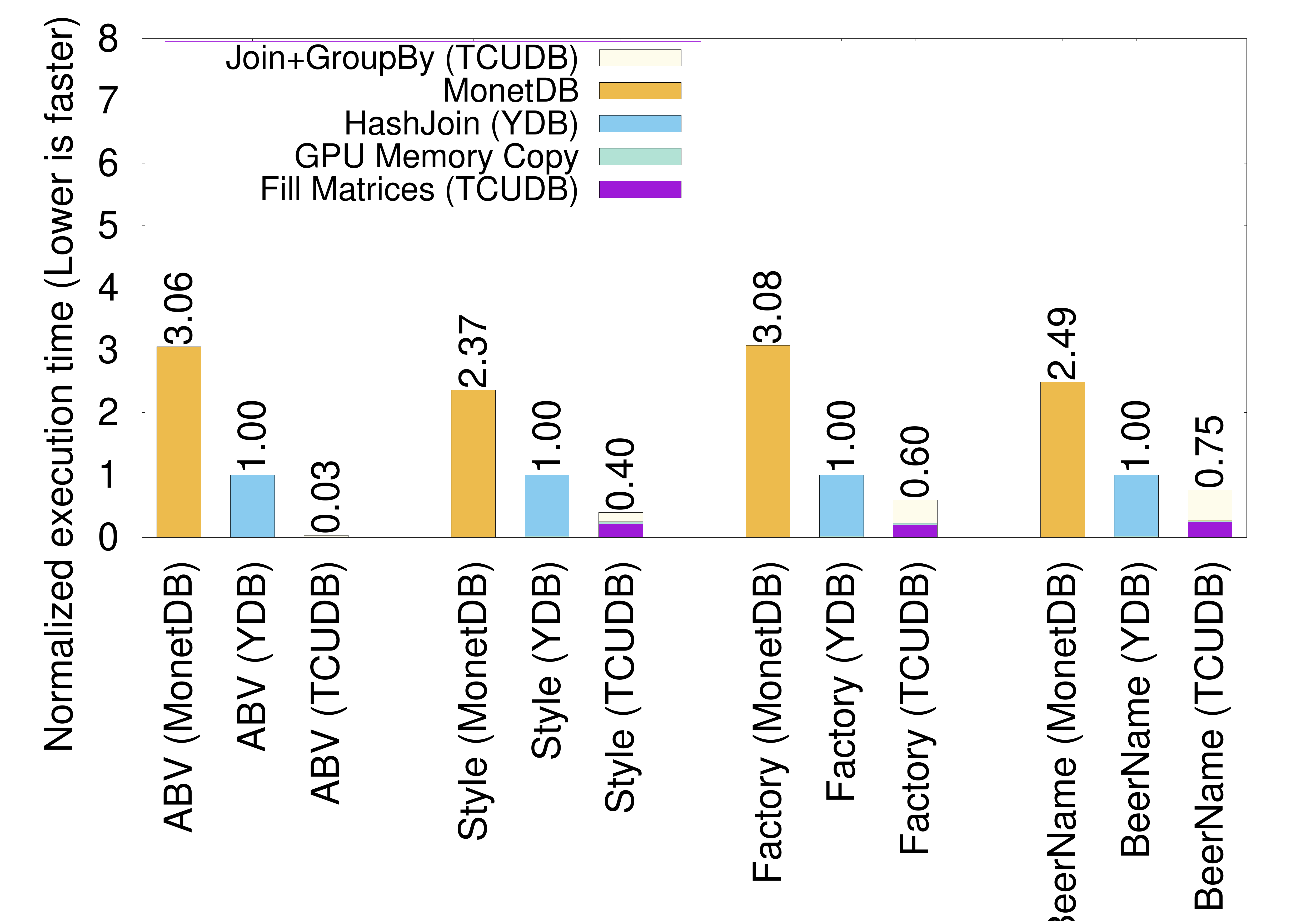} &
\hspace{-0.25in}\includegraphics[width=2.5in]{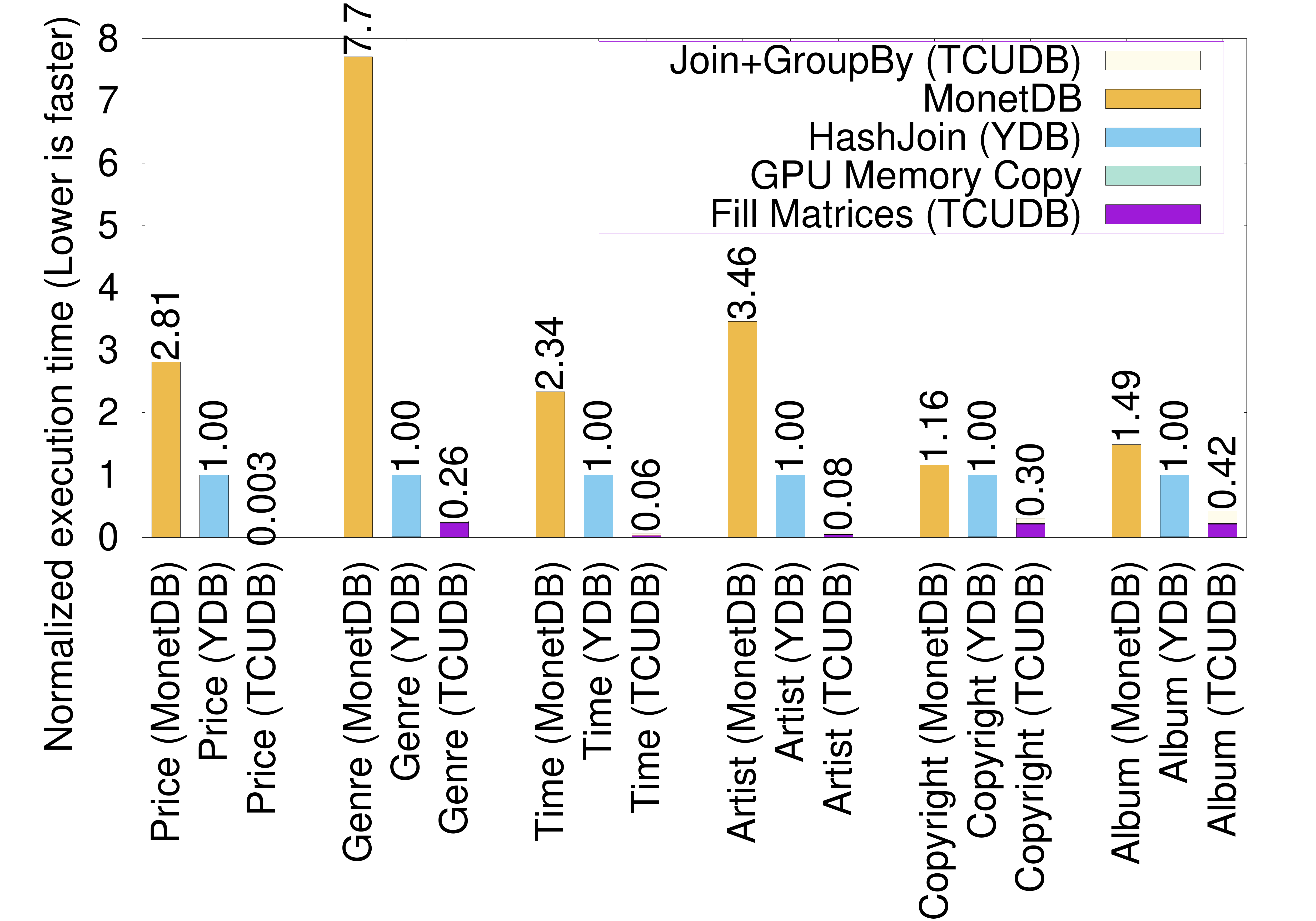} &
\hspace{-0.25in}\includegraphics[width=2.5in]{Figures/case_EM_scaleup.pdf} \\
(a) & (b) & (c)\\
\end{tabular}
\vspace*{-0.1in}
\caption{The relative runtime of 
the EM-blocking queries on \TCUDB{} using the default deepmatcher
datasets (a) BeerAdvo-RateBeer (b) iTunes-Amazon and (c) scaled
iTunes-Amazon, compared to MonetDB and YDB running the same query as the
baseline.}
\label{fig:case_EM}
\end{center}
\vspace*{-0.2in}
\end{figure*}

Due to the limited 16-bit precision of TCUs, they cannot generate 100\%
accurate results in some cases. Table~\ref{table:error} shows the mean absolute percentage error
(MAPE) rates in performing matrix multiplication queries. In the cases where
the values are only 0s and 1s -- similar to the cases of Q1 and Q2,
the generated \TCUDB{} operations can always produce accurate outputs.
Therefore, the result implies that \TCUDB{} never leads to incorrect
outcomes for sub-queries like Q1 and Q2. When we enlarge the value ranges,
we start to see errors in results, but with very limited imprecision -- even
in the worst case, the MAPE is lower than 0.01\%. We believe this error rate is
acceptable in most cases.
\hungwei{
This level of data error does not cause any inexact query results for the entity matching or the microbenchmark workloads.
For numerical analysis such as SSB, the result values can have minor error rates typically less than 0.001\% for cases with input values larger than $2^{15}$ or matrices with
a dimension larger than 8192 due to the 16-bit representation. However, the error rate is very insignificant and never results in misplacement of rankings and orderings of the query results.
}

\subsubsection{Entity Matching}
\label{sec:EM}
Entity matching (EM), also known as entity resolution, fuzzy join, and record
linkage, searches records correspond to the same real-world entities from 
different data sources~\cite{elmagarmid2006duplicate,dong2013big,christophides2015entity,konda2016magellan}.
%
A key component of EM is \emph{blocking}~\cite{konda2016magellan,gagliardelli2019sparker,papadakis2020blocking}. 
Given two tables of entity records, the goal of blocking is to apply matching heuristics to 
quickly generate candidate pairs of records that are likely to be real matches,
which are later processed by a more accurate pairwise classifier (aka the \emph{matcher}).
Scalability is the main challenge of blocking as the heuristics are typically natural join conditions 
(e.g., selecting products with the same brand) that often produce large join results.
Therefore, we expect that \TCUDB{} can provide significant performance gain for this
EM workload.

To validate this hypothesis, we evaluate \TCUDB{}'s performance on two real EM datasets BeerAdvo-RateBeer and iTunes-Amazon from 
the Deepmatcher benchmark~\cite{deepmatcher}. The BeerAdvo-RateBeer dataset contains two 
tables, where one of them contains 3,777 rows and the other contains 2,671 rows, from different sources. 
Each table has the same table schema 
with five attributes \texttt{\{ID, BEER\_NAME, FACTORY, STYLE, ABV\}}. Table~\ref{tab:beer_tab} 
reveals the number of distinct values of each attribute, which acts as one matrix dimension for \TCUDB{}
when performing join operation. We evaluate the following query on BeerAdvo-RateBeer
dataset to perform blocking:
\begin{lstlisting}[
        	language=SQL,
        	showspaces=false,
        	basicstyle=\ttfamily\small,
        	columns=fullflexible,
          frame=single,
  			  breaklines=true,
        	numbers=left,
          numberstyle=\tiny,
        	commentstyle=\color{gray}
        ]
-- EM-blocking query for BeerAdvo-RateBeer dataset:
SELECT TABLE_A.ID, TABLE_A.BEER_NAME,
       TABLE_B.ID, TABLE_B.BEER_NAME
FROM TABLE_A, TABLE_B
WHERE TABLE_A.ABV = TABLE_B.ABV; -- attributes may vary
\end{lstlisting}

\begin{table}[t]
\vspace*{0.1in}
\small
\centering
\begin{tabular}{|l|c|c|c|c|}
\hline
Attribute & ABV & Style & Factory & BeerName \\ \hline
\#distinct values & 20 & 71 & 3678 & 6228 \\ \hline
\end{tabular}
\caption{Distinct values in BeerAdvo-RateBeer dataset.}
\label{tab:beer_tab}
\vspace*{-0.15in}
\end{table}

\setlength{\tabcolsep}{3pt}
\begin{table}[t]
\centering
\small
\hspace*{-0.2in}
\begin{tabular}{|l|c|c|c|c|c|c|}
\hline
Attribute & Price & Genre & Time & Artist & Copyright & Album \\ \hline
\#distinct values & 12 & 813 & 908 & 2418 & 3197 & 6004 \\ \hline
\#distinct values & 25 & 1614 & 1208 & 6420 & 8199 & 11005 \\
  (scaled)	  &    &      &      &      &      &       \\
\hline
\end{tabular}
\caption{Distinct values in iTunes-Amazon dataset.}
\vspace*{-0.15in}
\label{tab:itunes_tab}
\end{table}

The iTunes-Amazon dataset contains two tables, where one of them has 6,907 rows and the other
has 55,923 rows, from iTunes and Amazon music. Both tables share the same table schema 
with seven attributes $\mathtt{ID}$, $\mathtt{PRICE}$, $\mathtt{GENRE}$, $\mathtt{TIME}$, $\mathtt{ARTIST}$, $\mathtt{COPYRIGHT}$, 
and $\mathtt{ALBUM}$. 
Table~\ref{tab:itunes_tab} shows the number of distinct values for each attribute in 
the iTunes-Amazon dataset. We perform the following query on the iTunes-Amazon dataset for blocking:
\begin{lstlisting}[
          language=SQL,
          showspaces=false,
          basicstyle=\ttfamily\small,
          columns=fullflexible,
          frame=single,
          breaklines=true,
          numbers=left,
          numberstyle=\tiny,
          commentstyle=\color{gray}
        ]
-- EM-blocking query for iTunes-Amazon dataset:
SELECT TABLE_A.ID, TABLE_A.SONG,
       TABLE_B.ID, TABLE_B.SONG
FROM TABLE_A, TABLE_B
WHERE TABLE_A.ARTIST = TABLE_B.ARTIST; -- attributes may vary
\end{lstlisting}

Figure~\ref{fig:case_EM} presents the result of running the above EM-blocking queries on the
two datasets and different attributes. As the execution time varies
significantly among different queries, we use \YDB{} running the same query as
the baseline and show the relative execution time. \TCUDB{}
outperforms \YDB{} in most cases, achieving a maximum speedup of 288\x{} among our
experiments.

\TCUDB{} is especially effective when the
number of distinct values is small.
For the BeerAdvo-RateBeer dataset in Figure~\ref{fig:case_EM}(a), \TCUDB{} is at most
33\x{} faster
than \YDB{} when
searching for matches on the ABV attribute where there are only 20 distinct
values. For the iTunes dataset in Figure~\ref{fig:case_EM}(b), \TCUDB{} further
shows 288\x{} speedup over \YDB{} when performing entity matchings on the Price
attribute that only has 12 distinct values. 
\ignore{We found in some cases, \YDB{}
underperforms MonetDB. This is because in EM queries, \YDB{} may need to
perform multiple }
When the number of distinct values becomes larger, the performance advantage of
\TCUDB{}'s operators relying on dense matrix operations over \YDB{} starts to shrink, for the reason we have described in
Section~\ref{sec:microbenchmark}. However, as \TCUDB{} uses TCU-spMM in
these cases,
\TCUDB{} still outperforms \YDB{} and MonetDB in all
cases.

\vspace*{-\lineskip}
\noindent
\textbf{Scaling up. }
To demonstrate the ability of \TCUDB{} and the query optimizer in dealing
with larger EM datasets, we synthesized an iTunes-Amazon
dataset 
by randomly duplicating each input table's entry values. 
The resulting dataset contains 111,846 records in the larger input source
and 13,814 in the smaller one. 
The \#distinct values (scaled) show the resulting distinct values in each
attribute field of this synthetic dataset. 

Figure~\ref{fig:case_EM}(c) shows the relative execution time of
\TCUDB{}, compared with \YDB{} running the same query. \TCUDB{} still
outperforms \YDB{} in most cases, by up to 216\x{} when performing matching on
the price field. When \TCUDB{} performs the query on artist, album and copyright
fields, the query optimizer detects that these cases contain way too many
distinct values and the pure TCU operator cannot
efficiently process the query since the input matrices are sparse. Therefore, \TCUDB{}
uses a TCU-SpMM operator for query processing and achieves
more than 6.67\x{} and 7.8\x{} speedup on Copyright and Album, respectively, over \YDB{} that essentially performs sparse matrix
multiplications using CUDA cores. 


\subsubsection{PageRank}
\label{sec:PageRank}
\begin{figure*}[t]
\vspace*{-0mm}
\begin{center}
\begin{tabular}{ccc}
\hspace{-0.15in}\includegraphics[width=2.5in]{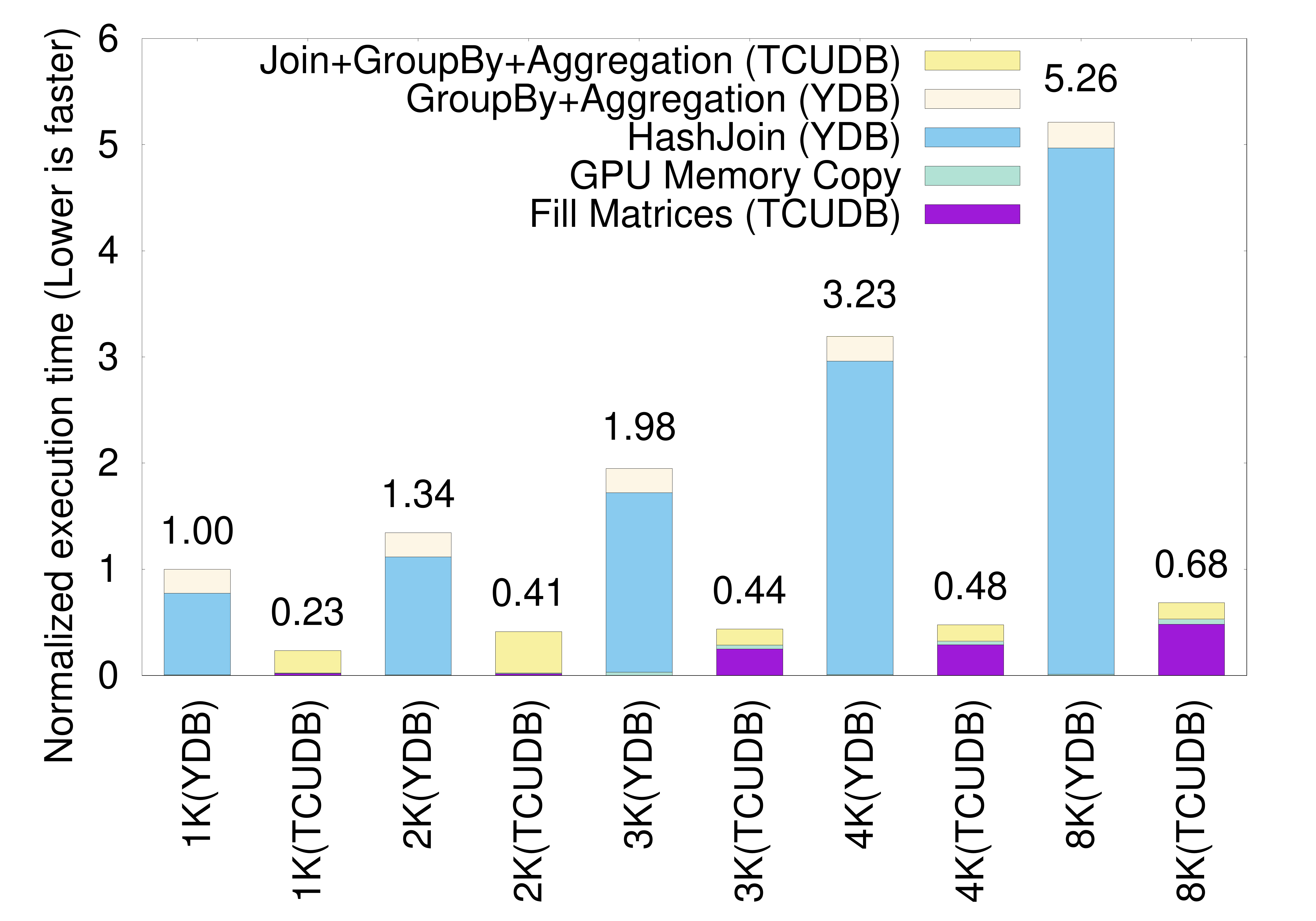} &
\hspace{-0.2in}\includegraphics[width=2.5in]{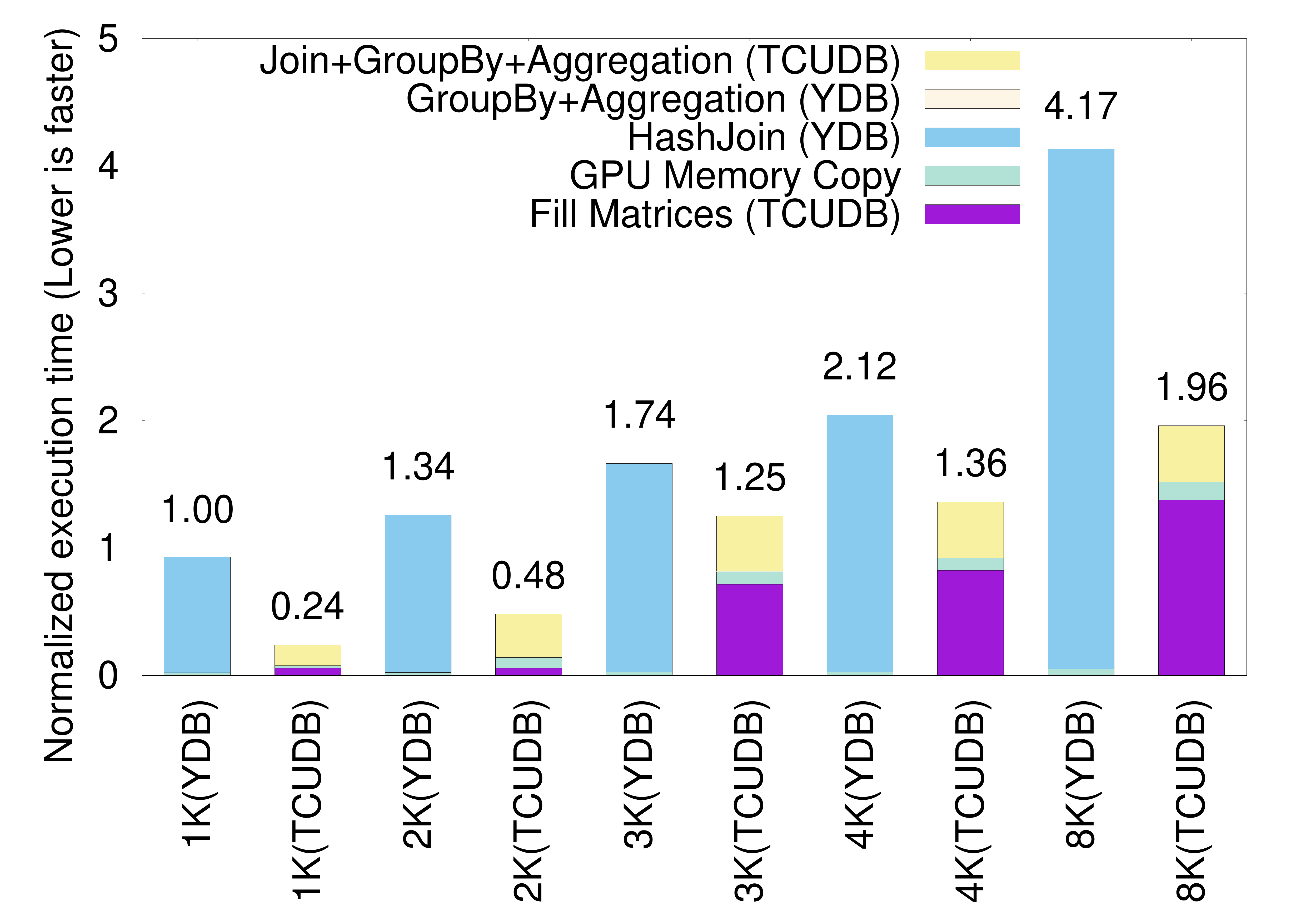} &
\hspace{-0.2in}\includegraphics[width=2.5in]{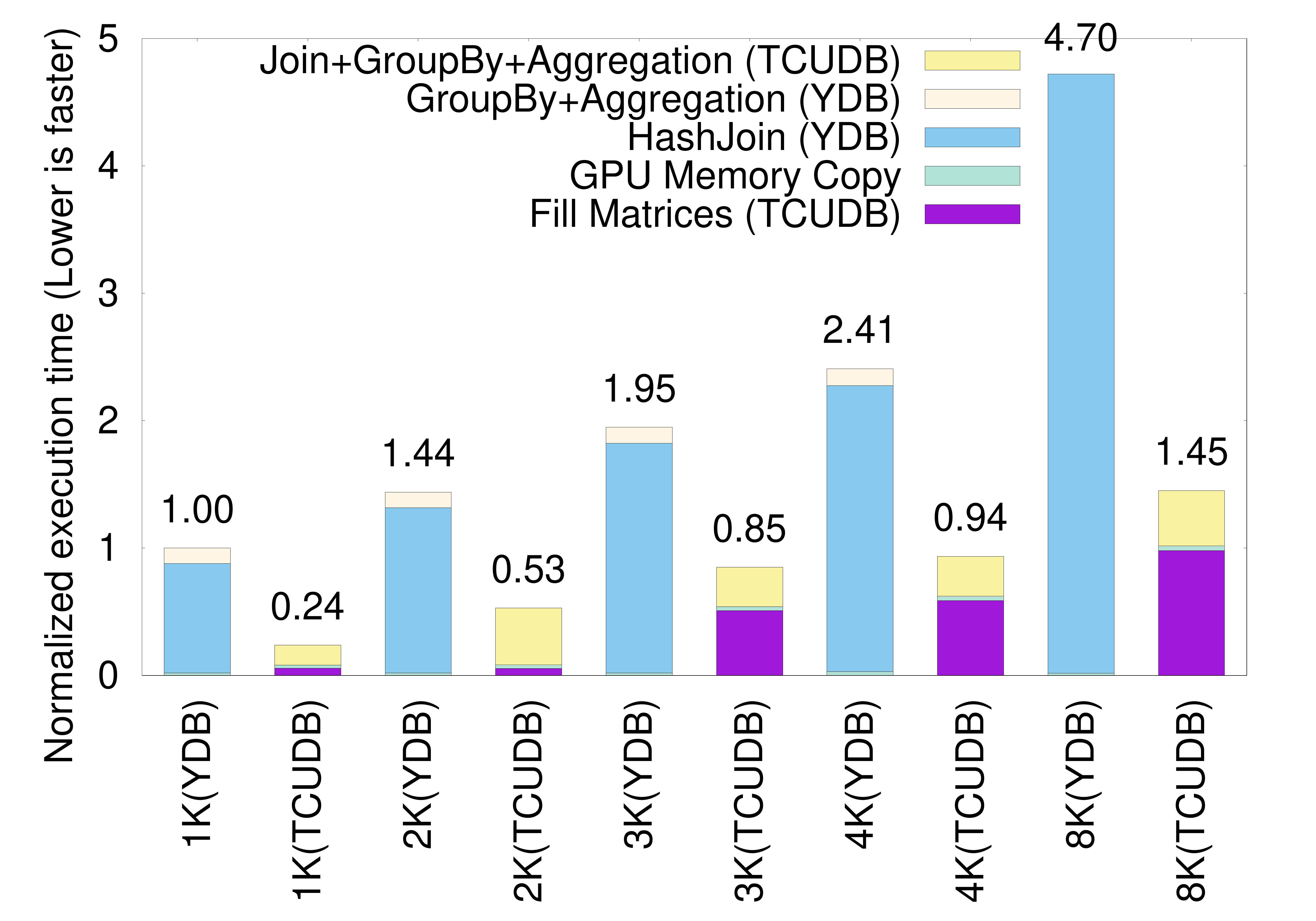} \\
(a) & (b) & (c)\\
\end{tabular}
\vspace*{-0.1in}
\caption{The relative execution time of
executing PageRank queries (a) Q1, (b) Q2, and (c) Q3 on \TCUDB{}, using \YDB{} running the same query as the
baseline. Each value equals the actual query time divided by \YDB{}'s runtime on the 1k
table.}
\label{fig:case_PR}
\end{center}
\vspace*{-0.2in}
\end{figure*}

To demonstrate \TCUDB{}'s ability in processing graph-related queries as
well as data analytics, we also evaluate \TCUDB{} in performing the PageRank
algorithm. PageRank algorithm consists of three steps: (1) computing the 
out-degree of each node, (2) initializing the value of each node, and
finally, (3) calculating the PageRank iteratively. 
The whole PageRank algorithm can be implemented as the following three 
queries:
\ignore{
Given PageRank will compute multiple rounds, we assume the data is cached. 
We evaluate the performance of \TCUDB{} and \YDB{} using three queries that represents 
the above three steps separately. Consider the following queries as examples:
}
\ignore{
We select PageRank (PR) algorithm to evaluate the performance of \TCUDB{} on a directed 
graph dataset. We used the Pennsylvania road network dataset from SNAP~\cite{leskovec2009snap} 
that contains 1.08M nodes and 1.54M edges. As the whole dataset is too large to fit in the device memory, we perform random clustering to shrink the graph for our experiment, the shrunken graph information in Table~\ref{tab:pr_tab}.
}

\ignore{
PageRank algorithm consists of three steps: (1) Compute the out-degree of each node. (2) Initialize the value of each node. (3) Calculate the pageranks in multiple iterations. Given PageRank will compute multiple rounds, we assume the data is cached. We evaluate the performance of \TCUDB{} and \YDB{} using three queries that represents the above three steps separately. Consider the following queries as
examples:}

\begin{lstlisting}[
          language=SQL,
          showspaces=false,
          basicstyle=\ttfamily\small,
          columns=fullflexible,
          frame=single,
          breaklines=true,
          numbers=left,
          numberstyle=\tiny,
          commentstyle=\color{gray}
        ]
-- PR Q1: compute out-degree
SELECT NODE.ID, 
       COUNT(EDGE.SRC)
FROM NODE, EDGE
WHERE NODE.ID = EDGE.SRC
GROUP BY NODE.ID;
\end{lstlisting}
\begin{lstlisting}[
          language=SQL,
          showspaces=false,
          basicstyle=\ttfamily\small,
          columns=fullflexible,
          frame=single,
          breaklines=true,
          numbers=left,
          numberstyle=\tiny,
          commentstyle=\color{gray}
        ]
-- PR Q2: initialize values
SELECT NODE.ID, 
       (1-@alpha)/@num_node as rank
FROM NODE, OUTDEGREE
WHERE NODE.ID = OUTDEGREE.ID;
-- @alpha is 0.85 by default
\end{lstlisting}
\begin{lstlisting}[
          language=SQL,
          showspaces=false,
          basicstyle=\ttfamily\small,
          columns=fullflexible,
          frame=single,
          breaklines=true,
          numbers=left,
          numberstyle=\tiny,
          commentstyle=\color{gray}
        ]
-- PR Q3: calculate the PageRank score
SELECT  
      SUM(@alpha * PAGERANK.rank / OUTDEGREE.DEGREE)
      + (1-@alpha)/@num_node
FROM PAGERANK, OUTDEGREE
WHERE PAGERANK.ID = OUTDEGREE.ID;
-- @alpha is 0.85 by default
\end{lstlisting}
\ignore{
\begin{figure}
\includegraphics[width=70mm,scale=0.9]{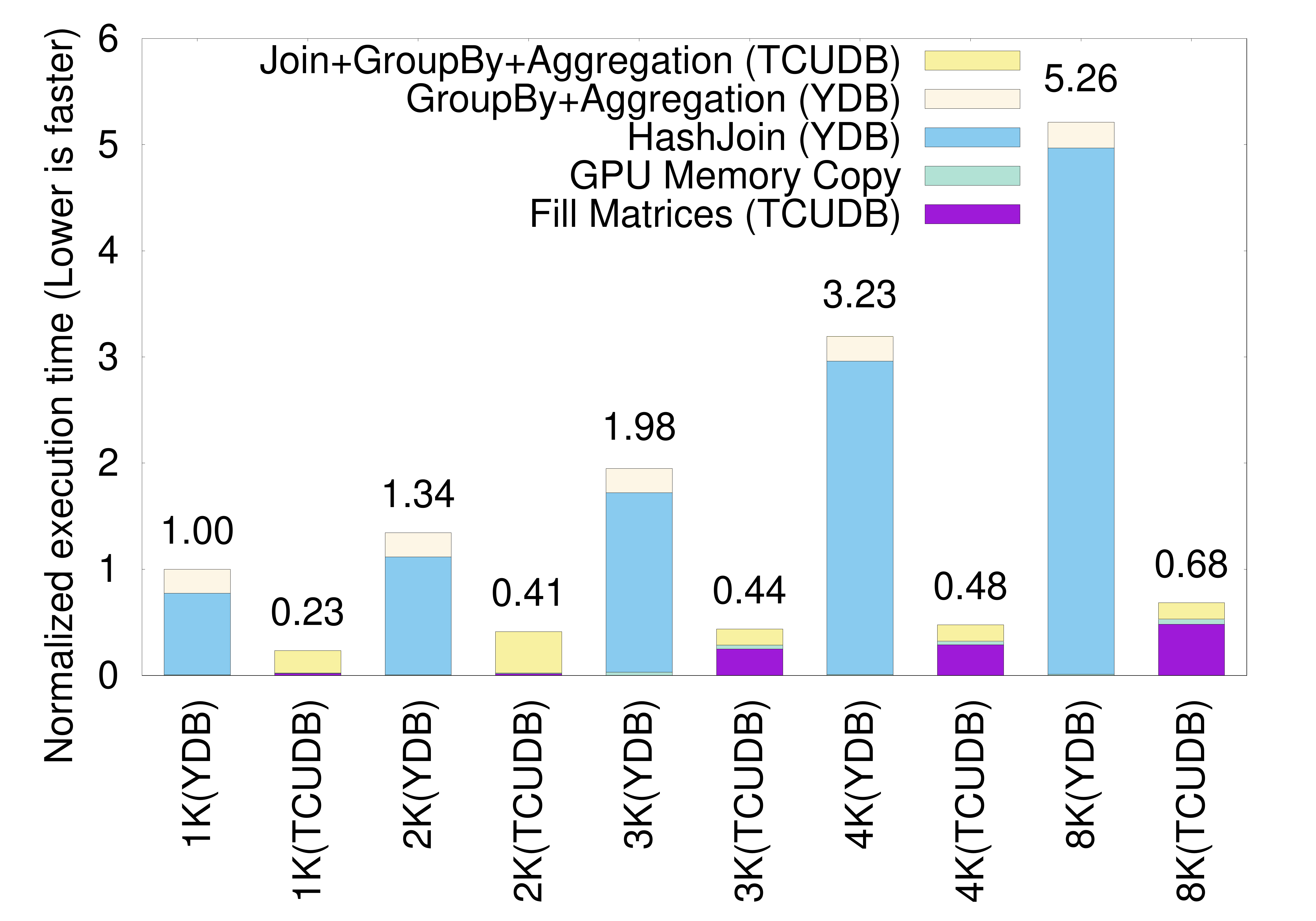}
\caption{\small{The performance of PageRank Q1.}}
\label{fig:case_PR_q1}
\end{figure}
}

Among these three queries, PR Q1 represents step 1, PR Q2 represents step 2 and PR Q3 represents
step 3. A complete run of the PageRank algorithm will invoke PR Q1 and PR Q2
once and execute PR Q3 several times until the PageRank scores converge or
reach the maximal number of 
iterations.

\setlength{\tabcolsep}{6pt}
\begin{table}
\small
\centering
\begin{tabular}{|l|c|c|c|c|c|c|c|}
\hline
 \#Nodes & 1024 & 2048 & 3072 & 4096 & 8192 & 16384 & 32768\\ \hline
\#Edges & 2058 & 4152 & 6280 & 8450 & 17444 & 37106 & 82070 \\ \hline
\end{tabular}
\caption{Reduced graph information.}
\vspace*{-0.3in}
\label{tab:pr_tab}
\end{table}

We used the Pennsylvania road network dataset from SNAP~\cite{leskovec2009snap} 
that contains 1.08M nodes and 1.54M edges as the input dataset. 
Evaluated \TCUDB{} under different sizes of graphs, we created a subset of
the original  graph for our 
experiments using the most popular $N$ nodes and preserving the
connectivity of selected nodes in the original graph. Table~\ref{tab:pr_tab}
describes the characteristics of the resulting graphs.
Figure~\ref{fig:case_PR} illustrates the relative execution time and the
breakdown of latency in each system component for all three queries. We
normalized the execution time to run the same query using the graph with
1K nodes on \YDB{}.  

Though the computation of out-degree using PR Q1 is a one-pass task
(Figure~\ref{fig:case_PR}(a)),
\TCUDB{}'s pure TCU Join/Aggregation/Groupby operator still has advantages when the graph is small, 
by up to 3.6\x{} speedup with 1K graph. For graphs with more than 3K nodes,
\TCUDB{} selects TCU-SpMM to exercise
the Join/Aggregation/Groupby operator due to the low density in their adjacency matrices. Compared with
a pure TCU Join/Aggregation/Groupby operator, a TCU-SpMM-based operator spends
more time in creating operator inputs. However, as the TCU-SpMM-based operator
skips submatrices with all 0s, TCU-SpMM significantly reduces the computation
time on matrix multiplications and allows \TCUDB{} to outperform \YDB{} that
essentially performs sparse matrix operations on CUDA cores by up to
7.69\x{}. 
\ignore{
As Figure~\ref{fig:case_PR}
points out, the main reason is because of the inefficiency of
TCU-accelerated DB operations in sparse matrix operations. However, the
performance difference is so minor that the cost estimation function cannot
easily catch up, so we still see a slight 4\% slowdown, compared to \YDB{}. }

PR Q2 is also a one-time process in the PageRank algorithm but requires additional arithmetic to
initialize the values for PR Q3. Figure~\ref{fig:case_PR}(b) shows that \TCUDB{} consistently performs better 
than \YDB{}. with speedup ranging from 1.40\x{} to 4.18\x{}. 
Similar to Q1, \TCUDB{} uses a dense TCU operator for graphs smaller than 2K
and uses TCU-SpMM's Join/Aggregation/Groupby to exercise queries for larger
graphs.
\ignore{
 as one of the input
dimensions of the input matrices in TCU-accelerated join grows proportional
to the number of distinct nodes in the graph. In contrast, though the
number of vector operations also grows in \YDB{} as the number of nodes
increases, the size of the vector is smaller due to the lower probability of
matching.} 

Figure~\ref{fig:case_PR}(c) shows the performance of \TCUDB{} and \YDB{} in
performing PR Q3, the core of the PageRank algorithm that the algorithm executes multiple times until values converge.
In our experiments, we performed PR Q3 for 50 iterations for each configuration. 
For PR Q3, \TCUDB{}'s Join/Aggregation/Group operator improves the
execution time of arithmetic calculations over the multi-step
process in \YDB{}. \TCUDB{} is 4.22\x{} faster than \YDB{} with 1K nodes in the
graph. Even with graphs containing 8K nodes, \TCUDB{} still outperforms
\YDB{} by 3.24\x{}, as TCU-SpMM's Join/Aggregation/Groupby skips
submatrices containing all 0s.

\subsection{Comparison with Graph Database Systems}
\label{sec:graphDB}
\cfigure[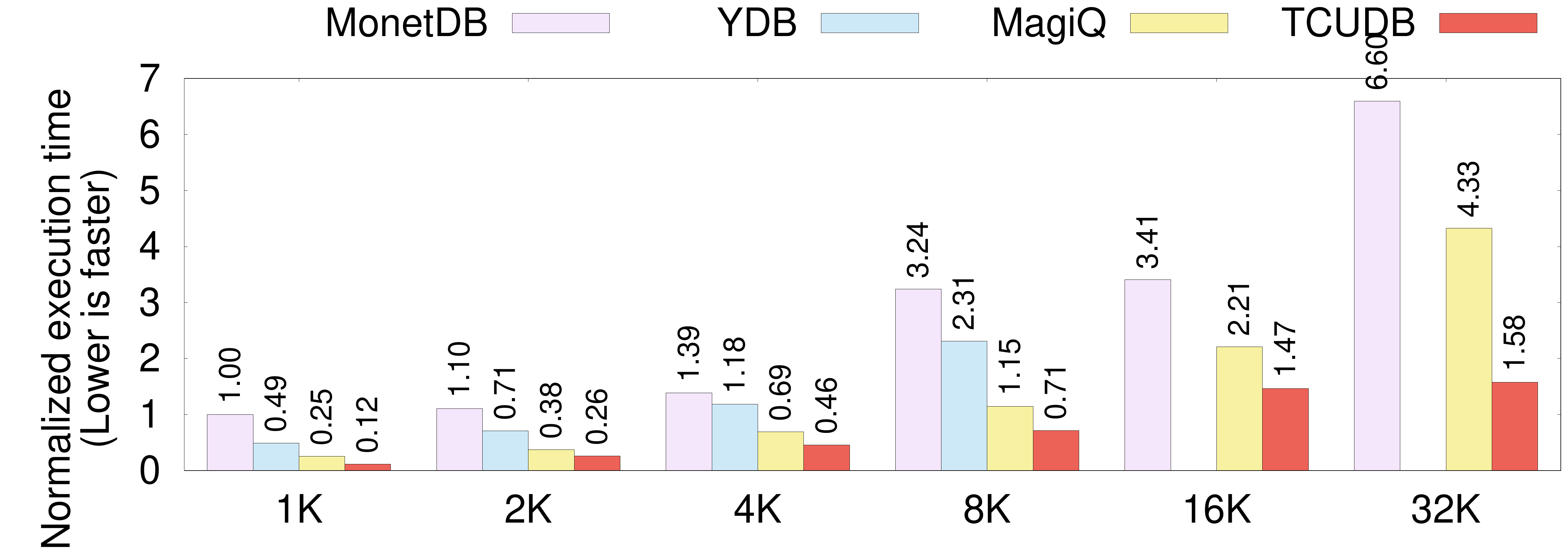,
{The relative latency of the core join and aggregation operation when running
PageRank Q3 in MonetDB, YDB, MAGiQ, and \TCUDB{}. \ignore{\yuliang{explain the missing bar?}}},fig:MagiQ]

\hungwei{\TCUDB{} demonstrates the potential of using relational database
engines to analyze datasets that are originally graphs through case studies
on PageRank. On the other hand, graph database systems provide more natural
representations and storage layouts to serve the same purpose. To investigate 
the strength and the implications of \TCUDB{} in the future
advancement of graph database systems, }
\hungwei{this section compares the performance of \TCUDB{} on the PageRank 
algorithm with the state-of-the-art graph query engine
MAGiQ~\cite{Jamour2019MAGiQ}.
In contrast to the table-style storage that relational database systems and
 \TCUDB{} use, MAGiQ's backend storage is organized as 2-dimensional
key-value pairs, typically already in some sparse matrix formats. MAGiQ translates 
the queries described by SPARQL into a set of GraphBLAS~\cite{GraphBLAS}
calls on these sparse matrices.} 


\hungwei{We use the same SNAP dataset as in Section~\ref{sec:PageRank} to evaluate 
the PageRank performance of MAGiQ with GPU and \TCUDB{}.
Figure~\ref{fig:MagiQ} compares the performance of MAGiQ
and \TCUDB{} with MonetDB and YDB as references. However, the released
version of YDB can only support these queries with datasets containing at
most 8,192 nodes. 
Due to the large overhead of retrieving
sparse matrices in MAGiQ compared to other counterparts, we only
present the latency of the core join and aggregation operations in each
experiment. 
The presented numbers are PageRank Q3's performance on the sub-sampled
graphs listed in Table~\ref{tab:pr_tab}.
MAGiQ outperforms YDB, the pure GPU query engine on
relational databases, in all cases, demonstrating that a customized graph
database engine does provide a more efficient platform for graph analytics on the
same architecture. Meanwhile, \TCUDB{} outperforms MAGiQ in all evaluated cases.
The main reason is that 
TCUs allow \TCUDB{} to more efficiently exercise these queries than GraphBLAS that uses
only conventional GPU cores at this moment. 
We observed that the difference is more significant as the graph becomes larger and more sparse.
These results help us generate two insights. First, with TCUs, graph
analytics can be efficient with existing relational databases. Second, graph
databases can also be more efficient if their backends can leverage TCUs as
\TCUDB{} does. }

\ignore{
The advantage of \TCUDB{} still decays as the input sizes
grow, because the distinct node IDs also increase as the input data grows. 
}
\ignore{
As we execute PR Q3 several times using TCUs, the initialization overhead of
TCUs is amortized and becomes insignificant. This also implies that \TCUDB{}
can potentially avoid the initialization overhead if CUDA provides an API to
keep TCU activated all the time or at least activate TCU hardware in
advance.} 


\subsection{\TCUDB{} on different GPU architectures}
\label{sec:3090vs2080}
\cfigure[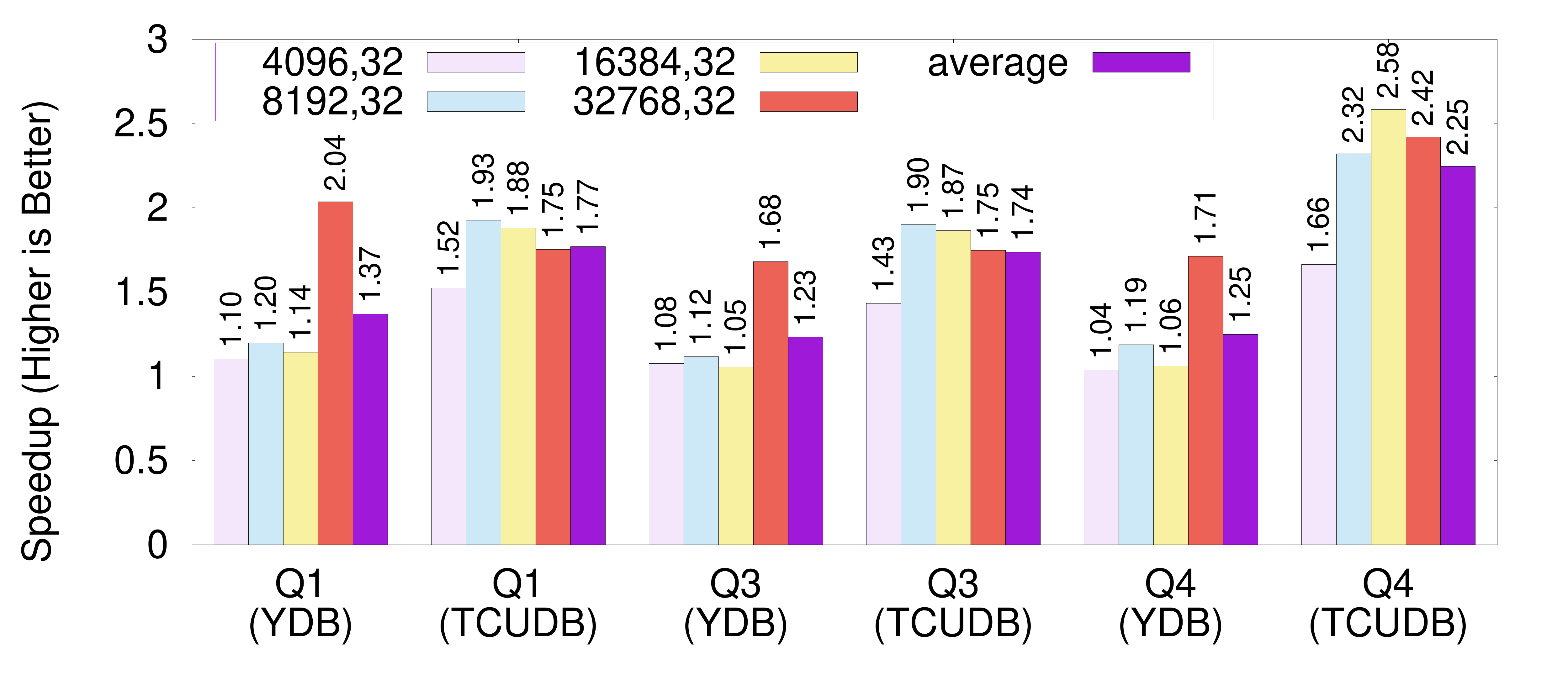,
{The microbenchmark speedup of using RTX 3090 over RTX 2080 for Q1, Q3, Q4 on \TCUDB{} and \YDB{}.
Each value equals RTX 2080 time divided by RTX 3090 time.},fig:2080vs3090]
To investigate the performance scaling on different GPU architectures and
their implications to the design of the TCU-accelerated DB engine, we
perform experiments on NVIDIA's 2080, which uses an earlier Turing GPU architecture
with the last generation TCU available. 

\ignore{
To demonstrate that the latest generation GPUs have better scalibility and computing capability, 
we profiled operator-level performance using using RTX 3090 GPU and RTX 2080 GPU. Compared to RTX 3090, 
the previous generation RTX 2080  GPU contains 8GBs of device memory and 468 2nd generation Tensor Cores. 
Given the performance of matrix multiplication query is dominated by data movement (cudaMemcpy), shown 
in Figure~\ref{fig:case_MM_breakdown}, it is hard to tell whether RTX3090 GPU can provide better performance. 
We run microbenchmarks on the same queries Q1, Q3, Q4 mentioned in Section~\ref{sec:microbenchmark} using 
RTX 3090 GPU and RTX 2080 GPU.}

Figure~\ref{fig:2080vs3090} compares the performance of microbenchmarks on the same queries Q1, Q3, Q4 
mentioned in Section~\ref{sec:microbenchmark} using both \YDB{} and \TCUDB{} on RTX 3090 GPU and RTX 2080
GPU. The baseline ran the same query using the same DB engine on RTX
2080. We observed that \TCUDB{} performs better
generation-over-generation -- when using RTX 3090 \TCUDB{} achieved an average speedup of
1.77\x{} on Q1, 1.74\x{} on Q3 and 2.25\x{} on Q4, but \YDB{} only achieved
1.37\x{} on Q1, 1.23\x{} on Q3 and 1.25\x{} on Q4. It is worth noting that
RTX 3090 contains only 328 Tensor Cores compared to 368 Tensor Cores in RTX 2080.
On the other hand, the RTX 3090 has 10496 conventional CUDA GPU cores for
vector processing while RTX 2080 only has 2944 of them. 
The results reveals that
the performance scaling of Tensor Cores in newer generations of GPU architectures
is stronger than conventional vector processing cores, given that RTX 3090 has
fewer Tensor Cores, 3.4\x{} more CUDA cores, but \TCUDB{}'s speedup is more
significant on RTX 3090. This result also 
indicates applications, including DB engines, with a larger portion relying on TCUs
will expect to receive more performance gains when new GPU
architectures are used. 
\section{Related Work}
\label{sec:related}

\noindent
\textbf{Hardware-accelerated DB's. }
Integrating advanced hardware accelerators into database systems has been an active line of research 
for the past few decades.
Commonly considered accelerators include 
GPUs~\cite{govindaraju2004fast,walkowiak2010exploring,wu2012kernel,yuan2013yin,bress2013time,Wang:2014:CAQ:2732967.2732976,wu2014multipredicate,li2016hippogriffdb,paul2016gpl,SparkGPU,GPUDBCodeGeneration,chrysogelos2019hardware,10.1145/3183713.3183734,shanbhag2020study} and
FPGAs~\cite{Mueller:2009:FWD:1559845.1559965,wang2016accelerating,mahajan2018rdbms,FPGADataProcessing,fang2020memory}.
Optimization techniques have been proposed for database operators including 
Select~\cite{sitaridi2013optimizing}, 
Join~\cite{he2008relational,he2013revisiting,8855452,HardwareConsciousGPUHashJoin},  
Sort~\cite{govindaraju2006gputerasort} and
Group-by Aggregate~\cite{karnagel2015optimizing}.
In particular, to support star schema queries,
\YDB{}~\cite{yuan2013yin} implements these operators into a data warehousing engine, 
which we used as a baseline for \TCUDB{}.
GPUs have also been incorporated into industrial DB engines such as OmnisciDB~\cite{omniscidb}, Kinetica~\cite{kinetica}, and BlazingSQL~\cite{blazingsql}. 
\ignore{
To the best of our knowledge, \TCUDB{} is the first database system that leverages
Tensor Core Units (TCUs) to accelerate database
queries while addressing the limitation of TCUs, but largely inspired by prior projects
in using TCUs for scan/reduction~\cite{TCUScan} as well as TensorFlow and
Google service's exclusive TPU platform for relational queries~\cite{TPUDB}.
}

With GPUs reducing the computation time but the increasing volume of
datasets, the data movement overhead becomes more significant to the degree
that DB engines must be aware~\cite{pelley2013storage,10.1145/2882903.2882936}. Several GPUDB
systems incorporate GPU RDMA techniques~\cite{GPUDirect, DirectGMA, kim2014gpunet,
Gullfoss, zhang2015nvmmu, hippogrifficcd} to directly access data on the storage
devices~\cite{zhang2015hetero,li2016hippogriffdb,9210197} or efficiently
exchange data among multiple GPUs~\cite{NVLinkGPUDB}, bypassing the host system's main memory.
This paper is orthogonal but will receive significant benefit from this line of research
projects.
To fundamentally address the data movement overhead, DB systems can push down the
computation of query processing into existing or additional hardware logic to
offload part of the computation instead of using computing resources on the
host system~\cite{JoinOnSSD,SmartSSD,BlueDBM,wang2016ssd,ACIS,MILC}. However, due to
the power and hardware budget of memory/storage devices, the computing
resources near data locations are typically limited. For the cases studied
in this paper, DB systems still have to rely on host computing resources
(i.e., GPUs, TCUs, FPGAs and TPUs) to
efficiently perform the received queries. 
\ignore{
In addition, if future processing
technologies allow storage devices to contain TCU/TPU-like accelerators,
the implementation of operators and query optimizations proposed by this
paper will still be valid for executing queries in these devices. 
}

\hungwei{
\smallskip
\noindent
\textbf{Matrix processors in relational databases. }
To the best of our knowledge, \TCUDB{} is the first database system that 
fully leverages Tensor Core Units (TCUs) as matrix processors to
accelerate compute-intensive database queries.
Prior work~\cite{TCUScan} leverages TCUs for scan/reduction operators by mapping
scan/reduction into matrix-vector products. However, \cite{TCUScan} 
only treats TCUs as wider vector processors
leveraging TCU's fused operations that can 
perform multiplications and accumulations in a single operation. 
In contrast, \TCUDB{} transforms queries into matrix-matrix operations so that 
it can fully utilize TCUs' nature as matrix processors. 
Prior work~\cite{TPUDB} investigated the feasibility of accelerating
relational queries using Google Cloud's closed-architecture TPU platform
and proprietary version of TensorFlow. 
However, due to limitations of the platform,
\cite{TPUDB} only accelerates vector-based operators such as 
reduced sum. Its implementation can only support single-table queries
(called Dimension Join in \cite{TPUDB}).
On the other hand, \TCUDB{} can support a wide range of queries include
two-way natural joins by leveraging TCUs for matrix operations.
}

\vspace*{\lineskip}
\noindent
\textbf{Join processing as matrix multiplication. } A key technical contribution of \TCUDB{}
is to cast the join operator as dense matrix multiplication. While being unconventional 
due to the high theoretical computational complexity, this idea was explored in \cite{amossen2009faster}
and more recently in \cite{deep2020fast}. 
In particular, \cite{deep2020fast} proposed a fast join algorithm that combines
worst-case optimal join algorithms~\cite{ngo2018worst} and fast matrix multiplication.
The authors also provide a CPU-based implementation highlighting performance gain
from the highly-optimized linear algebra framework such as Intel MKL~\cite{wang2014intel}.
The implementation achieves up to 50\x{} performance improvement compared to baselines.
In \TCUDB{}, we further push this trend by leveraging NVIDIA's TCUs that are specialized
for tensor processing, which commonly appears in deep learning workloads to achieve
up to 288\x{} performance gain. 

\hungwei{
\vspace*{\lineskip}
\noindent
\textbf{Graph queries as matrix operators. } 
Processing queries as matrix operators have also been
considered in the context of graph databases. 
In particular, MAGiQ~\cite{Jamour2019MAGiQ} accelerates SPARQL queries
on RDF graphs by translating queries into sparse matrix linear algebra programs.
We have discussed the key differences between \TCUDB{} and MAGiQ in Section
\ref{sec:graphDB}. Our experiment results also show that 
integrating \TCUDB{}'s strategy of executing those matrix operators in TCUs
can be an interesting optimization opportunity for graph query engines like MAGiQ.
}


\vspace*{\lineskip}
\noindent
\textbf{Advanced in-database analytics. }
To accommodate the exponential growth in data science and machine learning applications,
a recent line of
work~\cite{hellerstein2012madlib,aberger2018levelheaded,AIDA,scalaleLA,dolmatova2020relational,LaraDB,SystemsforLA,10.1145/3331445} 
focuses on supporting advanced analytics queries that involve
linear algebra (LA) operators. 
\TCUDB{} shares the goal of LevelHeaded~\cite{aberger2018levelheaded} in identifying the 
worst-case optimal join (WCOJ)~\cite{ngo2018worst} or LaraDB's rule-based translation 
between relational queries and parallel LA queries, but \TCUDB{}
additionally provides the capability of 
translating (parts of) the query to 
TCU-accelerated matrix multiplication operator(s) and different sets of opportunities
from the orders of magnitude speedup by TCUs in such operations.
\TCUDB{} also offers a better system architecture by making TCU-accelerated operators
as integral parts of the DB engine and thus incurs zero system overhead in
processing TCU-accelerated queries. In contrast, query analyzers like AIDA~\cite{AIDA}
that rely on external parallel libraries from different language frameworks
from the query engine always lead to redundant memory copies
that are especially significant in our use cases. 
Compared with proposals relying on SQL extensions that introduce data type labels
(e.g., vector and matrix) to support LA queries~\cite{scalaleLA} or new
query languages~\cite{10.1145/3331445}, \TCUDB{} does not require any change to the SQL.
\vspace*{\lineskip}
\noindent
\textbf{Entity Matching and PageRank. } 
A major challenge in EM~\cite{elmagarmid2006duplicate,dong2013big,christophides2015entity,konda2016magellan}  is in the blocking phase~\cite{konda2016magellan,gagliardelli2019sparker,papadakis2020blocking}
to reduce the number of candidate pairs to be matched by heuristics specified as natural joins.
Our case study demonstrates that \TCUDB{} delivers over 300\x{} speedup for blocking queries compared to a GPU-accelerated $\mathtt{HashJoin}$ implementation.
This indicates the potential of building scalable EM systems with \TCUDB{} as the backend.

PageRank is a graph-based ranking algorithm with applications from web searches to basic science
(see \cite{gleich2015pagerank} for a survey). PageRank is also commonly used in benchmarks of graph databases~\cite{mitliagkas2015frogwild,needham2019graph,deutsch2020aggregation}.
While there has been an effort to accelerate PageRank (and other graph analytic queries) using GPUs~\cite{wu2010efficient,rungsawang2012fast,shi2018graph},
to our knowledge, \TCUDB{} is the first to attempt to accelerate PageRank using TCUs.

\section{Conclusion}
\label{sec:conclude}
\ignore{
\vvspace{-0.5mm}
In this paper, we proposed \HDB, an efficient, scalable heterogenous data analytics system. \HDB{} is the first GPU-based data analytics that can scale up to support terabyte input. \HDB{} reaches high performance by fixing the huge imbalance between GPU kernel and I/O using compression and peer-to-peer transfer path.
\HDB~ uses a streaming execution model to process data sets larger than the GPU  memory. Our comprehensive experiments have demonstrated the superiority of \HDB~ in terms of both scalability and performance.
}
This paper proposes, implements and evaluates \TCUDB{}, an efficient database query engine with TCUs, an emerging
type of AI/ML hardware accelerator presented in modern GPU architectures. 
This paper identifies query patterns that match TCUs' acceleration model. 
Through solving technical difficulties such as remapping inputs and limited precision,
the resulting \TCUDB{} shows ours 
achieves up to 288\x{} speedup against the baseline GPU-accelerated DB
engine. 
The performance gain of \TCUDB{} over conventional
GPU-based DB engines indicates a strong performance scaling in new GPU architectures.
For future work, we plan to extend \TCUDB{} by exploring more potential workloads
and addressing the complex query optimization problem with multiple accelerators
of different types.

\section*{Acknowledgments}
The authors would like to thank the anonymous reviewers
for their helpful comments.  
This work was sponsored by the two National Science Foundation (NSF) awards,
CNS-1940048 and CNS-2007124.
This work was also supported by new faculty start-up funds from University of California, Riverside. 



\bibliographystyle{ACM-Reference-Format}
\bibliography{paper}


%
\end{document}